\documentclass[twocolumn,preprintnumbers,amsmath,amssymb]{revtex4}   

\usepackage{graphicx}
\usepackage{dcolumn}
\usepackage{bm}

\usepackage{esint}

\begin{document}

\title{Larkin-Ovchinnikov state of superconducting Weyl metals: Fundamental differences between pairings restricted and extended in the $\emph{\textbf{k}}$-space}

\author{Lei Hao$^{1,2}$, Rui Wang$^{3}$, Pavan Hosur$^{1}$, and C. S. Ting$^{1}$}
\address{$^1$Department of Physics and Texas Center for Superconductivity, University of Houston, Houston, Texas 77204, USA  \notag \\
$^2$Department of Physics, Southeast University, Nanjing 210096, China \notag \\
$^3$National Laboratory of Solid State Microstructures, Collaborative Innovation Center of Advanced Microstructures,
and Department of Physics, Nanjing University, Nanjing 210093, China}

\date{\today}

\begin{abstract}
Two common approaches of studying theoretically the property of a superconductor are shown to have significant differences, when they are applied to the Larkin-Ovchinnikov state of Weyl metals. In the first approach the pairing term is restricted by a cutoff energy to the neighborhood of the Fermi surface, whereas in the second approach the pairing term is extended to the whole Brillouin zone. We explore their difference by considering two minimal models for the Weyl metal. For a model giving a single pair of Weyl pockets, both two approaches give a partly-gapped (fully-gapped) bulk spectrum for small (large) pairing amplitude. However, for very small cutoff energy, a portion of the Fermi surface can be completely unaffected by the pairing term in the first approach. For the other model giving two pairs of Weyl pockets, while the bulk spectrum for the first approach can be fully gapped, the one from the second approach has a robust line node, and the surface states are also changed qualitatively by the pairing. We elucidate the above differences by topological arguments and analytical analyses. A factor common to both of the two models is the tilting of the Weyl cones which leads to asymmetric normal state band structure with respect to the Weyl nodes. For the Weyl metal with two pairs of Weyl pockets, the band folding leads to a double degeneracy in the effective model, which distinguishes the pairing of the second approach from all others.
\end{abstract}


\maketitle

\section{\label{sec:Introduction}Introduction}

In the Bardeen-Cooper-Schrieffer (BCS) theory of superconductivity, the attractive interaction responsible for the formation of Cooper pairs is mediated by electron-phonon interactions \cite{bcs}. Since the effective interaction is attractive only within a thin shell around the Fermi surface (the width of which is of the order the Debye frequency $\omega_{D}$), BCS approximated the pairing interaction as a constant within the pairing shell
\begin{equation}
H_{int}=V\sum\limits_{\mathbf{k}_{1},\mathbf{k}_{2},\mathbf{q}} c^{\dagger}_{\mathbf{k}_{1}\sigma}c^{\dagger}_{\mathbf{k}_{2}\sigma'}c_{\mathbf{k}_{2}-\mathbf{q},\sigma'}c_{\mathbf{k}_{1}+\mathbf{q},\sigma}, \notag
\end{equation}
where $\sigma$ and $\sigma'$ label the two spin states of an electron and are summed over, the four electron operators are all subject to the restriction of $|\epsilon-\mu|<\omega_{D}$, a unit volume of the system is assumed. This is the starting point of standard weak-coupling analysis of the superconducting transition. By further restricting $\mathbf{k}_{2}=-\mathbf{k}_{1}$ and making mean-field decoupling to the interaction term, the weak-coupling analysis of the BCS transition temperature and condensation energy can be performed in an analytical manner. The BCS pairing interaction, however, is not gauge-invariant \cite{anderson58,nambu60}. A simple recipe for restoring the gauge invariance in studying the electromagnetic properties of superconductors is to remove the constraint in the summation over single-particle states and extend the pairing interaction to the full Brillouin zone (BZ), which then takes the form
\begin{equation}
\tilde{H}_{int}=V\int c^{\dagger}_{\sigma}(\mathbf{r})c^{\dagger}_{\sigma'}(\mathbf{r})c_{\sigma'}(\mathbf{r})c_{\sigma}(\mathbf{r})d\mathbf{r}. \notag
\end{equation}
Shifting from $H_{int}$ to $\tilde{H}_{int}$ was known to induce only minor errors (e.g., the correction to the gap function is of order $T_{c}/\omega_{D}$, where $T_{c}$ is the superconducting transition temperature, see Sec.34.2 of Ref. \cite{abrikosov63}), for the BCS state. Therefore, the two interaction terms and the resulting mean-field superconducting states were usually regarded as equivalent and the choices among them were made depending on the nature of the problem: E.g., in terms of $H_{int}$ when comparing condensation energy of different pairing channels and in terms of $\tilde{H}_{int}$ in studying electromagnetic properties and performing numerical calculations on a lattice. There are also many works which transfer freely between them in a single study \cite{abrikosov63,cho12,li15,rwang16}. In intrinsically strongly correlated systems where the pairing interaction usually takes a form similar to $\tilde{H}_{int}$, the similar trick is to adopt a pairing interaction of the form $H_{int}$ in comparing mean-field energies of various pairing channels.

The above qualitative equivalence between $H_{int}$ and $\tilde{H}_{int}$ and the corresponding mean-field theories were also assumed to be true implicitly in studying superconducting states other than the BCS state, such as the Fulde-Ferrell-Larkin-Ovchinnikov (FFLO) state \cite{ff,lo}. The aim of the present work is to point out that fundamental differences can arise when studying the LO state in terms of the two different forms of pairing interactions and the resulting mean-field theories. The mean-field superconducting state arising from a pairing interaction like $H_{int}$ contains only pairing correlations that couple states in the neighborhood of the Fermi surface. It will be named as \emph{restricted in $\mathbf{k}$-space (i.e., the BZ) pairing}, or RBZP for short. The mean-field superconducting state corresponding to a pairing interaction like $\tilde{H}_{int}$, on the other hand, gives rise to real-space pairing uniformly distributed in the whole lattice. After making Fourier transformation to the momentum space, it corresponds to a pairing extended throughout the whole BZ \cite{baskaran87,kotliar88,si08,seo08,zhu01,zhou10}. This pairing will be named as \emph{extended in $\mathbf{k}$-space pairing}, or EBZP. While the RBZP seems appropriate for pairings mediated by the electron-phonon interaction, the EBZP could be better suited for pairings in strongly correlated systems such as cuprates and iron pnictides \cite{baskaran87,kotliar88,si08,seo08,zhu01,zhou10}.

In contrast to the BCS state where the Cooper pairs have zero center-of-mass momentum, in the FFLO state the Cooper pairs have nonzero center-of-mass momentum. If all Cooper pairs of the FFLO state share a single center-of-mass momentum, the FFLO state is also called the Fulde-Ferrel state (FF state) \cite{ff}. In other systems, the Cooper pairs of the FFLO state can be grouped into one or several sets, the center-of-mass momenta of the Cooper pairs within each set take values from two nonzero opposite wave vectors. This latter form of the FFLO state is also called the Larkin-Ovchinnikov state (LO state) \cite{lo}. Proposed more than $50$ years ago and has long been elusive from direct detection, the FFLO state has gained enormous attention recently in several different research fields. One is the experimental signal in heavy fermion superconductors and organic superconductors under magnetic field, which were claimed to be consistent with the FFLO state \cite{bianchi03,radovan03,singleton00,mayaffre14,shimozawa16}. Similar earlier works include two-dimensional superconductors under an in-plane exchange field \cite{wang06,zhou09} and heterostructures consisting of an $s$-wave superconductor and a ferromagnetic metal \cite{buzdin05,demler97}. The second is the theoretical proposal in cold atom systems, in which the interplay between spin-orbit interaction and magnetic field (exchange field) were argued to give FFLO states \cite{chen09,qu13,zhang13,wu13,xu14,hu14,chan14}. The third is the superconducting state of Weyl metal (doped Weyl semimetal), in which the FFLO state is regarded as a strong competitor to the conventional BCS state \cite{cho12,wei14,bednik15,zhou15,kim16,meng12,hosur14,li15,lu15,rwang16,shivamoggi13,maciejko14,yang14,kobayashi15}.

In the Weyl metals that we focus on in this work, the Fermi surface consists of an even number of disconnected Fermi pockets (Weyl pockets) distributed symmetrically with respect to the center of the BZ \cite{murakami07,wan11,yang11,burkov11}. In these systems, the LO state is favored when the intra-pocket pairing interaction outweighs the inter-pocket (i.e., BCS) pairing interaction. This mechanism has also been discussed in the context of graphene \cite{roy10,roy13,murray14}. Several theoretical calculations have found the LO state as the ground state of the superconducting phase of Weyl metal \cite{cho12,wei14,kim16}. But there are also works claiming that the BCS state is the true ground state \cite{bednik15,zhou15}.

The main purpose of this work is not to identify the leading pairing instability of the Weyl metal. Instead, we take the LO state of the Weyl metal as a model system to illustrate the fundamental differences between the two kinds of mean-field pairings, the RBZP-LO (RLO) state versus the EBZP-LO (ELO) state. We consider two different Weyl metals, the Fermi surfaces of which consist of a single pair and two pairs of Weyl pockets, respectively. For the Weyl metal with a single pair of Weyl pockets, the quasiparticle spectrum is fully gapped (partly gapped) for both the RLO state and the ELO state when the pairing amplitude is large (small). On the other hand, for very small cutoff energy, a portion of the Fermi surface can be completely unaffected by the pairing term in the RLO state, which we call a momentum-space phase separation in the states on the Fermi surface. For the LO state of the Weyl metal with two pairs of Weyl pockets, while the bulk quasiparticle spectrum can be fully gapped for the RLO state, there is a robust line of nodes in the bulk quasiparticle spectrum of the ELO state, persisting independent of the pairing amplitudes.

The remaining part of the paper is organized as follows. In Sec.II, we define the two models and the corresponding LO pairings. In Sec. III, through a general analysis, we identify two cases where the RBZP and EBZP pictures could be different. In one case, the tilting of the Weyl cone leads to a Fermi pocket asymmetric with respect to the Weyl node, which is responsible for the momentum-space phase separation in the RLO state. In the other case, by utilizing the connection between the nodal structure of the superconducting gap and the total Berry flux of the particle and hole states on the Fermi surface, we show that the ELO state of the second model can host robust nodes in the quasiparticle spectrum. The numerical results of the quasiparticle spectra are then presented in Sec.IV. The differences between the two pictures and between the two models are elucidated further. Finally, we summarize the main results and discuss possible implications of the present work in Sec.V. Several technical details omitted in the main text are added as appendices.

\section{\label{sec:model}Models and pairings}

Two models of Weyl (semi-)metal will be considered in this work to illustrate the physics. One model breaks time-reversal symmetry and has a single pair of Weyl nodes. The other is time-reversal symmetric but breaks the inversion symmetry and has two pairs of Weyl nodes.

\emph{Time-reversal-symmetry broken Weyl metal with a single pair of Weyl nodes.{\textemdash}} We consider the following minimal two-band model\cite{cho12,yang11}
\begin{eqnarray}
h_{0}(\mathbf{k})&=&t(s_{x}\sin k_{x}+s_{y}\sin k_{y})+t_{z}(\cos k_{z}-\cos Q)s_z   \notag \\
&&+m(2-\cos k_{x}-\cos k_{y})s_z-\mu s_0.
\end{eqnarray}
$s_{0}$ is the $2\times2$ unit matrix and $s_{\alpha}$ ($\alpha=x,y,z$) are the Pauli matrices. The basis is taken as $\phi^{\dagger}_{\mathbf{k}}=[c^{\dagger}_{\uparrow}(\mathbf{k}),c^{\dagger}_{\downarrow}(\mathbf{k})]$. $\uparrow$ and $\downarrow$ label the two degrees of freedom and are understood as the two spins of the electron.
The model can be written compactly as
\begin{equation}
h_{0}(\mathbf{k})=\sum\limits_{\alpha=x,y,z}d_{\alpha}(\mathbf{k})s_{\alpha}-\mu s_0,
\end{equation}
where $d_{x}(\mathbf{k})=t\sin k_{x}$, $d_{y}(\mathbf{k})=t\sin k_{y}$, and $d_{z}(\mathbf{k})=t_{z}(\cos k_{z}-\cos Q)+m(2-\cos k_{x}-\cos k_{y})$. The eigenenergies of $h_{0}(\mathbf{k})$ are
\begin{equation}
E_{\alpha}(\mathbf{k})=\alpha\sqrt{d_{x}^{2}(\mathbf{k})+d_{y}^{2}(\mathbf{k})+d_{z}^{2}(\mathbf{k})}-\mu,
\end{equation}
where $\alpha=\pm$. When $m$ is large, only one pair of Weyl nodes exists in the BZ, at $\mathbf{P}_{\pm}=(0,0,\pm Q)$. As a typical set of parameters, we take $t=-1$, $t_{z}=-2$, $m=1$, and $Q=\pi/4$ \cite{zhou15}, unless otherwise specified. More discussions on the symmetries and topological properties of the model can be found in Appendix A.

The model can be considered as describing a single pair of Weyl nodes in a magnetic Weyl semimetal, such as the pyrochlore iridate YIr$_{2}$O$_{7}$ \cite{wan11,yang11}. It is also topologically equivalent to the model for the Weyl semimetal realized in a multilayer consisting of thin films of magnetically doped topological insulators and ordinary insulators \cite{burkov11}. A property of the band structure, which is important to our following analysis, is that $E_{\alpha}(q_{x},q_{y},\pm Q+q_{z})\ne E_{\alpha}(-q_{x},-q_{y},\pm Q-q_{z})$ for $q_{z}\ne0$. This relation leads to the tilting of the Weyl cones along $k_{z}$. In a loose sense, we may call this band structure which is asymmetric with respect to the Weyl nodes as a band structure with broken local (in momentum space) inversion symmetry.

Now we turn on the LO pairing term in the model. For small chemical potential $\mu$ that we focus on, the Fermi surface consists of two disconnected spheroidal pockets centering at $\mathbf{P}_{+}$ and $\mathbf{P}_{-}$, respectively. For states close to $\mathbf{P}_{\alpha}$ ($\alpha=\pm$), introducing the relative momenta $\mathbf{q}$, the basis vector can be denoted as $\phi^{\dagger}_{\mathbf{k}}=\phi^{\dagger}_{\mathbf{P}_{\alpha}+\mathbf{q}}=\phi^{(\alpha)\dagger}_{\mathbf{q}}$. Consider the simplest realization of the LO state, the $s$-wave singlet pairing. It is written as\cite{cho12,bednik15}
\begin{equation}
H^{FFLO}=\frac{1}{2}\sum\limits_{\mathbf{q},\alpha}\Delta_{\alpha} \phi^{(\alpha)\dagger}_{\mathbf{q}}is_{y}[\phi^{(\alpha)\dagger}_{-\mathbf{q}}]^{\text{T}}+\text{H.c.},
\end{equation}
where $\Delta_{\alpha}$ ($\alpha=\pm$) is the pairing amplitude for states close to the $\alpha$-th Weyl pocket (or, Fermi pocket) and will be taken as a real constant number. $\text{H.c.}$ means taking the Hermitian conjugate of the terms written out explicitly.

In Eq.(4), there is an ambiguity in the $\mathbf{q}$-summation. In the RBZP picture \cite{bcs}, the $\mathbf{q}$-summation is restricted to the neighborhood of the Fermi surface, within an energy cutoff $\omega_{c}$. On the other hand, in the EBZP picture, the $\mathbf{q}$-summation in Eq.(4) is extended over the whole BZ ($\pm\mathbf{q}+\mathbf{P}_{\alpha}$ are also extended over the whole BZ, by the periodicity of the BZ) \cite{zhou15,cho12,bednik15}. While in the first interpretation (RLO) the pairing can be considered as intra-node pairing in the strict sense, in the second interpretation (ELO) a single wave vector in the BZ in fact couples by the pairing interaction to two wave vectors, which are symmetrical to it with respect separately to the two Weyl nodes, $\mathbf{P}_{+}$ and $\mathbf{P}_{-}$.

\emph{Time-reversal-symmetric and inversion-asymmetric Weyl metal with two pairs of Weyl nodes.{\textemdash}} In the presence of time-reversal symmetry, the inversion symmetry must be broken to get a Weyl semimetal. In this case, the minimal number of Weyl nodes is four. One relevant model was proposed by Hosur \emph{et al} as a description of Na$_{3}$Bi with broken inversion symmetry \cite{hosur14}. Regularizing the model to a cubic lattice and defining the basis as $\psi^{\dagger}_{\mathbf{k}}=[c^{\dagger}_{1\uparrow}(\mathbf{k}),c^{\dagger}_{2\uparrow}(\mathbf{k}) ,c^{\dagger}_{1\downarrow}(\mathbf{k}),c^{\dagger}_{2\downarrow}(\mathbf{k})]$ (1 and 2 label the two orbitals, $\uparrow$ and $\downarrow$ label the two spin states), it is written as
\begin{equation}
H_{1}=\sum\limits_{\mathbf{k}}\psi^{\dagger}_{\mathbf{k}}h_{1}(\mathbf{k})\psi_{\mathbf{k}} =\sum\limits_{\mathbf{k}}\psi^{\dagger}_{\mathbf{k}}\begin{pmatrix} h_{\uparrow}(\mathbf{k}) & 0 \\
0 & h_{\downarrow}(\mathbf{k}) \end{pmatrix}\psi_{\mathbf{k}},
\end{equation}
where\cite{footnote1}
\begin{equation}
h_{s}(\mathbf{k})=h_{s}'(\mathbf{k})+M(2-\cos k_{x}-\cos k_{y})\sigma_{z},
\end{equation}
and
\begin{eqnarray}
h_{s}'(\mathbf{k})&=&\xi_{\mathbf{k}}\sigma_{0}+t_{z}'(\cos k_{z}-\cos Q)\sigma_{z}+\alpha_{s}t_{z}''\sin k_{z}\sigma_{z}   \notag \\
&&+t'(\alpha_{s}\sin k_{y}\sigma_{x}-\sin k_{x}\sigma_{y}),
\end{eqnarray}
$\alpha_{s}=1$ for $s=\uparrow$, and $\alpha_{s}=-1$ for $s=\downarrow$. $\sigma_{0}$ and $\sigma_{i}$ ($i=x,y,z$) are the unit matrix and pauli matrices in the orbital subspace. $\xi_{\mathbf{k}}=-2t_{1}(\cos k_{x}+\cos k_{y})-2t_{2}\cos k_{z}-\mu$. The four bands of the model are
\begin{equation}
E_{s\beta}=\xi_{\mathbf{k}}+\beta\sqrt{f^{2}_{xy}(\mathbf{k})+g^{2}_{sz}(\mathbf{k})},
\end{equation}
where $\beta=\pm$, $f_{xy}(\mathbf{k})=\sqrt{t'^{2}(\sin^{2}k_{x}+\sin^{2}k_{y})}$, and $g_{sz}(\mathbf{k})=t_{z}'(\cos k_{z}-\cos Q)+\alpha_{s}t_{z}''\sin k_{z}+M(2-\cos k_{x}-\cos k_{y})$. The Weyl nodes are determined by $\sin k_{x}=\sin k_{y}=g_{sz}(\mathbf{k})=0$. The term proportional to $M$ is added to ensure that we only have Weyl nodes along $(0,0,k_{z})$. For sufficiently large $M$, the above condition leads to
\begin{equation}
k_{x}=k_{y}=g_{sz}(0,0,k_{z})=0.
\end{equation}
When $t_{z}''=0$, the model has inversion symmetry, we have a pair of Dirac nodes at $(0,0,\pm Q)$. As we turn on the term proportional to $t_{z}''$, the inversion symmetry of the model is broken, each Dirac node is split into two Weyl nodes. One particular set of parameters is
\begin{equation}
Q=\frac{\pi}{2},   \hspace{0.5cm}   t_{z}''=t_{z}',
\end{equation}
for which the two pairs of Weyl nodes are at $\pm\mathbf{Q}_{1}=\pm(0,0,\frac{\pi}{4})$ and $\pm\mathbf{Q}_{2}=\pm(0,0,\frac{3\pi}{4})$. Among the four Weyl nodes, the $-\mathbf{Q}_{1}$ and $\mathbf{Q}_{2}$ nodes are associated with spin-$\uparrow$ electrons, and the $\mathbf{Q}_{1}$ and $-\mathbf{Q}_{2}$ nodes are associated with spin-$\downarrow$ electrons. For convenience of later reference, we also label the four Weyl nodes consecutively in the order of increasing $k_{z}$ as $\mathbf{P}_{1}=-\mathbf{Q}_{2}$, $\mathbf{P}_{2}=-\mathbf{Q}_{1}$, $\mathbf{P}_{3}=\mathbf{Q}_{1}$, and $\mathbf{P}_{4}=\mathbf{Q}_{2}$.

Without losing generality, we take the following set of parameters of the above model in all subsequent calculations unless otherwise stated, $t_{1}=0$, $t_{2}=0$, $t'=2$, $t'_{z}=t''_{z}=\sqrt{2}$, $M=2$, and $Q=\pi/2$. In the absence of the $t_{1}$ and $t_{2}$ terms, the band structure has an accidental particle-hole symmetry for $\mu=0$, which also holds for the previous model. Symmetries and topological properties of the model can be found in Appendix A.

Introducing the relative momentum $\mathbf{q}$, the basis vector for states close to $\mathbf{P}_{\alpha}$ ($\alpha$=1, $...$, 4) can be denoted as $\psi^{\dagger}_{\mathbf{k}}=\psi^{\dagger}_{\mathbf{P}_{\alpha}+\mathbf{q}}=\psi^{(\alpha)\dagger}_{\mathbf{q}}$. Since states close to each Weyl node are fully spin-polarized, the intra-node FFLO pairing can only occur in the parallel-spin channel. For the present two-orbital model, there are two approaches of fulfilling the Fermi statistics of the two electrons forming a pair. One kind of pairing is written as
\begin{equation}
H_{1\alpha}^{FFLO}=\frac{1}{2}\sum\limits_{\mathbf{q}}\psi^{(\alpha)\dagger}_{\mathbf{q}}\Delta_{\alpha}[\mathbf{d}(\mathbf{q})\cdot\mathbf{s}]is_{y}\sigma_{0} [\psi^{(\alpha)\dagger}_{-\mathbf{q}}]^{\text{T}}+\text{H.c.},
\end{equation}
where $\alpha=1,...,4$ labels the Weyl nodes, $\mathbf{d}(-\mathbf{q})=-\mathbf{d}(\mathbf{q})$ [$d_{3}(\mathbf{q})=0$] is odd in the relative wave vector. Another kind of pairing can be written as
\begin{equation}
H_{2\alpha}^{FFLO}=\frac{1}{2}\sum\limits_{\mathbf{q}}\psi^{(\alpha)\dagger}_{\mathbf{q}}\Delta_{\alpha}[f_{\mathbf{q}}s_{0}+g_{\mathbf{q}}s_{z}]i\sigma_{y} [\psi^{(\alpha)\dagger}_{-\mathbf{q}}]^{\text{T}}+\text{H.c.},
\end{equation}
where $f_{-\mathbf{q}}=f_{\mathbf{q}}$ and $g_{-\mathbf{q}}=g_{\mathbf{q}}$ are both even functions of the relative wave vector. While the first kind of pairing appears naturally for parallel-spin pairing, the second kind of pairing can also be the most stable state if interorbital pairing interaction outweighs the intraorbital pairing interaction. Therefore, as the simplest possible pairing for the above model, we consider the following FFLO pairing
\begin{equation}
H_{\nu\alpha}^{FFLO}=\frac{1}{2}\sum\limits_{\mathbf{q}}\psi^{(\alpha)\dagger}_{\mathbf{q}}\Delta_{\nu\alpha}s_{\nu}i\sigma_{y} [\psi^{(\alpha)\dagger}_{-\mathbf{q}}]^{\text{T}}+\text{H.c.},
\end{equation}
where $\nu=0$ or $3$, and $\Delta_{\nu\alpha}$ ($\alpha$=1, $...$, 4) will be treated as constants with no wave-vector dependence.

In the RLO state, the summation over $\mathbf{q}$ in Eq.(13) is restricted by the cutoff energy $\omega_{c}$ to the neighborhood of the Fermi surface. In the ELO state, however, the $\mathbf{q}$-summation in Eq.(13) for each Weyl pocket is extended to the whole BZ. Again, because of the periodicity of the BZ, the wave vectors $\pm\mathbf{q}+\mathbf{P}_{\alpha}$ ($\alpha=1,...,4$) are also extended through the whole BZ. Therefore, states of the four Weyl pockets are coupled together in a highly nontrivial manner by the pairing term of the ELO state.

\section{general analyses}

Before embarking on explicit calculations of the quasiparticle spectrum, we try to figure out based on general considerations why the RBZP and EBZP pictures for the same BCS pairing can be regarded as equivalent, and how this equivalence can be broken.

To answer the above question, a crucial observation is that the physical properties of a superconductor are determined only by the low-energy quasiparticle states. Conventional BCS pairing forms in a centrosymmetric band structure in which $E_{n}(\mathbf{k})=E_{n}(-\mathbf{k})$ is satisfied, where $n$ is the index for the energy bands and the energies are measured relative to the chemical potential. In the Nambu basis, the formation of Cooper pairs in the BCS channel amounts to the emergence of a pairing term coupling the particle states with wave vector $\mathbf{k}$ and energy $E_{n}(\mathbf{k})$ to hole states with wave vector $-\mathbf{k}$ and energy $-E_{n}(-\mathbf{k})$. The influence of the pairing term to the spectrum is most significant when $E_{n}(\mathbf{k})=-E_{n}(-\mathbf{k})$. In these crossing points between the particle band and the hole band, a gap opens and separates the two low-energy quasiparticle bands. For a conventional BCS superconductor with a normal state band structure satisfying $E_{n}(\mathbf{k})=E_{n}(-\mathbf{k})$, the condition $E_{n}(\mathbf{k})=-E_{n}(-\mathbf{k})$ leads to $E_{n}(\mathbf{k})=0$, which is simply the definition of the Fermi surface. In these prototypical examples of pairings, to which the BCS theory was initially applied, the RBZP and EBZP pictures indeed agree with each other on the low-energy quasiparticle states. The EBZP picture simply extends the pairing correlations already existing in the RBZP picture to high-energy states far away from the Fermi surface, which do not influence low-energy properties of the superconductor. This underlies the previous general consensus that EBZP and RBZP can be considered as equivalent and chosen based on the convenience of study for the specific property under consideration \cite{abrikosov63}.

\begin{figure}[!htb]\label{fig1}
\centering
\hspace{-5cm} {\textbf{(a)}}\\
\includegraphics[width=6.5cm,height=4.2cm]{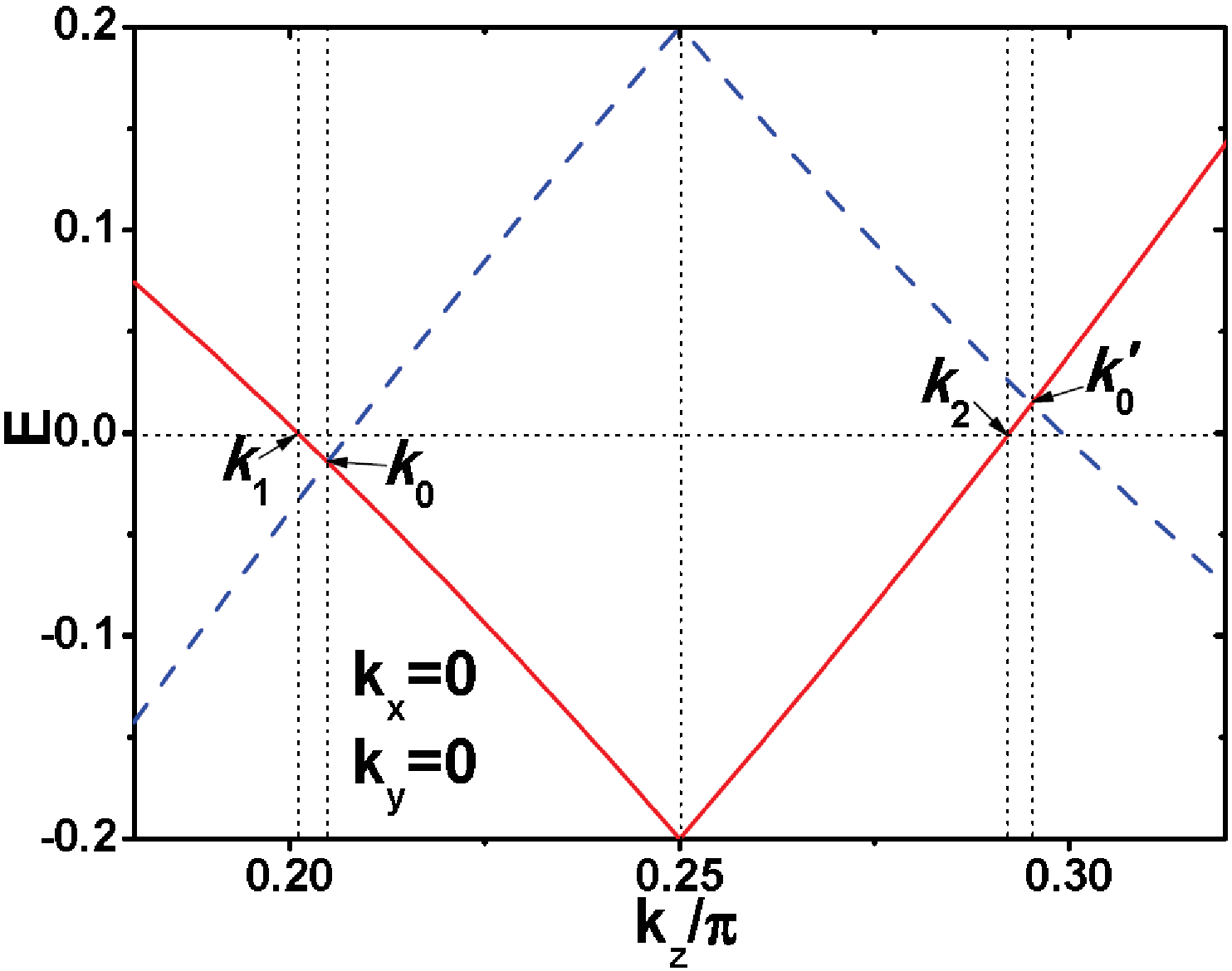} \\ \vspace{-0.05cm}
\hspace{-5cm} {\textbf{(b)}}\\
\includegraphics[width=6.5cm,height=4.2cm]{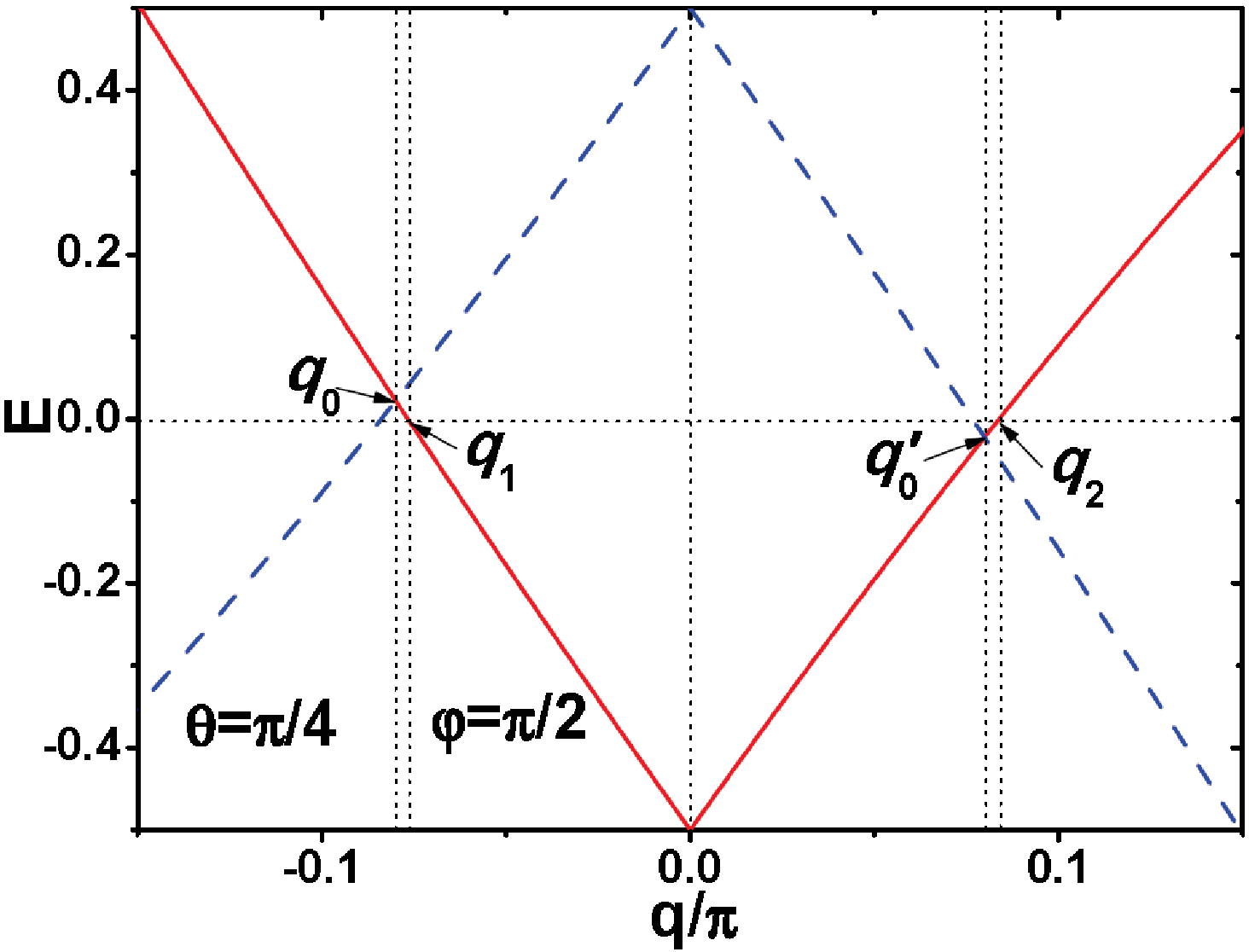}  \\
\caption{(Color online) (a) Plot of the particle band $E_{+}(\mathbf{k})$ (red solid curve) and the hole band $-E_{+}(2\mathbf{P}_{+}-\mathbf{k})$ (blue dashed curve), according to Eq.(3), along $k_{z}$ and close to $\mathbf{P}_{+}$. (b) Plot of the particle band $E_{\downarrow+}(\mathbf{q}+\mathbf{P}_{3})$ (red solid curve) and the hole band $-E_{\downarrow+}(-\mathbf{q}+\mathbf{P}_{3})$ (blue dashed curve), according to Eq.(8), along a path defined by $(q_{x},q_{y},q_{z})=q(\sin\theta\cos\varphi,\sin\theta\sin\varphi,\cos\theta)$ with $\theta=\pi/4$ and $\varphi=\pi/2$. $\mu=0.2$ for (a) and $\mu=0.5$ for (b). Other parameters are as defined in Sec.II. Two common features of (a) and (b) are, (1) at the Fermi points ($\mathbf{k}_{1}$, $\mathbf{k}_{2}$, $\mathbf{q}_{1}$, $\mathbf{q}_{2}$) where the particle bands cross the Fermi level, the corresponding hole states have nonzero energies; (2) the crossing points between the particle band and the hole band ($\mathbf{k}_{0}$, $\mathbf{k}_{0}'$, $\mathbf{q}_{0}$, $\mathbf{q}_{0}'$) are away from the Fermi level $E=0$.}
\end{figure}

From the above analysis, an obvious deviation from the above picture for conventional BCS pairing occurs when $E_{n}(\mathbf{k})\ne E_{n}(-\mathbf{k})$.
Namely, when the band structure is not centrosymmetric \cite{note2}. Consider the BCS pairing formed in such a noncentrosymmetric band structure, the pairing term still couples the particle states with wave vector $\mathbf{k}$ and energy $E_{n}(\mathbf{k})$ to hole states with wave vector $-\mathbf{k}$ and energy $-E_{n}(-\mathbf{k})$. However, a crucial difference from the conventional BCS state is that the wave vectors where the particle band and the hole band crosses are most generally not on the Fermi surface. Explicitly, combining $E_{n}(\mathbf{k}_{0})=-E_{n}(-\mathbf{k}_{0})$ with  $E_{n}(\mathbf{k}_{0})\ne E_{n}(-\mathbf{k}_{0})$, we have $E_{n}(\mathbf{k}_{0})\ne0$ for a wave vector $\mathbf{k}_{0}$ where the crossing happens. (For pictorial understanding, see Fig.1 and consider the Weyl nodes in the LO state as the center of the BZ in the BCS state.) Therefore, the corresponding gap opening occurs at nonzero energy. Consider the cases of very small pairing amplitudes, close to the transition temperature or for superconductors with very low transition temperature. In the RBZP picture and for fixed pairing interaction, the pairing amplitude scales linearly with $\omega_{c}$ \cite{bcs,abrikosov63}. Therefore, there are cases where $\omega_{c}$ is also very small. For very small $\omega_{c}$, there can exist states on the Fermi surface of the noncentrosymmetric band structure which satisfies $E_{n}(\mathbf{k})=0$ and $|E_{n}(-\mathbf{k})|>\omega_{c}$ simultaneously. Because in the RBZP picture the pairing correlation exists only when both the particle state and the hole state are within the pairing shell, a portion of the Fermi surface featured by states satisfying $E_{n}(\mathbf{k})=0$ and $|E_{n}(-\mathbf{k})|>\omega_{c}$ is completely unaffected by the pairing term. In the EBZP picture, however, the pairing is extended throughout the whole BZ and so all states are influenced by the pairing term. But for pairing amplitude smaller than $|E_{n}(\mathbf{k}_{0})|$, the gap opening at $\mathbf{k}_{0}$ defined above would be smaller than $|E_{n}(\mathbf{k}_{0})|$, and the Fermi level is outside of the gap region. There are thus zero-energy quasiparticle states close to $\mathbf{k}_{0}$. Therefore, on one hand the bulk quasiparticle spectra for both the RBZP and EBZP pictures are partly gapped for small pairing amplitude. On the other hand, the RBZP picture differs from the EBZP picture by the possible presence of Fermi surface portions that are completely unaffected by the pairing term, which can be considered as a \emph{phase separation in the momentum space} into superconducting and normal parts. Finally, when the pairing amplitude and the cutoff energy are both large, the two pictures should give qualitatively the same low-energy bulk spectrum.

The LO state of Weyl metal provides an ideal testing-ground for the above idea. In this state, the Cooper pairs are formed relative to the wave vector of the Weyl node, which is the center of the Fermi pocket surrounding that Weyl node. Then, most importantly, the Weyl cones are very generally tilted and therefore the band structure is not centrosymmetric with respect to the Weyl nodes. This tilting is related to the fact that Weyl nodes usually appear at ordinary points of the BZ, where there are no enough symmetries to \emph{prohibit} the tilting along all directions. In another word, the Weyl nodes situating at ordinary points of the BZ are always allowed to tilt along a certain direction, along which there is no symmetry to prohibit the tilting. As shown in Fig.1 are the particle and hole bands responsible for the intranode pairing of a certain Weyl pocket, for both the first model [Fig.1(a)] and the second model [Fig.1(b)]. It is clear from the two figures that, at the two Fermi points ($\mathbf{k}_{1}$ and $\mathbf{k}_{2}$ for the first model, $\mathbf{q}_{1}$ and $\mathbf{q}_{2}$ for the second model) of the particle band, the energies of the corresponding states on the hole band are nonzero. On the other hand, the crossing points between the particle band and the hole band ($\mathbf{k}_{0}$, $\mathbf{k}_{0}'$, $\mathbf{q}_{0}$, $\mathbf{q}_{0}'$) are at nonzero energies.

The difference identified above between the RBZP and EBZP pictures of a certain pairing depends crucially on the magnitude of the cutoff energy used for the RBZP picture. It thus appears that the above difference induced by the tilting of the Weyl nodes is quantitative by nature. From the above analysis, the same physics applies to the BCS pairing realized in a band asymmetric with respect to the center of the BZ. Following a heuristic argument, we could claim that there can be more fundamental differences between the two pictures, particularly, when they are applied to the LO state of the Weyl metals. For this purpose, we compare the topological properties of the RLO and ELO states of the two models defined in Sec.II.

The topological property of a normal metal can be determined by studying the Berry phase of the single-particle states constituting the Fermi surface \cite{haldane04}. Similarly, in the mean-field approximation, the topological property of a superconductor is determined by its quasiparticle spectrum. In terms of the Nambu basis and the associated Bogoliubov-de Gennes Hamiltonian, the low-energy quasiparticle states come from the particle states on the Fermi surface, hybridized with hole states through the pairing term. For pairings without internal structures, e.g., the spin-singlet pairing with constant pairing amplitude, the topological property of the superconducting state was known to be determined completely by the topological property of the states on the Fermi surface in the normal phase \cite{murakami03,li15}. In particular, the nodal structure of the superconducting gap was shown to be determined completely by the total Berry flux of the particle and hole states on the Fermi surface \cite{li15,murakami03}. Our LO pairings defined in Sec.II belong to the above case if we take the pairing amplitudes to be the same for all Weyl pockets. Therefore, we expect the nodal property of the quasiparticle spectrum in the LO state to be determined solely by the states on the Fermi surface. What follows we focus on this scenario for simplicity.

Different from previous works concentrating on the pairing term \cite{murakami03,li15}, we focus on the low-energy quasiparticle states. By trying to construct the eigenoperators for the corresponding quasiparticle bands in terms of the equation of motion method, we get an effective model which is directly related to one of the Fermi pockets and determines the ensuing quasiparticle states. The nodal structure of the quasiparticle spectrum is then inferred from the total Berry flux calculated from the effective model. This method is more straightforward and gives the same conclusion when applied to the BCS states studied in previous works \cite{murakami03,li15}.

First consider the LO state of the Weyl metal defined by Eqs.(1) and (3) with broken time reversal symmetry. The quasiparticle state for a wave vector $\mathbf{q}+\mathbf{P}_{\alpha}$ on the Weyl pocket around $\mathbf{P}_{\alpha}$ ($\alpha=\pm$) is constructed by starting from $\phi^{(\alpha)\dagger}_{\mathbf{q}}$, which is accompanied by a model $h_{0}(\mathbf{q}+\mathbf{P}_{\alpha})$. In the normal phase, the Fermi pocket around $\mathbf{P}_{\alpha}$ is determined by this model Hamiltonian alone. Turning on the LO pairing defined by Eq.(4), we see that $\phi^{(\alpha)\dagger}_{\mathbf{q}}$ is coupled to hole states. By calculating the commutator of $\phi^{(\alpha)\dagger}_{\mathbf{q}}$ with the Hamiltonian, we see that in the RLO state and in the ELO state $\phi^{(\alpha)\dagger}_{\mathbf{q}}$ is coupled separately to one and two hole operators. Specifically, in the RLO state, it is coupled to $\phi^{(\alpha)\text{T}}_{-\mathbf{q}}$. Introducing the basis $[\phi^{(\alpha)\dagger}_{\mathbf{q}}, \phi^{(\alpha)\text{T}}_{-\mathbf{q}}]$, the effective model determining the quasiparticle states derived from $\phi^{(\alpha)\dagger}_{\mathbf{q}}$ is
\begin{equation}
\begin{pmatrix} h_{0}(\mathbf{q}+\mathbf{P}_{\alpha}) & i\Delta_{\alpha}s_{y} \\
-i\Delta_{\alpha}s_{y} &  -h_{0}^{\text{T}}(-\mathbf{q}+\mathbf{P}_{\alpha}) \end{pmatrix}.
\end{equation}
In the ELO state, $\phi^{(\alpha)\dagger}_{\mathbf{q}}$ is coupled not only to $\phi^{(\alpha)\text{T}}_{-\mathbf{q}}$ but also to $\phi^{(\tilde{\alpha})\text{T}}_{-\mathbf{q}+2\mathbf{P}_{\tilde{\alpha}}}$, where $\tilde{\alpha}=-\alpha$ and $\mathbf{P}_{\tilde{\alpha}}=-\mathbf{P}_{\alpha}$. Introducing the basis $[\psi^{(\alpha)\dagger}_{\mathbf{q}}, \phi^{(\alpha)\text{T}}_{-\mathbf{q}}, \phi^{(\tilde{\alpha})\text{T}}_{-\mathbf{q}+2\mathbf{P}_{\tilde{\alpha}}}]$, the effective model determining the corresponding quasiparticle state is
\begin{equation}
\begin{pmatrix} h_{0}(\mathbf{q}+\mathbf{P}_{\alpha}) & i\Delta_{\alpha}s_{y}  & i\Delta_{\tilde{\alpha}}s_{y} \\
-i\Delta_{\alpha}s_{y} &  -h_{0}^{\text{T}}(-\mathbf{q}+\mathbf{P}_{\alpha}) & 0  \\
-i\Delta_{\tilde{\alpha}}s_{y} & 0 & -h_{0}^{\text{T}}(-\mathbf{q}+3\mathbf{P}_{\tilde{\alpha}}) \end{pmatrix}.
\end{equation}
According to our assumption, $\Delta_{\alpha}=\Delta$ is $\alpha$-independent and real. Since the spin-singlet pairing with constant pairing amplitude is trivial, nontrivial topological properties of the quasiparticle states are determined by the particle and hole states in the diagonal blocks of the above models. To focus on the topological property of the Fermi surface, we take $\mathbf{q}$ to span the Fermi pocket surrounding $\mathbf{P}_{\alpha}$. By diagonalizing $h_{0}(\mathbf{q}+\mathbf{P}_{\alpha})$, the single-particle eigenstates on the Fermi pocket surrounding $\mathbf{P}_{\alpha}$ ($\alpha=\pm$) can be defined as $a^{\dagger}_{\alpha}(\mathbf{q})=\phi^{(\alpha)\dagger}_{\mathbf{q}}u_{\alpha}(\mathbf{q})$, where $u_{\alpha}(\mathbf{q})$ is a two-component column vector representing the eigenfunction of the single-particle state. From the single-particle states at the Fermi surface, we can define the Berry connection in the normal phase as $\mathbf{A}_{\alpha}(\mathbf{q})=iu^{\dagger}_{\alpha}(\mathbf{q})\boldsymbol{\triangledown}_{\mathbf{q}}u_{\alpha}(\mathbf{q})$ ($\alpha=\pm$). The nontrivial topology of the normal phase is embodied in the quantized Berry flux through each Fermi pocket, namely
\begin{equation}
\oiint_{FS_{\alpha}}d\mathbf{q}\cdot\boldsymbol{\triangledown}_{\mathbf{q}}\times\mathbf{A}_{\alpha}(\mathbf{q})=4\pi C_{\alpha},
\end{equation}
where $C_{\alpha}$ is the Chern number for the Fermi pocket ($FS_{\alpha}$) around $\mathbf{P}_{\alpha}$, which is also the monopole charge of the Weyl node surrounded by it. $|C_{\alpha}|=1$ and $C_{+}=-C_{-}$ (see Appendix A). The total Berry flux for the corresponding quasiparticle bands has the additional contribution from the hole bands. For the RLO state with a quasiparticle spectrum determined by Eq.(14), the total Berry flux contributed by the hole band is
\begin{equation}
\oiint_{FS_{\alpha}}d\mathbf{q}\cdot\boldsymbol{\triangledown}_{\mathbf{q}}\times[-\mathbf{A}_{\alpha}(-\mathbf{q})]=-4\pi C_{\alpha}.
\end{equation}
Use has been made of the relation $i[\boldsymbol{\triangledown}_{\mathbf{q}}u^{\dagger}_{\alpha}(\mathbf{q})]u_{\alpha}(\mathbf{q}) =-iu^{\dagger}_{\alpha}(\mathbf{q})\boldsymbol{\triangledown}_{\mathbf{q}}u_{\alpha}(\mathbf{q})$.
Adding Eqs. (16) and (17), we get a null result \cite{li15}. For the ELO state with a quasiparticle spectrum determined by Eq.(15), the total Berry flux contributed by the hole band is
\begin{equation}
\oiint_{FS_{\alpha}}d\mathbf{q}\cdot\boldsymbol{\triangledown}_{\mathbf{q}}\times [-\mathbf{A}_{\alpha}(-\mathbf{q})-\mathbf{A}_{\tilde{\alpha}}(-\mathbf{q}+2\mathbf{P}_{\tilde{\alpha}})]=-4\pi C_{\alpha}.
\end{equation}
The contribution from the hole band $-h_{0}^{\text{T}}(-\mathbf{q}+3\mathbf{P}_{\tilde{\alpha}})$ vanishes, because as $\mathbf{q}+\mathbf{P}_{\alpha}$ and $-\mathbf{q}+\mathbf{P}_{\alpha}$ wrap around $\mathbf{P}_{\alpha}$ along $FS_{\alpha}$, $-\mathbf{q}+3\mathbf{P}_{\tilde{\alpha}}$ does not wrap around $\mathbf{P}_{\tilde{\alpha}}$ (of course, it neither wrap around $\mathbf{P}_{\alpha}$). Therefore, from a topological point of view, the low-energy quasiparticle states of the RLO state and the ELO state of the Weyl metal with a single pair of Weyl pockets are the same. In both of them, the two low-energy quasiparticle bands close to the Fermi level can be fully separated by a (local) gap.

The above analysis is extended straightforwardly to the RLO and ELO states of the Weyl metal defined by Eqs.(5) to (13). We still consider the simplest case of $\Delta_{\nu\alpha}=\Delta$ and real for $\nu$=$0$ and $\alpha$=$1,...,4$. The quasiparticle states for wave vectors on the Weyl pocket around $\mathbf{P}_{\alpha}$ ($\alpha$=$1,...,4$) is constructed by starting from $\psi^{(\alpha)\dagger}_{\mathbf{q}}$, which accompanies a model Hamiltonian $h_{1}(\mathbf{q}+\mathbf{P}_{\alpha})$ defined by Eq.(5). In the normal phase, the Fermi pocket around $\mathbf{P}_{\alpha}$ is determined by this model alone. Turning on the LO pairing defined by Eq.(13), the $\psi^{(\alpha)\dagger}_{\mathbf{q}}$ states are coupled to hole states. The corresponding eigenoperator is constructed by calculating the equation of motion of the $\psi^{(\alpha)\dagger}_{\mathbf{q}}$ basis operator. From the commutator of $\psi^{(\alpha)\dagger}_{\mathbf{q}}$ with the Hamiltonian, we see that $\psi^{(\alpha)\dagger}_{\mathbf{q}}$ couples to one set (four sets) of hole states in the RLO (ELO) state. Specifically, in the RLO state, it is coupled to $\psi^{(\alpha)\text{T}}_{-\mathbf{q}}$. In the basis $[\psi^{(\alpha)\dagger}_{\mathbf{q}}, \psi^{(\alpha)\text{T}}_{-\mathbf{q}}]$, the effective model determining the corresponding quasiparticle state is
\begin{equation}
\begin{pmatrix} h_{1}(\mathbf{q}+\mathbf{P}_{\alpha}) & \Delta_{0\alpha}s_{0}i\sigma_{y} \\
\Delta_{0\alpha}s_{0}(-i\sigma_{y}) &  -h_{1}^{\text{T}}(-\mathbf{q}+\mathbf{P}_{\alpha}) \end{pmatrix}.
\end{equation}
In the ELO state, $\psi^{(\alpha)\dagger}_{\mathbf{q}}$ couples to four sets of hole states through the four pairing terms in Eq.(13). Defining $h_{1\alpha}(\mathbf{q})=h_{1}(\mathbf{q}+\mathbf{P}_{\alpha})$, the effective model determining the quasiparticle states for the ELO state is a $5\times5$ block matrix. It can be obtained by supplementing Eq.(19) with three additional sets of pairing correlations to the remaining three Weyl pockets, and three corresponding diagonal block matrices $-h^{\text{T}}_{1\beta}(-\mathbf{q}+\mathbf{P_{\beta}}-\mathbf{P_{\alpha}})$ ($\beta=1,...,4$ and $\beta\ne\alpha$).

In regard to the low-energy quasiparticle states, the model defined by Eqs.(5)-(7) and the pairing term defined by Eq.(13) have three important features. Firstly, the two spin components ($\uparrow$ and $\downarrow$) are completely decoupled in both the model and the pairing term. Secondly, for each spin component, the pairing term is an orbital-singlet with constant pairing amplitude. Thirdly, as was shown in Sec.II, only a single spin component contributes to each Fermi pocket, with the other spin component contributing only to high-energy states far away from the Fermi surface. In combination of these features, the orbital-singlet LO pairing defined by Eq.(13) can be regarded as equivalent to the spin-singlet LO pairing of the previous model. Therefore the pairing term itself is trivial, the topological charge (total Berry flux) of the low-energy quasiparticle states obtained from solving Eq.(19) (and the corresponding effective model for the ELO state) is also determined completely by the corresponding topological charge of the normal phase.

The single-electron eigenstates on the Fermi pocket surrounding $\mathbf{P}_{\alpha}$ ($\alpha$=$1,...,4$) can be defined as $b^{\dagger}_{\alpha}(\mathbf{q})=\psi^{(\alpha)\dagger}_{\mathbf{q}}v_{\alpha}(\mathbf{q})$, where $v_{\alpha}(\mathbf{q})$ is a four-component column vector storing the state vector. The Berry connection in the normal phase is defined as $\mathbf{A}_{\alpha}(\mathbf{q})=iv^{\dagger}_{\alpha}(\mathbf{q})\boldsymbol{\triangledown}_{\mathbf{q}}v_{\alpha}(\mathbf{q})$. Similar to Eq.(16), we define the Chern number for the Fermi pocket surrounding the Weyl node $\mathbf{P}_{\alpha}$ as $C_{\alpha}$, which turns out to satisfy $|C_{\alpha}|=1$, $C_{1}=-C_{3}$ and $C_{2}=-C_{4}$ (see Appendix A). Defining $\bar{1}=3$ and $\bar{2}=4$, we can write $C_{\bar{\alpha}}=-C_{\alpha}$. For the RLO state, the total Berry flux for the low-energy quasiparticle states close to $\mathbf{P}_{\alpha}$ is
\begin{equation}
\oiint_{FS_{\alpha}}d\mathbf{q}\cdot\boldsymbol{\triangledown}_{\mathbf{q}}\times [\mathbf{A}_{\alpha}(\mathbf{q})-\mathbf{A}_{\alpha}(-\mathbf{q})]=4\pi C_{\alpha}-4\pi C_{\alpha}=0.
\end{equation}
For the ELO state, the total Berry flux for the low-energy quasiparticle states close to $\mathbf{P}_{\alpha}$ has three additional contributions and is
\begin{equation}
\oiint_{FS_{\alpha}}d\mathbf{q}\cdot\boldsymbol{\triangledown}_{\mathbf{q}}\times [\mathbf{A}_{\alpha}(\mathbf{q})-\sum\limits_{\beta=1}^{4}\mathbf{A}_{\beta}(-\mathbf{q}+\mathbf{P}_{\beta}-\mathbf{P}_{\alpha})].
\end{equation}
Same as Eq.(20), the net contribution from the particle and hole pockets for $\mathbf{P}_{\alpha}$ vanishes identically. Among the three additional contributions, two are from the opposite-spin Fermi pockets. These two contributions sum to zero, because of the relation $C_{\bar{\alpha}}=-C_{\alpha}$. The remaining contribution is from the pocket which has the same spin as the $\alpha$-th pocket but has the opposite Chern number. As a result, the total Berry flux determined by Eq.(21) is $-4\pi C_{\bar{\alpha}}=4\pi C_{\alpha}$, which is nonzero and is thus qualitatively different from the null result for the RLO state.

In conclusion, by considering the simplest case for which the pairing amplitudes on all Fermi pockets share the same value, we see that the total Berry flux for the low-energy quasiparticle states can have qualitative difference between the RBZP picture and the EBZP picture. From the connection between the nonvanishing total Berry flux and the existence of nodes in the quasiparticle spectrum \cite{li15}, it is thus natural to expect that, while the bulk quasiparticle spectrum for the RLO states of the two models and the ELO state of the model with one pair of Weyl pockets can be fully gapped, there should be robust bulk nodes in the quasiparticle spectrum of the ELO state of the second model with two pairs of Weyl pockets. As we will show later, this expectation is confirmed by explicit calculations.

Note that, the above topologically nontrivial result for Eq.(21) is closely related to the highly symmetric distributions of the four Weyl nodes in the BZ. However, even if we displace slightly the positions of the Weyl nodes by changing Eq.(10), the total Berry flux can still keep unchanged. This happens when the deviations in the positions of the four Weyl nodes from the ideal positions used above are compensated by a sufficiently large Fermi momentum of each pocket. Clearly, to validate the above arguments, the deviations of the Weyl nodes from the ideal positions should not be too large, in order to get a Fermi surface consisting of four disconnected Fermi pockets. Also note that, in the above topological arguments we have made a cutoff to the full equation of motion by retaining only states coupled to the initial basis operator up to linear order of the pairing amplitude (the first commutator with the Hamiltonian). As it will become clear from the more complete analyses in the next section, this approximate treatment captures correctly the qualitative differences between the various LO states.

\section{\label{sec:spectra}Quasiparticle spectrum for the RLO state and the ELO state}

In this section, we present the numerical results and more detailed analytical analyses of the quasiparticle energy spectrum for the LO (RLO and ELO) states of the two models. We focus mainly on the quasiparticle spectra of the two models in the configuration of films, which contain simultaneously the information of the bulk spectrum and the spectrum of possible surface states. While some of the results for the time-reversal-symmetry broken Weyl metal are already known \cite{cho12,zhou15,rwang16}, more interesting properties are revealed in this work. The results for the LO phase of the time-reversal-symmetric but inversion-asymmetric Weyl metal are new.

Note that the unit cell for the ELO state is four times (corresponding to the common minimal wave vector of $Q=\pi/4$ for the Weyl nodes) that of the RLO state and the normal phase. Accordingly, the BZ for the ELO state is $1/4$ of the BZ for the RLO state, along $k_{z}$. However, by setting the lattice parameters as units, $k_{z}$ for both RLO and ELO run through $[-\pi,\pi)$. To avoid confusion, we denote hereafter the wave vector along $z$ direction as $\tilde{k}_{z}$ for the ELO states. The folding of the states from the larger BZ of the normal phase and the RLO state to the smaller BZ of the ELO state is important to understand the quasiparticle spectra of the two cases. Specifically, single-particle states with $k_{z}\in[\frac{n-3}{2}\pi,\frac{2n-5}{4}\pi]$ ($n=1,2,3,4$) in the normal phase fold to $\tilde{k}_{z}\in[0,\pi]$ in the ELO state, and single-particle states with $k_{z}\in[\frac{2n-5}{4}\pi,\frac{n-2}{2}\pi]$ ($n=1,2,3,4$) in the normal phase fold to $\tilde{k}_{z}\in[-\pi,0]$ in the ELO state.

\begin{figure}\label{fig2} \centering
\includegraphics[width=8.5cm,height=14.5cm]{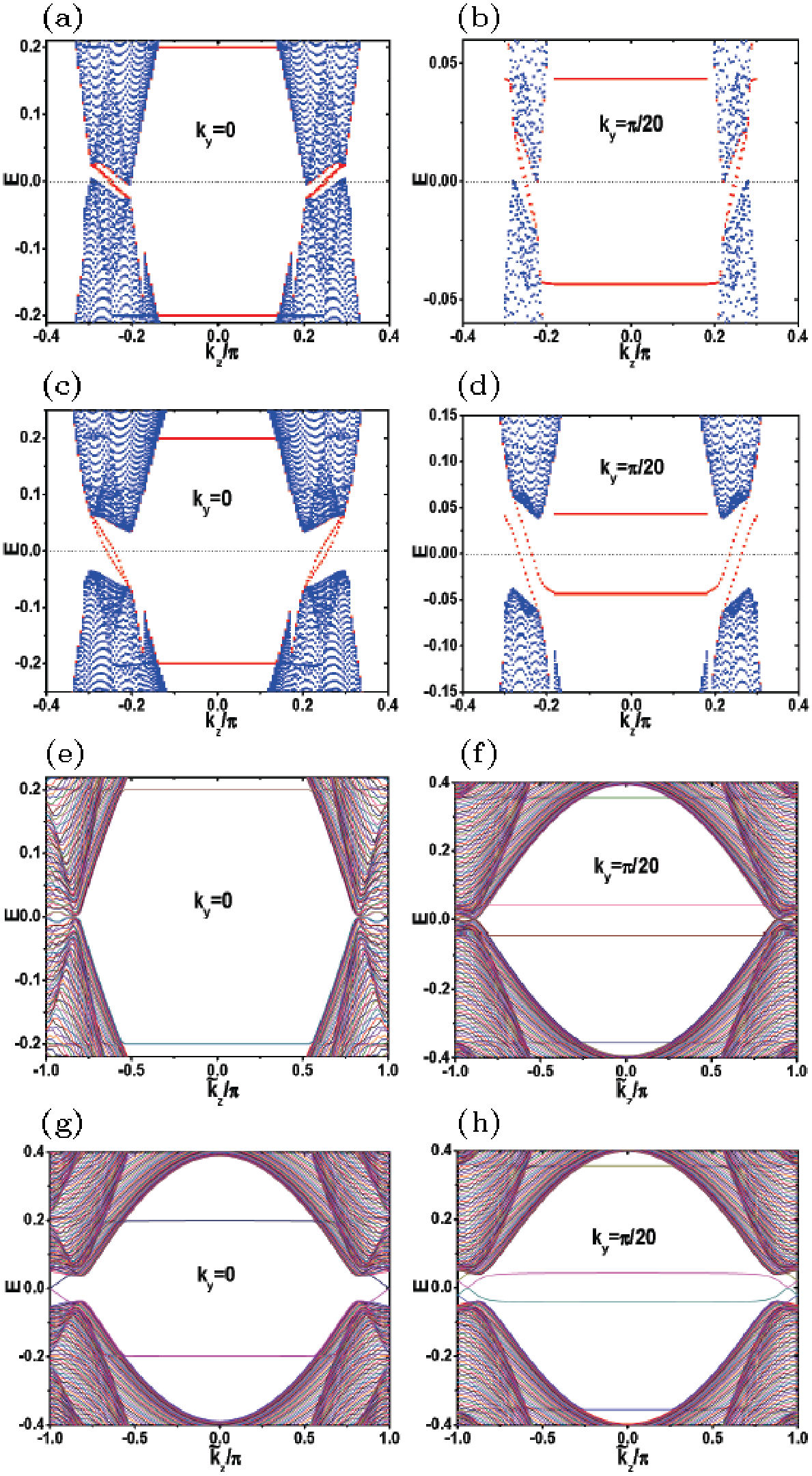} \\
\caption{(Color online) Energy spectra for a film of the LO state of the time-reversal-symmetry broken Weyl semimetal. The film grows along the $x$ axis and has $200$ layers ($N_{x}=200$). (a) to (d) are for the RLO state, with $\omega_{c}=0.1$. (e) to (h) are for the ELO state. For (a), (b), (e), and (f), $\mu=0.2$, $\Delta_{+}=\Delta_{-}=0.01$. For (c), (d), (g), and (h), $\mu=0.2$, $\Delta_{+}=\Delta_{-}=0.05$. Energy spectra for $k_{y}=0$ and $0.05\pi$ are illustrated.}
\end{figure}

\emph{Time-reversal-symmetry broken Weyl metal with a single pair of Weyl pockets.{\textemdash}} Consider a film of the Weyl metal defined by Eq.(1), with two surfaces perpendicular to the $x$ axis. As shown in Fig.2 are typical quasiparticle spectrum for the LO phase of the film. Figs.2(a) to 2(d) are results for the RLO state (see Appendix B for the formulae). In the calculations, we have fixed the value of $\omega_{c}$. Another choice, by fixing the ratio of $\omega_{c}/|\Delta_{\alpha}|$, is verified not to bring about any qualitative changes to the energy spectra and the following discussions. Figs.2(e) to 2(h) are results corresponding to the ELO state (see Appendix C for the formulae).

First consider the low-energy bulk spectrum. Both the RLO state and the ELO state have zero-energy quasiparticles for small pairing amplitude [e.g., $\Delta_{+}=\Delta_{-}=0.01$, for 2(a), 2(b), 2(e), and 2(f)], which disappear as the pairing amplitude is increased [e.g., $\Delta_{+}=\Delta_{-}=0.05$, for 2(c), 2(d), 2(g), and 2(h)]. Therefore, for the parameters considered, the low-energy bulk spectrum for the RLO state is qualitatively the same as that for the ELO state. By implementing the folding of the BZ explained in the beginning of this section to the results in Figs.2(a)-2(d), the correspondence between the low-energy bulk spectra in Figs.2(a)-2(d) and those in Figs.2(e)-2(h) becomes even clearer.

As was explained in Sec.III, for very small $\omega_{c}$, we expect to see the momentum-space phase separation in the states on the Fermi surface, into portions influenced and uninfluenced by the pairing term. What follows we determine the critical $\omega_{c}$ (denoted as $\tilde{\omega}_{c}$) for the present model. For $\mu>0$, the Fermi surface is determined by $E_{+}(\mathbf{k})=0$. According to the definition of the RLO state, a state $\mathbf{k}=\mathbf{q}+\mathbf{P}_{\alpha}$ ($\alpha=\pm$) on the Fermi surface is completely not influenced by the pairing term if $E_{+}(\mathbf{q}+\mathbf{P}_{\alpha})=0$ and $|E_{+}(-\mathbf{q}+\mathbf{P}_{\alpha})|>\omega_{c}$ are fulfilled simultaneously. Suppose we increase $\omega_{c}$ from zero, $\tilde{\omega}_{c}$ is determined as the $\omega_{c}$ above which there is no $\mathbf{k}$ satisfying simultaneously the two conditions. Since the Weyl nodes tilt along $q_{z}$, it is enough to focus on the four Fermi points lying on the $q_{z}$ axis. For the model defined by Eq.(1) and Eq.(3), we define the two Fermi points on the Fermi pocket centering at $\mathbf{P}_{\alpha}$ ($\alpha=\pm$) as $\mathbf{q}_{\alpha\pm}=(0,0,q_{\alpha\pm})$. For small $\mu$ which creates two small Fermi pockets around $\mathbf{P}_{+}$ and $\mathbf{P}_{-}$, we have $q_{++}\simeq\sqrt{1+2\tilde{\mu}}-1$, $q_{+-}\simeq\sqrt{1-2\tilde{\mu}}-1$, $q_{-+}=-q_{+-}$, and $q_{--}=-q_{++}$. $\tilde{\mu}=\mu/|t_{z}\sin Q|$ and we have used $\sin Q=\cos Q$ for $Q=\pi/4$. From the symmetry of the two nodes, we focus on the two Fermi points around the Fermi pocket associated with $\mathbf{P}_{+}$. We have
\begin{equation}
E_{+}(-\mathbf{q}_{++}+\mathbf{P}_{+})\simeq-|t_{z}\cos Q|q_{++}^{2},
\end{equation}
\begin{equation}
E_{+}(-\mathbf{q}_{+-}+\mathbf{P}_{+})\simeq|t_{z}\cos Q|q_{+-}^{2}.
\end{equation}
For $\mu>0$, we have $|q_{+-}|>|q_{++}|$. Therefore, the critical value of the cutoff energy is
\begin{equation}
\tilde{\omega}_{c}=|E_{+}(-\mathbf{q}_{+-}+\mathbf{P}_{+})|\simeq|t_{z}\cos Q|(1-\sqrt{1-2\tilde{\mu}})^{2}.
\end{equation}
For the parameters defined in Sec.II and used for Fig.2, we have $\tilde{\omega}_{c}$$\simeq$$0.033$. As we decrease $\omega_{c}$ from $\tilde{\omega}_{c}$, the phase separation of the states on the Fermi pocket centering at $\mathbf{P}_{+}$ occur in sequence as follows. Firstly, for $|E_{+}(-\mathbf{q}_{++}+\mathbf{P}_{+})|<\omega_{c}<\tilde{\omega}_{c}$, a ring with states $\mathbf{q}_{0}$ ($q_{0z}<0$) determined by $E_{+}(\mathbf{q}_{0}+\mathbf{P}_{+})=0$ and $|E_{+}(-\mathbf{q}_{0}+\mathbf{P}_{+})|=\omega_{c}$ can be defined. All the states to the left of the ring are completely uninfluenced by the pairing term. Secondly, as $\omega_{c}$ decreases further to $\omega_{c}<|E_{+}(-\mathbf{q}_{++}+\mathbf{P}_{+})|$, another ring with states $\mathbf{q}_{1}$  ($q_{1z}>0$) emerges from the positive side of the Fermi pocket, which is determined by $E_{+}(\mathbf{q}_{1}+\mathbf{P}_{+})=0$ and $|E_{+}(-\mathbf{q}_{1}+\mathbf{P}_{+})|=\omega_{c}$. In this case, while all the states on the Fermi pocket in between the two rings (i.e., $q_{0z}<q_{z}<q_{1z}$) are subject to the influence of the pairing term, the remaining two patches of states are completely uninfluenced by the pairing term.

The momentum-space phase separation of the states on the Fermi surface is a unique feature of the RBZP pairing realized in a band structure asymmetric with respect to the reference point of the pairing. In the present case of the RLO state in Weyl metal, the reference point of the pairing is the Weyl node. For conventional BCS pairing, the reference point of the pairing is the center of the BZ. Therefore, the same momentum-space phase separation can occur in the RBZP BCS state realized in a noncentrosymmetric band structure, for which $E_{n}(-\mathbf{k})\ne E_{n}(\mathbf{k})$ with $n$ labeling the energy band. An example is the BCS paring induced by the proximity effect in a ferromagnetic metal with Rashba spin-orbit coupling \cite{hao1617}.

We next study more carefully the bulk quasiparticle spectrum of the ELO state, to verify that the zero-energy bulk quasiparticle states in Figs. 2(e) and 2(f) are caused by the tilting of the Weyl nodes and show explicitly how this happens. Let us focus on the spectrum in the region $\tilde{k}_{z}\in[0,\pi]$, the analysis over which applies also to the spectrum in the region $\tilde{k}_{z}\in[-\pi,0]$. The states in this region inherit from four disconnected $k_{z}$ regions of the normal state list above, which couple through the ELO pairing term to four sections of the hole states (which map to $\tilde{k}_{z}\in[-\pi,0]$). To be clear, let us redefine the group of single-particle states with $k_{z}\in[\frac{2n-5}{4}\pi,\frac{n-2}{2}\pi]$ as $S_{n}$ ($n=1,2,3,4$) and define the group of single-particle states with $k_{z}\in[\frac{n-3}{2}\pi,\frac{2n-5}{4}\pi]$ as $S_{n+4}$ ($n=1,2,3,4$). See Figs. 3(a) and 3(b) for a pictorial understanding of the above partition of the states and the BZ, and the energy dispersions in different regions. The hole states in the corresponding regions are to be denoted as $\tilde{S}_{n}$ ($n=1,...,8$). To be specific, let us consider a pairing term in Eq.(4), written as $\phi^{(\alpha)\dagger}_{\mathbf{q}}is_{y}[\phi^{(\alpha)\dagger}_{-\mathbf{q}}]^{\text{T}}$. Interpreting $\phi^{(\alpha)\dagger}_{\mathbf{q}}$ as the creation operator for particle state at $\mathbf{q}+\mathbf{P}_{\alpha}\equiv\mathbf{k}_{1}$, and $[\phi^{(\alpha)\dagger}_{-\mathbf{q}}]^{\text{T}}$ as the annihilation operator for hole state at $-\mathbf{q}+\mathbf{P}_{\alpha}\equiv\mathbf{k}_{2}$. Then the value of $k_{2z}$ determines the group to which the hole state annihilated by $[\phi^{(\alpha)\dagger}_{-\mathbf{q}}]^{\text{T}}$ belongs, in the same manner as the value of $k_{1z}$ determines the group to which the particle state created by $\phi^{(\alpha)\dagger}_{\mathbf{q}}$ belongs.

In the normal phase, by quadruplicating the unit cell, states in $S_{n}$ and $S_{n+4}$ ($n=1,2,3,4$) fold separately to $\tilde{k}_{z}\in[-\pi,0]$ and $\tilde{k}_{z}\in[0,\pi]$. In the ELO state, for which the wave vector summation in Eq.(4) is throughout the full BZ, the bulk spectrum for $\tilde{k}_{z}\in[0,\pi]$ is determined by the particle states in $S_{n+4}$ ($n=1,2,3,4$), the hole states in $\tilde{S}_{n}$ ($n=1,2,3,4$), and the pairing correlations among them. Since there are two pairing channels, the center-of-mass momentum of which are separately $2\mathbf{P}_{+}$ and $2\mathbf{P}_{-}$, each $S_{n+4}$ section is coupled to two sections of $\tilde{S}_{n}$ ($n=1,2,3,4$), and vice versa. The pairing correlations responsible for the energy spectrum in the region $\tilde{k}_{z}\in[0,\pi]$ are shown explicitly in Table I.

\begin{table}[ht]
\caption{The pairing correlations between particle states in $S_{n}$ ($n=5,...,8$) and hole states in $\tilde{S}_{m}$ ($m=1,...,4$). These couplings determine the bulk spectrum of the ELO state of the Weyl metal with broken time reversal symmetry, in the region $\tilde{k}_{z}\in[0,\pi]$. $\Delta_{+}$ and $\Delta_{-}$ are separately the pairing amplitudes for the pairing components with center-of-mass momenta $2\mathbf{P}_{+}$ and $2\mathbf{P}_{-}$. The first row lists the four groups of particle states. The second and the third rows list the groups of hole states that they are coupled to by the pairing correlations with a center-of-mass momenta $2\mathbf{P}_{+}$ and $2\mathbf{P}_{-}$, respectively.} \centering
\begin{tabular}{c c c c c}
\hline\hline
$$ \hspace{0.3cm} & $S_{5}$ \hspace{0.9cm} & $S_{6}$ \hspace{0.9cm} & $S_{7}$ \hspace{0.9cm} & $S_{8}$ \hspace{0.2cm} \\ [0.2ex]
\hline
$\Delta_{+}$ \hspace{0.3cm} & $\tilde{S}_{1}$ \hspace{0.9cm} & $\tilde{S}_{4}$ \hspace{0.9cm} & $\tilde{S}_{3}$ \hspace{0.9cm} & $\tilde{S}_{2}$ \hspace{0.2cm} \\
\hline
$\Delta_{-}$ \hspace{0.3cm} & $\tilde{S}_{3}$ \hspace{0.9cm} & $\tilde{S}_{2}$ \hspace{0.9cm} & $\tilde{S}_{1}$ \hspace{0.9cm} & $\tilde{S}_{4}$ \hspace{0.2cm} \\
\hline
\hline
\end{tabular}
\end{table}

Notice that, among the eight groups of states in Table I, four groups ($S_{5}$, $S_{8}$, $\tilde{S}_{1}$, and $\tilde{S}_{4}$) only give high-energy single-particle states far away from the Fermi surface, and the remaining four groups ($S_{6}$, $S_{7}$, $\tilde{S}_{2}$, and $\tilde{S}_{3}$) contain low-energy single-particle states crossing the chemical potential $\mu$. Taking into account this fact, the pairing correlations in Table I have the following important properties. Firstly, in the four sets of correlations mediated by $\Delta_{+}$ ($\Delta_{-}$), one set is between high-energy particle states $S_{5}$ ($S_{8}$) and high-energy hole states $\tilde{S}_{1}$ ($\tilde{S}_{4}$), one set is between low-energy particle states $S_{7}$ ($S_{6}$) and low-energy hole states $\tilde{S}_{3}$ ($\tilde{S}_{2}$), and the remaining two sets are both between high-energy particle states and low-energy hole states. The pairing correlations involving high-energy single-particle states is one of the central features of the ELO state differentiating it from the RLO state. Secondly, the pairing correlation divide the eight groups of states into two tetrad, ($S_{5}$, $S_{7}$, $\tilde{S}_{1}$, $\tilde{S}_{3}$) and ($S_{6}$, $S_{8}$, $\tilde{S}_{2}$, $\tilde{S}_{4}$). In each tetrad, each particle (hole) group couple to the two hole (particle) groups through $\Delta_{+}$ and $\Delta_{-}$, whereas the two particle (hole) groups are not coupled. An important consequence of this property is that, although each group of low-energy states ($S_{6}$, $S_{7}$, $\tilde{S}_{2}$, and $\tilde{S}_{3}$) couple to a group of high-energy states, they also couple to a group of low-energy states. And it is the pairing correlation between the low-energy particle group and the low-energy hole group that determine the low-energy quasiparticle spectrum of each tetrad. Thirdly, in conclusion, we can focus on the pairing correlation between the low-energy particle group and the low-energy hole group in each tetrad to determine the qualitative behavior of the low-energy quasiparticle spectra of that tetrad. These include the coupling between $S_{7}$ and $\tilde{S}_{3}$ by $\Delta_{+}$, and the coupling between $S_{6}$ and $\tilde{S}_{2}$ by $\Delta_{-}$.

As shown in Fig. 3(c) [3(d)] is the low-energy part of the crossing between $S_{6}$ and $\tilde{S}_{2}$ ($S_{7}$ and $\tilde{S}_{3}$) along the $k_{z}$ axis, for $\Delta_{-}=0$ ($\Delta_{+}=0$). Turning on the pairing correlation, a gap will open at the crossing points, which occurs at $E\ne0$. As a result, when the pairing amplitude is very small, there is a gap region along the $k_{z}$ axis which is away from the $E=0$ line. Therefore, some quasiparticle are left ungapped at $E=0$ and the spectrum is gapless. Upon increasing the pairing amplitude, the pairing gap enlarges and finally spans the $E=0$ line. The quasiparticle spectrum then becomes fully gapped. This is just what the quasiparticle spectrum behaves when we go from Figs.2(e) and 2(f) to Figs.2(g) and 2(h). The above picture is confirmed by explicit numerical calculations of the bulk spectrum.

Therefore, the absence of a full gap in the quasiparticle spectrum of the ELO state for small pairing amplitudes results from the fact that the low-energy particle and low-energy hole bands make crossings at nonzero energies. This, as we have pointed out in Sec.III, is a direct result of the tilting of the Weyl cones along the $k_{z}$ axis, as can be seen from Fig.1 and Fig.3. Let us focus on the two groups of states $S_{7}$ and $\tilde{S}_{3}$ which are coupled by $\Delta_{+}$. For $\mu>0$ the crossing on Fig.3(d) is between $E_{+}(0,0,k_{z0})=E_{+}(0,0,Q-q_{z0})$ and $-E_{+}(0,0,2Q-k_{z0})=-E_{+}(0,0,Q+q_{z0})$, with $q_{z0}$ determined by $E_{+}(0,0,Q-q_{z0})=-E_{+}(0,0,Q+q_{z0})$. Since the energy spectrum is locally inversion asymmetric along $k_{z}$ with respect to $\mathbf{P}_{+}$, that is
\begin{equation}
E_{+}(0,0,Q+q_{z})\ne E_{+}(0,0,Q-q_{z})
\end{equation}
for $q_{z}\ne0$, the band crossing cannot occur at $E=0$ unless $\mu=0$. As regards the presence of low-energy quasiparticle states for small pairing amplitude, the quasiparticle spectrum of the RLO state and the ELO state are similar, which as we have illustrated both arise from the tilting of the Weyl nodes. We reemphasize that tilting of the Weyl cones is a very general property of Weyl (semi-)metals. Therefore, the properties of the quasiparticle spectrum reported in Fig.2 and the above discussions for them apply generally to the LO state of all Weyl metals.

\begin{figure}\label{fig3} \centering
\hspace{-2.95cm} {\textbf{(a)}} \hspace{3.8cm}{\textbf{(b)}}\\
\hspace{0cm}\includegraphics[width=4.2cm,height=3.5cm]{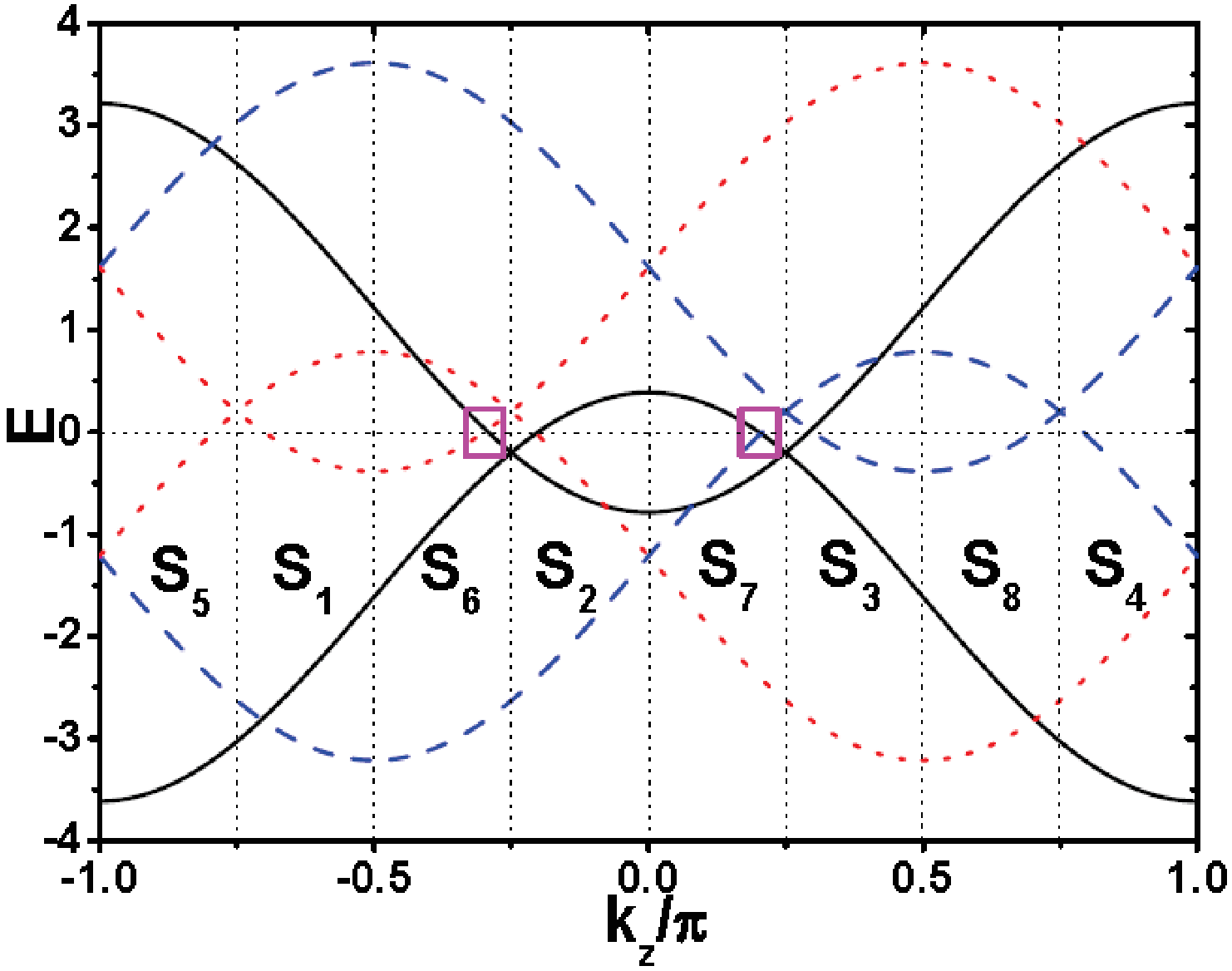}
\includegraphics[width=4.2cm,height=3.5cm]{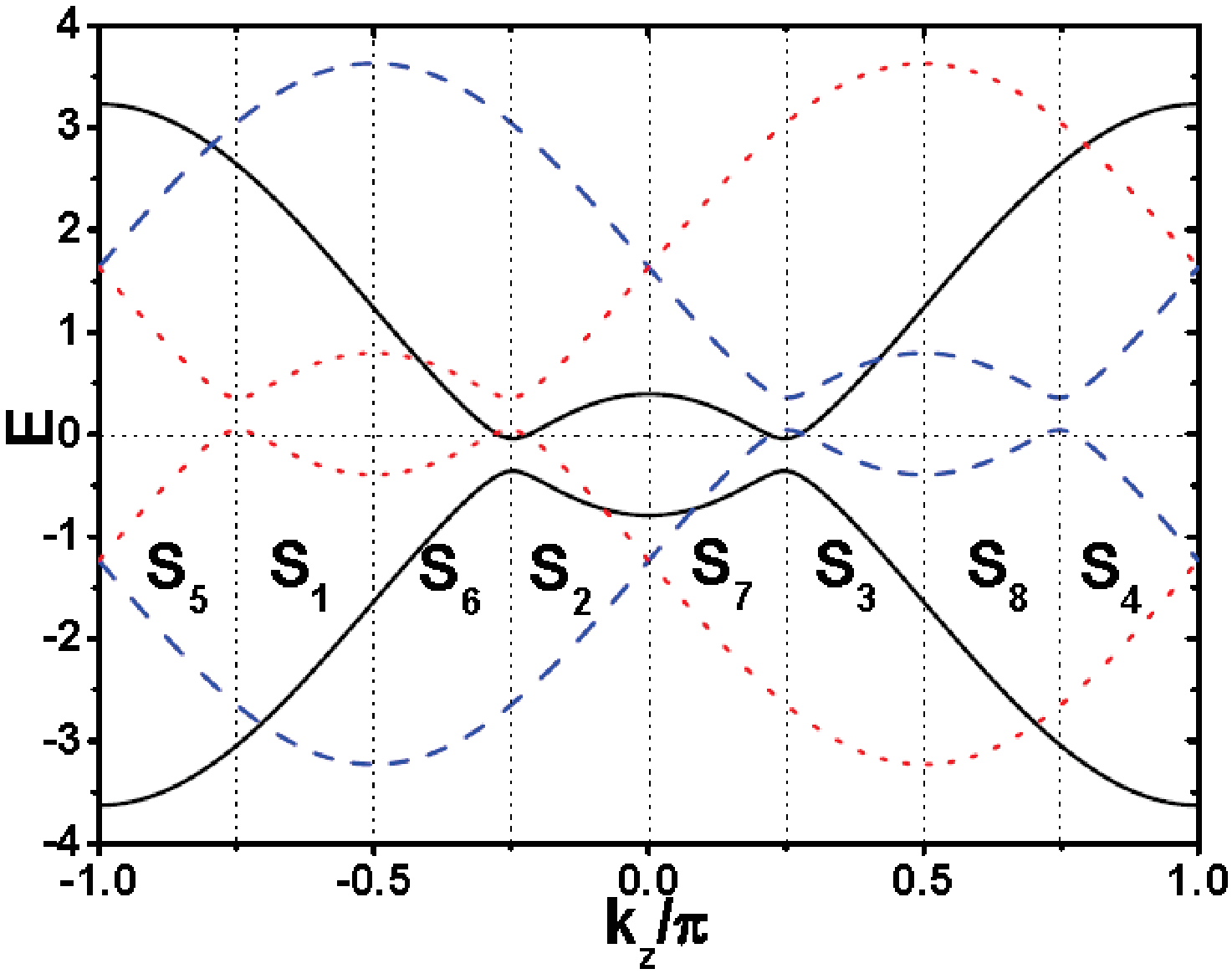}\\
\vspace{-0.10cm}
\hspace{-2.95cm} {\textbf{(c)}} \hspace{3.8cm}{\textbf{(d)}}\\
\hspace{0cm}\includegraphics[width=4.2cm,height=3.5cm]{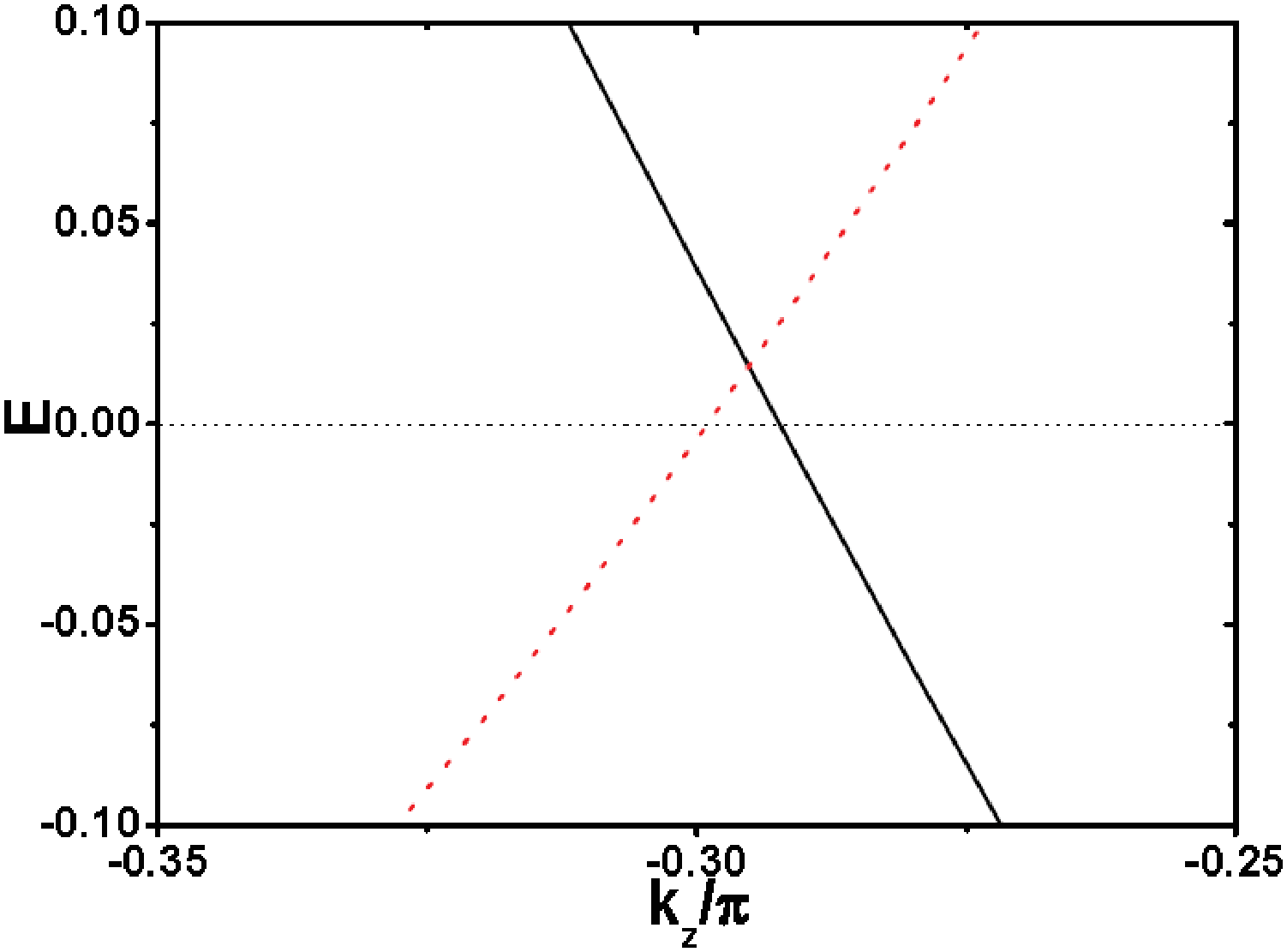}
\includegraphics[width=4.2cm,height=3.5cm]{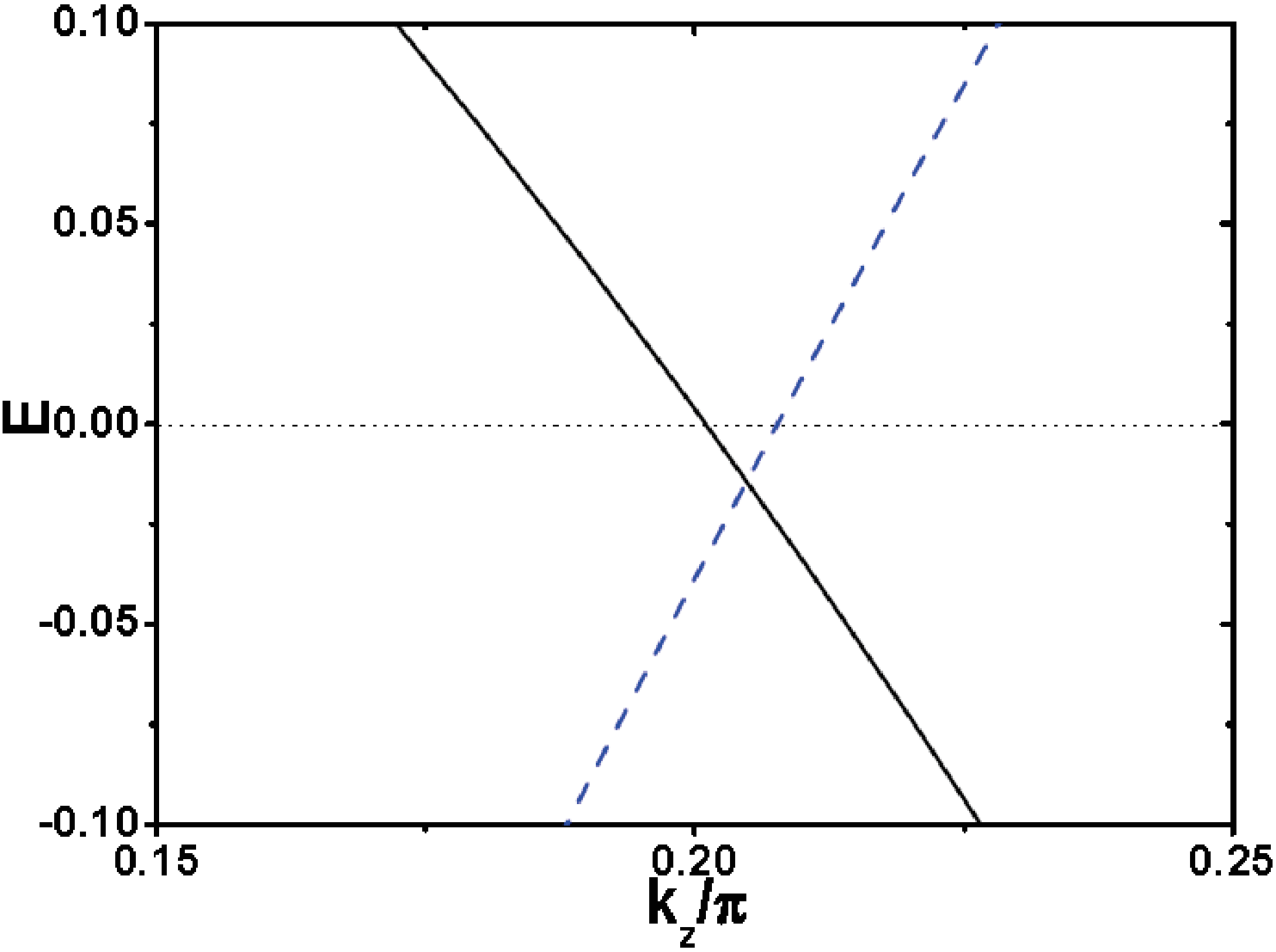} \\
\caption{(Color online) The particle and hole bands of the time-reversal-symmetry broken Weyl metal, as a function of $k_{z}$. $k_{x}=k_{y}=0$ for (a). $k_{x}=0$ and $k_{y}=\pi/20$ for (b). $\mu=0.2$. In (a) and (b), the black solid lines are the particle bands defined as eigenvalues of $h_{0}(\mathbf{k})$, the blue dashed lines are hole bands defined as eigenvalues of $-h_{0}^{\text{T}}(-\mathbf{k}+2\mathbf{P}_{+})$, and the red dotted lines are hole bands defined as eigenvalues of $-h_{0}^{\text{T}}(-\mathbf{k}+2\mathbf{P}_{-})$. The BZ is partitioned according to the value of $k_{z}$ into eight regions, $S_{i}$ ($i=1,...,8$). (c) and (d) are partially enlarged drawings, for the portions of (a) which are separately within the left and right magenta rectangular boxes. The horizontal and vertical dotted straight lines are guides to the eyes.}
\end{figure}

Finally, we compare the surface states in Fig.2 between the ELO state and the RLO state. The Fermi arcs inherited from the normal phase are essentially the same for the RLO state and the ELO state. The only difference between them and the Fermi arcs in the normal phase is a two-fold redundancy due to the use of the Nambu basis. On the other hand, the surface Andreev bound states (SABSs) inside the superconducting gap seems to be different in the two states. This has two aspects. Firstly, in the neighborhood of each Weyl node, the SABSs for the RLO state disperses along the same direction on the two surfaces, which is in contrast to the crossing dispersion obtained for the ELO state. This discrepancy is however not real and is easily eliminated once we imagine to fold the large BZ for the RLO state to the small reduced BZ for the ELO state. (see Appendix D for more discussions on the SABSs). Secondly, after accounting for the effect of BZ folding, a remaining difference lies in the additional splitting of the SABSs for the RLO state. This feature, which is apparent when comparing Fig.2(c) and Fig.2(g), is understood as arising from the long-range pairing correlation intrinsic to the RLO state (see Appendix B for more analysis). As a result of this long-range pairing correlation, the concept of a very thick film is not well defined. The states on the two surfaces are unavoidably coupled together which leads to hybridization and energy splitting.

In light of the above analyses, besides the splitting of the SABSs in the RLO state and the momentum-space phase separation for small cutoff energy in the RLO state, there are no further essential differences between the RLO and ELO states of the time-reversal-symmetry broken Weyl metal defined by Eq.(1). This is consistent with the analyses in Sec. III.

\begin{figure}\label{fig4} \centering
\includegraphics[width=8.5cm,height=12.5cm]{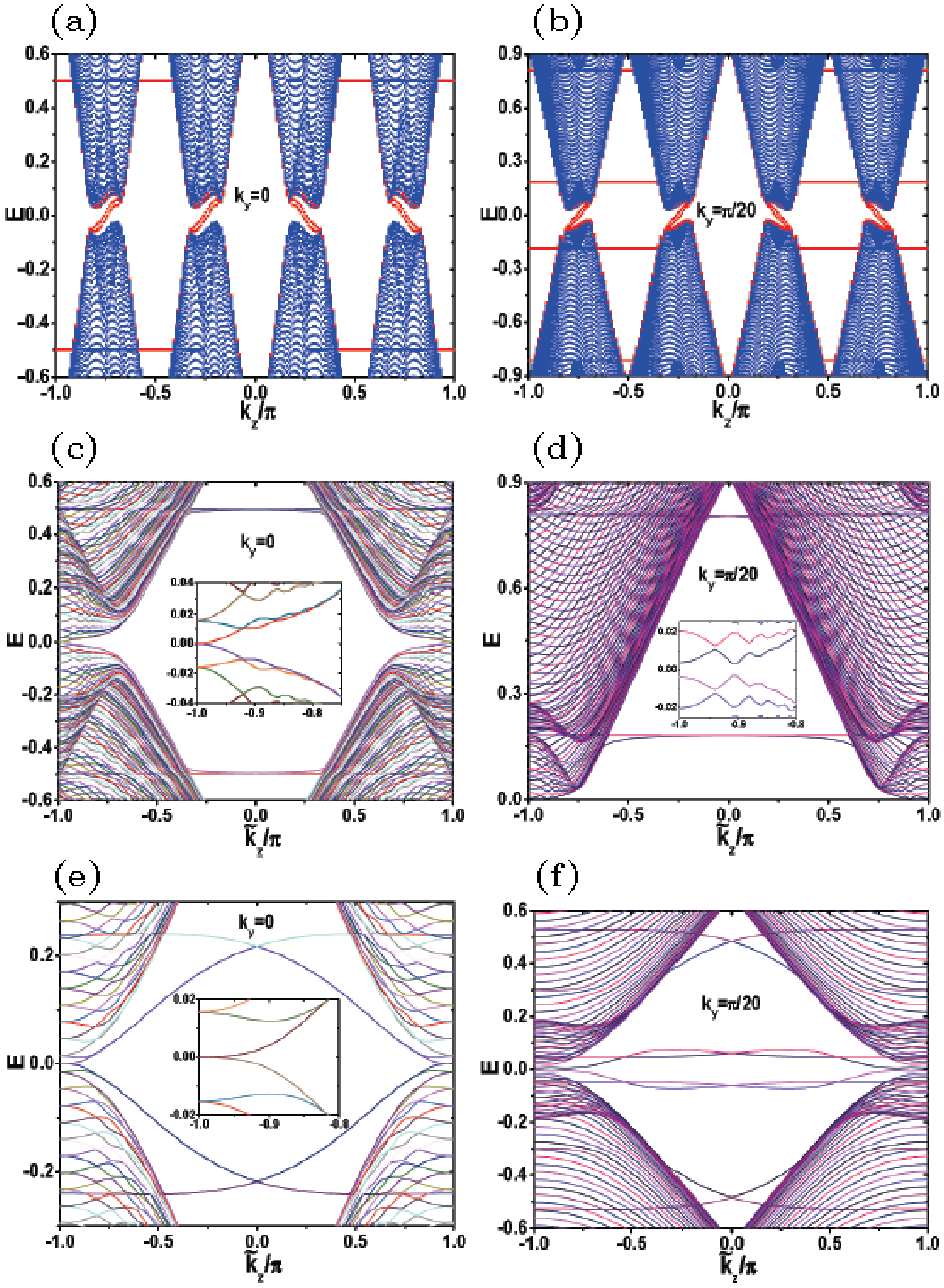} \\
\caption{(Color online)Energy spectra for a film of the LO state of the time-reversal-symmetric and inversion asymmetric Weyl metal. The film grows along the $x$ axis and has $200$ layers ($N_{x}=200$). (a) and (b) are for RLO state, with $\omega_{c}=0.1$. (c) to (f) are for ELO state. We have taken $\Delta_{3\alpha}=0$ for $\alpha=1,...,4$. For (a) to (d), $\mu=0.5$, $\Delta_{0\alpha}=0.05$ ($\alpha=1,...,4$). For (e) and (f), $\mu=0.5$, $\Delta_{0\alpha}=0.5$ ($\alpha=1,...,4$). The energy spectrum for two special $k_{y}$ values, $0$ and $\pi/20$, are illustrated.}
\end{figure}

\emph{Time-reversal-symmetric and inversion-asymmetric Weyl metal with two pairs of Weyl pockets.{\textemdash}} As shown in Fig. 4 are typical quasiparticle spectra for a film of the LO phase of the Weyl metal defined by Eqs.(5) to (7). The film is also assumed to be grown along the $x$ axis and has two $yz$ surfaces. Comparing to Fig.2, the spectra for the RLO state and for the ELO state (see Appendices B and C for the formulae used in calculations) show new qualitative differences. Firstly, the low-energy parts of the bulk spectra for the ELO state and the RLO state are qualitatively different, even though the pairing amplitude and the cutoff energy are both large, in a manner that the spectrum for the ELO state clearly cannot be constructed by folding that of the RLO state from the original BZ to the reduced BZ. Though we defer the detailed analysis, the bulk spectrum of the ELO state have nodes, which is consistent with the analysis in Sec.III. Secondly, the surface states show new qualitative differences between the two pictures, in both the low-energy region (i.e., the SABSs) and the high-energy region (the Fermi arcs). To see the properties of the surface states more clearly, we have shown the zero-energy spectral function on the leftmost layer ($n_{x}=1$) in Figs. 5(a) to 5(c), for both the normal phase [5(a)] and the ELO state [5(b) and 5(c)]. Noticeably, the surface states inherited from the normal phase, the Fermi arcs, are also strongly modified in the ELO state. These are to be compared to the corresponding zero-energy surface spectral function (on the leftmost layer of a film) for the previous Weyl metal with broken time-reversal-symmetry. As shown on Figs. 5(d) to 5(f), by increasing the pairing amplitude, the Fermi surface of the time-reversal-symmetry broken Weyl metal is continuously gapped out and the SABSs connect smoothly to the Fermi arc inherited from the normal phase. Therefore, confirming our conjecture based on the qualitative topological analysis in Sec.III, the ELO state is fundamentally different from the RLO state for the time-reversal-symmetric and inversion-asymmetric Weyl metal with two pairs of Weyl pockets. In this section, we will focus on the new features pointed out above and will not discuss the features due to the same physics explained for the previous model (e.g., the momentum-space phase separation and the energy splitting in the SABSs for the RLO state).

\begin{figure}\label{fig5} \centering
\includegraphics[width=8.5cm,height=12.5cm]{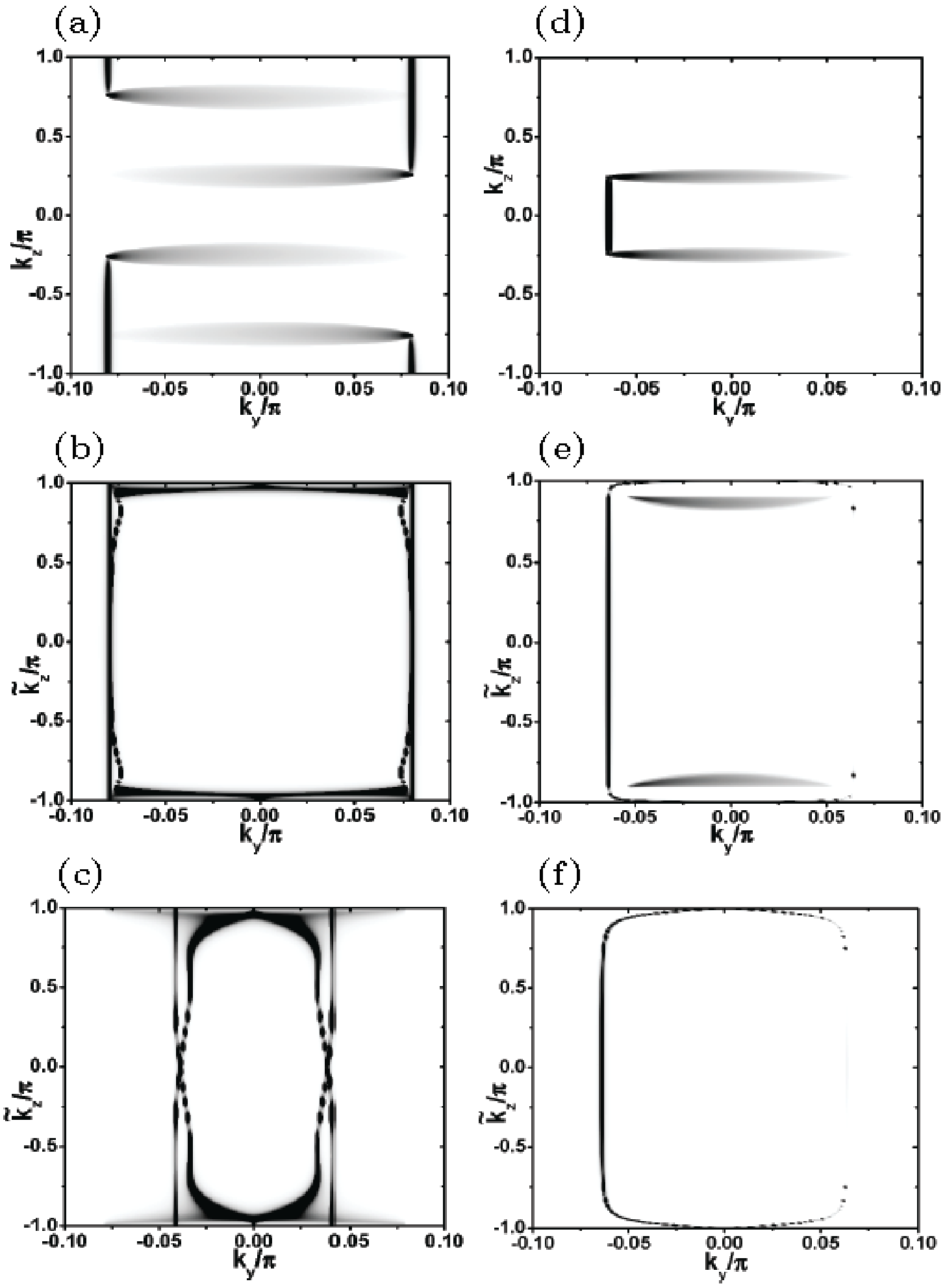} \\
\caption{(Color online) Zero-energy spectral function on the left surface of a film with two surfaces perpendicular to the $x$ axis. (a) to (c) are for the Weyl metal with two pairs of Weyl pockets, (d) to (f) are for the Weyl metal with one pair of Weyl pockets. (a) and (d) are for the normal phase. (b), (c), (e) and (f) are for the ELO state. $\mu=0.5$ for (a) to (c). We have taken $\Delta_{3\alpha}=0$ ($\alpha=1,...,4$) for both (b) and (c). For (b),  $\Delta_{0\alpha}=0.05$ ($\alpha=1,...,4$). For (c), $\Delta_{0\alpha}=0.5$ ($\alpha=1,...,4$). $\mu=0.2$ for (d) to (f). For (e), $\Delta_{+}=\Delta_{-}=0.01$. For (f), $\Delta_{+}=\Delta_{-}=0.05$.}
\end{figure}

To see the origin of the distinctions found above between the ELO state and the RLO state, we examine in more detail the folding of the BZ in the ELO state and the pairing correlations between different portions of the BZ. For the parameters considered (see Sec.II), the folding of the BZ for the ELO state of the time-reversal-symmetric and inversion-asymmetric Weyl metal is identical to the folding of the BZ for the ELO state of the time-reversal-symmetry broken Weyl metal. This is shown in the beginning of this section. However, the presence of four nodes (rather than two nodes for time-reversal-symmetry broken Weyl metal) and correspondingly four sets of pairing correlations separately with center-of-mass momenta $2\mathbf{P}_{\alpha}$ ($\alpha=1,...,4$) doubles the number of channels of pairing correlations. Again, let us focus on the spectrum of the ELO state in the region $\tilde{k}_{z}\in[0,\pi]$. Similar to the analysis for the time-reversal-symmetry broken Weyl metal, we redefine the group of single-particle states with $k_{z}\in[\frac{2n-5}{4}\pi,\frac{n-2}{2}\pi]$ as $S'_{n}$ ($n=1,2,3,4$) and define the group of single-particle states with $k_{z}\in[\frac{n-3}{2}\pi,\frac{2n-5}{4}\pi]$ as $S'_{n+4}$ ($n=1,2,3,4$). See Fig.6 for an illustration of this partition of the BZ and the states. The hole states in the corresponding regions are denoted as $\tilde{S}'_{n}$ ($n=1,...,8$), and are defined in the same manner as that for the previous model.

In the ELO state, where the unit cell is quadruplicated, states in $S'_{n}$ ($\tilde{S}'_{n}$) and $S'_{n+4}$ ($\tilde{S}'_{n+4}$) ($n=1,2,3,4$) fold separately to $\tilde{k}_{z}\in[-\pi,0]$ and $\tilde{k}_{z}\in[0,\pi]$. In the ELO state defined by Eq.(13), with the $\mathbf{q}$-summation extended over the whole BZ, the bulk quasiparticle spectrum for $\tilde{k}_{z}\in[0,\pi]$ is determined by the particle states in $S'_{n+4}$ ($n=1,2,3,4$), and the hole states in $\tilde{S}'_{n}$ ($n=1,2,3,4$), and the pairing correlations among them. Since there are now four sets of pairing correlations separately with center-of-mass momentum $2\mathbf{P}_{\alpha}$ ($\alpha=1,...4$), while each $S'_{n+4}$ section still couples to two sections of $\tilde{S}'_{n}$ ($n=1,2,3,4$), the coupling occurs simultaneously in two different channels. The pairing correlations are shown explicitly in Table II.

\begin{figure}[!htb]\label{fig6}
\centering
\includegraphics[width=7.5cm,height=5.0cm]{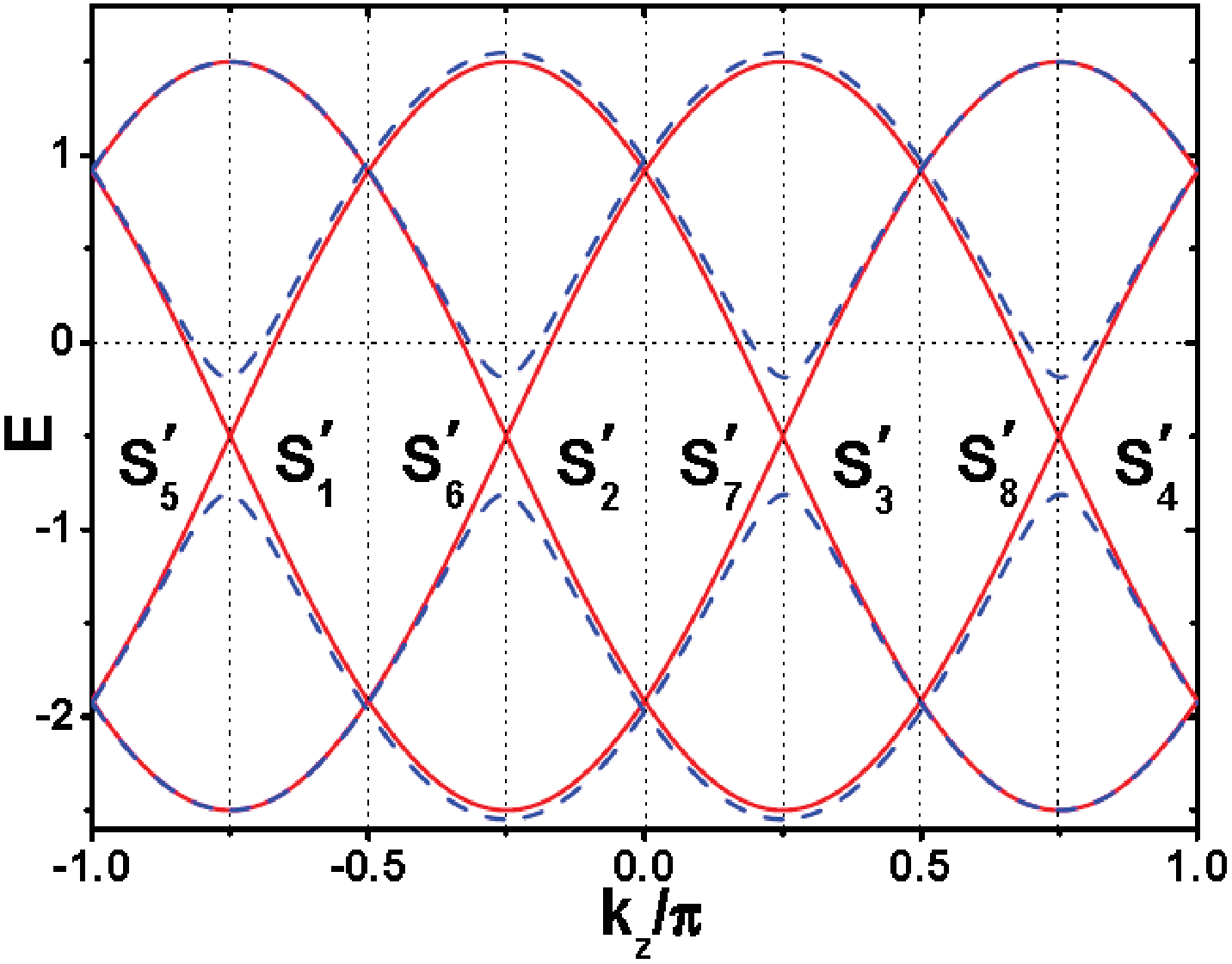}
\caption{(Color online) Plot of the normal state band structures defined by Eq.(8) as a function of $k_{z}$, for fixed $(k_{x},k_{y})=(0,0)$ (the red solid lines) and $(k_{x},k_{y})=(0,\pi/20)$ (the blue dashed lines). The eight regions of the BZ, $S_{i}'$ ($i=1,...,8$), are also shown. $\mu=0.5$. The other parameters are defined in Sec.II. The horizontal and vertical dotted straight lines are guides to the eyes.}
\end{figure}

\begin{table}[ht]
\caption{The pairing correlations between particle states in $S'_{n}$ ($n=5,...,8$) and hole states in $\tilde{S}'_{m}$ ($m=1,...,4$). These couplings determine the bulk spectrum of the ELO state of the time-reversal-symmetric and inversion-asymmetric Weyl metal, in the region $\tilde{k}_{z}\in[0,\pi]$. $\Delta_{\alpha}\equiv\Delta_{0\alpha}$ is the pairing amplitude for the pairing channel with center-of-mass momentum $2\mathbf{P}_{\alpha}$ ($\alpha=1,...,4$). The first row lists the four groups of particle states. The second to the fifth rows beginning with $\Delta_{\alpha}$ list the groups of hole states that they are coupled to by the pairing correlations with a center-of-mass momentum $2\mathbf{P}_{\alpha}$ ($\alpha=1,...,4$).} \centering
\begin{tabular}{c c c c c}
\hline\hline
$$ \hspace{0.3cm} & $S'_{5}$ \hspace{0.9cm} & $S'_{6}$ \hspace{0.9cm} & $S'_{7}$ \hspace{0.9cm} & $S'_{8}$ \hspace{0.2cm} \\ [0.2ex]
\hline
$\Delta_{1}$ \hspace{0.3cm} & $\tilde{S}'_{1}$ \hspace{0.9cm} & $\tilde{S}'_{4}$ \hspace{0.9cm} & $\tilde{S}'_{3}$ \hspace{0.9cm} & $\tilde{S}'_{2}$ \hspace{0.2cm} \\
\hline
$\Delta_{2}$ \hspace{0.3cm} & $\tilde{S}'_{3}$ \hspace{0.9cm} & $\tilde{S}'_{2}$ \hspace{0.9cm} & $\tilde{S}'_{1}$ \hspace{0.9cm} & $\tilde{S}'_{4}$ \hspace{0.2cm} \\
\hline
$\Delta_{3}$ \hspace{0.3cm} & $\tilde{S}'_{1}$ \hspace{0.9cm} & $\tilde{S}'_{4}$ \hspace{0.9cm} & $\tilde{S}'_{3}$ \hspace{0.9cm} & $\tilde{S}'_{2}$ \hspace{0.2cm} \\
\hline
$\Delta_{4}$ \hspace{0.3cm} & $\tilde{S}'_{3}$ \hspace{0.9cm} & $\tilde{S}'_{2}$ \hspace{0.9cm} & $\tilde{S}'_{1}$ \hspace{0.9cm} & $\tilde{S}'_{4}$ \hspace{0.2cm} \\
\hline
\hline
\end{tabular}
\end{table}

Similar to Table I, the pairing correlations in Table II divide the eight groups of states in it into two tetrad, ($S'_{5}$, $S'_{7}$, $\tilde{S}'_{1}$, $\tilde{S}'_{3}$) and ($S'_{6}$, $S'_{8}$, $\tilde{S}'_{2}$, $\tilde{S}'_{4}$), in which the spin components relevant to the low-energy degrees of freedom are respectively $\downarrow$ and $\uparrow$. However, a salient new feature of Table II is that the pairing correlations between particle and hole states always occur in the combination of $\Delta_{13}\equiv\Delta_{1}+\Delta_{3}$ and $\Delta_{24}\equiv\Delta_{2}+\Delta_{4}$. This can be seen more clearly by writing down the model explicitly. Let us take the tetrad ($S'_{5}$, $S'_{7}$, $\tilde{S}'_{1}$, $\tilde{S}'_{3}$) as an example, which are responsible for the spin-$\downarrow$ part of the low-energy quasiparticle spectrum in the region of $k_{z}\in[0,\frac{\pi}{4}]$ ($\tilde{k}_{z}\in[0,\pi]$). Since the two spin degrees of freedom are decoupled in both the model [see Eq.(5)] and the pairing term [see Eq.(13), and we have taken $\nu=0$ in all actual calculations], we can simplify further by focusing solely on the spin-$\downarrow$ degrees of freedom and get
\begin{equation}
\begin{pmatrix} h_{\downarrow}(\mathbf{k}+4\mathbf{P}_{2}) & 0_{2} & \underline{\Delta}_{13} &  \underline{\Delta}_{24} \\
0_{2} & h_{\downarrow}(\mathbf{k}) &  \underline{\Delta}_{24} &  \underline{\Delta}_{13}  \\
\underline{\Delta}^{\dagger}_{13} &  \underline{\Delta}^{\dagger}_{24} & -h^{\text{T}}_{\downarrow}(-\mathbf{k}+2\mathbf{P}_{2}) & 0_{2} \\
\underline{\Delta}^{\dagger}_{24} &  \underline{\Delta}^{\dagger}_{13} & 0_{2} & -h^{\text{T}}_{\downarrow}(-\mathbf{k}+2\mathbf{P}_{1}) \end{pmatrix},
\end{equation}
where $k_{z}\in[0,\frac{\pi}{4}]$, $0_{2}$ is a $2\times2$ matrix with only zero entries, $\underline{\Delta}_{13}=\Delta_{13}i\sigma_{y}$, and $\underline{\Delta}_{24}=\Delta_{24}i\sigma_{y}$. One dramatic ensuing consequence is that, when $\Delta_{13}=\Delta_{24}=0$ the pairing correlation vanishes for both of the two tetrad. That is, the pairing can be completely hidden even if the pairing amplitudes are all nonzero. This is in stark contrast to the RLO state, where the bulk quasiparticle spectra within the Weyl pocket centering around $\mathbf{P}_{\alpha}$ ($\alpha=1,...,4$) depend only on the pairing amplitude $\Delta_{\alpha}$ ($\alpha=1,...,4$), and the pairing never vanishes independent of the pairing amplitudes in other Weyl pockets. The difference between the ELO state and RLO state in the time-reversal-symmetric and inversion-asymmetric Weyl metal thus occurs in a completely different and more profound manner, consistent with the analysis in Sec.III.

As shown in Fig.4 and Figs.5(a)-5(c), though less striking than the extreme cases with $\Delta_{13}=\Delta_{24}=0$, the results for the parameter set chosen (i.e., $\Delta_{1}=\Delta_{2}=\Delta_{3}=\Delta_{4}$) are also qualitatively different between the ELO state and the RLO state. Compared to the bulk spectra for the RLO state which can be fully gapped [e.g., Figs.4(a) and 4(b)], the bulk spectra for the ELO state is featured by the existence of \emph{a line node} on the $\tilde{k}_{z}=\pi$ (or equivalently, $\tilde{k}_{z}=-\pi$) plane. Away from the $\tilde{k}_{z}=\pi$ plane, the bulk gap opens in a very slow pace. The line of nodes is shown explicitly on Fig.7 and is determined by
\begin{equation}
\sqrt{f^{2}_{xy}(k_{x},k_{y})+M^{2}(2-\cos k_{x}-\cos k_{y})^{2}}=\mu.
\end{equation}
Remarkably, it persists independent of the pairing amplitudes. On the $yz$ surface of a film grown along $x$, this line node projects to a line segment with $\tilde{k}_{z}=\pi$ and $|k_{y}|\le k_{y0}$. $k_{y0}=\arccos{\frac{1}{2}[1+\frac{t'^{2}-\mu^2}{M^2}]}\simeq0.08\pi$ for the parameters considered. For a film with two surfaces perpendicular to the $x$ axis, as a result of the broken translational invariance along $x$, the nodes with $|k_{y}|<k_{y0}$ pair annihilate and open a gap. Therefore, only a single pair of point nodes $(0,\pm k_{y0},\pi)$ of the bulk line node are not gapped out on the surface of a film grown along $x$.

\begin{figure}[!htb]\label{fig7} \centering
\includegraphics[width=7.0cm,height=6.50cm]{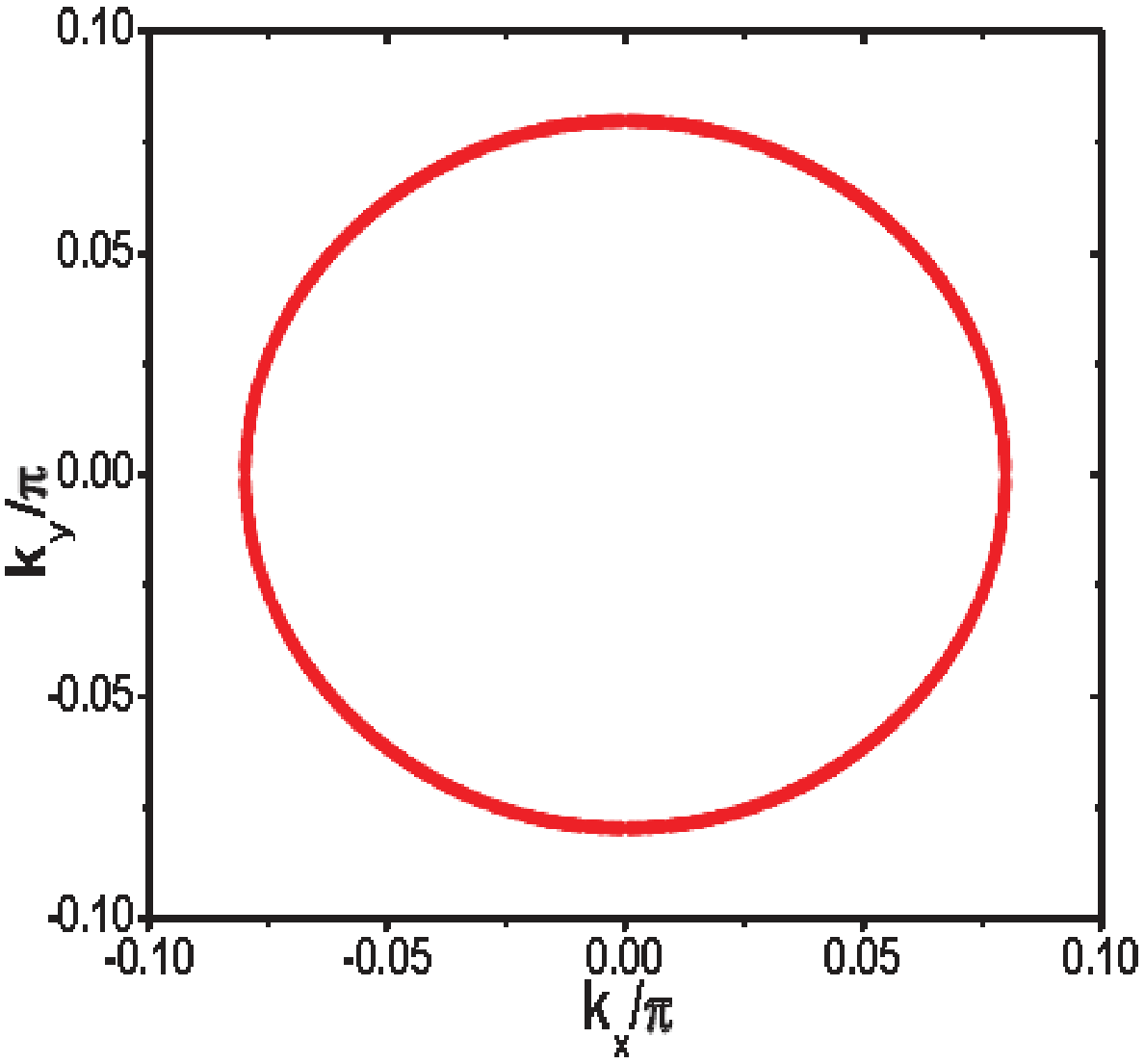}  \\
\caption{(Color online) The line of nodes in the bulk quasiparticle spectrum on the $\tilde{k}_{z}=\pi$ plane of the reduced BZ, for the model and parameters the same as those for Fig.4(c) or Fig.4(e).}
\end{figure}

Clearly, the bulk line node inherits directly from the Fermi surface of the normal phase. Naively, one would expect that it can be understood from the vanishing of the pairing matrix elements between low-energy particle and hole states for $k_{z}=\frac{\pi}{4}$ ($\tilde{k}_{z}=\pi$)  \cite{wang12,wang13,yang14nc,chiu14,chan16,yamakage16,shiozaki14,matsuura13,zhao13,hou13,micklitz09,kobayashi16}. However, it turns out not the case. Consider the minimal model defined by Eq.(26). For $\mu>0$, the low-energy states consist of two particle bands $E_{\downarrow+}(\mathbf{k}+4\mathbf{P}_{2})$ and $E_{\downarrow+}(\mathbf{k})$ and two hole bands $-E_{\downarrow+}(-\mathbf{k}+2\mathbf{P}_{2})$ and $-E_{\downarrow+}(-\mathbf{k}+2\mathbf{P}_{1})$. The pairing correlations between the above particle bands and hole bands are all nonzero, a fact that we have verified both analytically and numerically.

The above analysis reveals one major difference between the ELO state of the second model and all three other LO states. Namely, the low-energy quasiparticle spectrum for all three other LO states are determined by an effective model containing only one low-energy particle band and one low-energy hole band, whereas the low-energy quasiparticle spectrum for the ELO state of the second model is determined by an effective model containing two low-energy particle bands and two low-energy hole bands. We now study in more details the consequences of this emergent double degeneracy in the low-energy effective model. For the spectrum determined by the minimal model of Eq.(26), the two particle [hole] bands are $E_{\downarrow+}(\mathbf{k}+4\mathbf{P}_{2})\equiv E_{1}(\mathbf{k})$ and $E_{\downarrow+}(\mathbf{k})\equiv E_{2}(\mathbf{k})$ [$-E_{\downarrow+}(-\mathbf{k}+2\mathbf{P}_{2})=-E_{2}(\mathbf{k})$ and $-E_{\downarrow+}(-\mathbf{k}+2\mathbf{P}_{1})=-E_{1}(\mathbf{k})$], where $k_{z}\in[0,\frac{\pi}{4}]$. Let us interpret the subindices $1$ and $2$ in $E_{1}(\mathbf{k})$ and $E_{2}(\mathbf{k})$ as a pseudospin index and define $2\times2$ unit matrix $\tilde{\sigma}_{0}$ and Pauli matrices $\tilde{\sigma}_{\alpha}$ ($\alpha=1,2,3$) in this pseudospin subspace. On the $k_{z}=\frac{\pi}{4}$ ($\tilde{k}_{z}=\pi$) plane, $E_{1}(\mathbf{k})=E_{2}(\mathbf{k})$. Consider a general pairing $\underline{\Delta}(\mathbf{k})=[d_{0}(\mathbf{k})\tilde{\sigma}_{0}+\mathbf{d}(\mathbf{k})\cdot\boldsymbol{\tilde{\sigma}}]i\tilde{\sigma}_{2}$ realized in this two-fold degenerate band, where $d_{0}(\mathbf{k})$ and $\mathbf{d}(\mathbf{k})=(d_{1}(\mathbf{k}),d_{2}(\mathbf{k}),d_{3}(\mathbf{k}))$ are pseudospin-singlet and pseudospin-triplet components of the pairing. It is easy to see that, on the Fermi loop determined by $E_{1}(\mathbf{k})=E_{2}(\mathbf{k})=0$ which coincides with Eq.(27), we have $E=0$ eigenstates when
\begin{equation}
\sum_{\alpha=1}^{3}d_{\alpha}^{2}(\mathbf{k})-d_{0}^{2}(\mathbf{k})=0,
\end{equation}
which has $d_{\alpha}(\mathbf{k})=0$ ($\alpha=0,1,2,3$) as a special case \cite{yang14nc,chiu14,chan16,yamakage16,shiozaki14,micklitz09,kobayashi16}. Explicitly, by diagonalizing the four diagonal blocks of Eq.(26), the pairing matrix turns out to be $[d_{2}\tilde{\sigma}_{2}+d_{3}\tilde{\sigma}_{3}]i\tilde{\sigma}_{2}$, where $d_{2}/d_{3}=-i\Delta_{24}/\Delta_{13}$, $d_{2}$ and $d_{3}$ share a phase factor determined by $\sin k_{y}$$-$$i\sin k_{x}$. Applying the above criterion of Eq.(28), zero-energy quasiparticle state exists when $\Delta_{13}^{2}=\Delta_{24}^{2}$, which is clearly satisfied by the parameters used for Figs.4 and 5.

Away from the $k_{z}=\frac{\pi}{4}$ ($\tilde{k}_{z}=\pi$) plane, $E_{1}(\mathbf{k})\ne E_{2}(\mathbf{k})$ and the line node is lifted. However, the gap opening away from the $k_{z}=\frac{\pi}{4}$ ($\tilde{k}_{z}=\pi$) plane scales with the splitting between $E_{1}(\mathbf{k})$ and $E_{2}(\mathbf{k})$ on the Fermi surface, which is determined by the tilting of the Weyl node and is independent of the pairing amplitude. As a result, even for very large pairing amplitudes, the gap opening on the Fermi surface is small for parameters satisfying $\Delta_{13}^{2}=\Delta_{24}^{2}$. This is another qualitative difference compared to the RLO state, which also explains the remnant quasiparticle weight on Figs.5(b) and 5(c) as corresponding to the slightly gapped Fermi surface. In addition, this effective model with an emergent double degeneracy also explains naturally the changes of the surface states in Figs.5(b) and 5(c) compared to that in Fig.5(a), for the part corresponding to the Fermi arc of the normal phase. Qualitatively, since the two particle bands are both coupled simultaneously to the two hole bands, they are effectively coupled to each other and thus the Fermi arcs originating from the two bands are also hybridized. This argument clearly applies also to the Fermi arcs associated with the two hole bands. Finally, for pairing amplitudes leading to $\Delta_{13}^{2}\ne\Delta_{24}^{2}$, we have confirmed through extensive numerical calculations that the line node on $\tilde{k}_{z}=\pi$ disappears. For these parameter sets, for example when $\Delta_{13}=0$ and $\Delta_{24}\ne0$, the bulk quasiparticle spectrum of the ELO state can be fully gapped even if the pairing amplitudes are all very small. This is also qualitatively different from the bulk spectrum of the corresponding RLO state, which has zero-energy quasiparticle states for small pairing amplitudes. All in all, for the Weyl metal with two pairs of Weyl nodes, the ELO state is qualitatively different from the RLO state.

Before ending this section, we want to emphasize another difference between the ELO states of the two models. While the difference between the RLO state and the ELO state shown in Fig.2 for the Weyl metal with a single pair of Weyl pockets is rooted in the local inversion asymmetry of the band structure with respect to the Weyl node, the Weyl pockets of the time-reversal-symmetric and inversion-asymmetric Weyl metal is more symmetric with respect to the Weyl nodes. In this latter case, the breaking of local inversion symmetry around each Weyl node occurs along ordinary directions other than the $q_{x}=q_{y}=0$ axis (see Fig.1(b) for an example). Besides, the effect of the tilting of the Weyl cones is undermined by the effect of the emergent double degeneracy of the effective model for the ELO state. For example, the robust persistence of the bulk line node shown in Fig.7 cannot be understood from the tilting of the Weyl cones. Combining with the foregoing discussions, we conclude that the two examples considered in this work are two qualitatively different examples of the same idea that we intend to convey: The RBZP picture and the EBZP picture of the same pairing might exhibit non-negligible differences and should in general be regarded as different states. The usual BCS pairing realized in centrosymmetric band structures is a special case (rather than the general situation) where the two pictures do give qualitatively consistent low-energy quasiparticle spectrum \cite{bcs,anderson58,nambu60,abrikosov63}.

\section{summary and discussions}

In summary, by studying the LO states of two Weyl metals, we show that the RBZP picture (the RLO state) and the EBZP picture (the ELO state) of the same pairing can have significant differences in their low-energy quasiparticle spectra. Two situations where the difference could emerge are illustrated. In the first situation, the band structure is asymmetric with respect to the Weyl node for the LO phase of Weyl metal (or, center of the BZ for the BCS state of a metal). The most salient difference between the two pictures is the existence of momentum-space phase separation in the states on the Fermi surface, for very small cutoff energy in the RBZP picture. This mechanism can be realized both in the LO state of Weyl metal and in the BCS state of a band asymmetric with respect to the center of the BZ. In the second situation, the folding of the BZ for the ELO state increases the degeneracy of the bands where the pairing is formed. The degeneracy is increased from one to two when the approach is changed from RBZP to EBZP. If no order other than the superconducting order is formed, the folding of the BZ is unique to the ELO state of Weyl metals or the ELO state of other metals whose Fermi surface consists of several disconnected pockets arranged in a manner similar to the Fermi surfaces of the second model with two pairs of Weyl pockets. On the other hand, it is an interesting open question whether the second mechanism can be realized in superconductors with coexisting orders, such as the antiferromagnetic order.

From a theoretical point of view, due caution should be exercised when making theoretical studies on the LO state or BCS state in noncentrosymmetric band structures. On one hand, in trying to construct the mean-field phase diagram, the phases corresponding to the two pictures should be regarded as different, and studied separately. For a specific system, once the pairing mechanism is determined, only one picture is relevant. For example, the RBZP picture might be more appropriate for conventional superconductors mediated by the electron-phonon interaction \cite{bcs}, whereas the EBZP picture seems more suitable for the pairings realized in strongly correlated systems. Systems belonging to the latter case include cuprates \cite{baskaran87,kotliar88}, iron pnictides \cite{si08,seo08}, and cold atoms on an optical lattice tuned to the strongly correlated region \cite{chen09,qu13,zhang13}. On the other hand, the physical picture of sinusoidally varying (in real space) order parameter for the LO phase is a feature specific to the ELO state and may be misleading when applied to consider the properties of the RLO state.

Finally, we make some comments on the relevancy of the present work to actual experiments. Although there is presently no superconducting Weyl metal that can be simulated by the two models considered, and the actual symmetry of the pairing realizable in the superconducting phase of a Weyl metal is still to be determined, the present discussions are interesting for at least two reasons. Firstly, it points out a new character of the LO phase unexplored in all previous works, which can make the ELO state qualitatively different from the RLO state. Secondly, even if not realized in actual superconducting Weyl metal, it is still possible to find applications in fermionic cold atom systems. Besides, we note that the superconducting state formed at the point contact on Dirac semimetal Cd$_{3}$As$_{2}$ might be relevant to the present study \cite{aggarwal16,wang16}. In this system, the inversion symmetry of the bulk material is broken explicitly by the point contact and the local electronic structure should more properly be considered as an inversion-asymmetric Weyl semimetal.

\begin{acknowledgments}
We thank Tao Zhou and Yi Li for helpful discussions. This work was supported by the Texas Center for Superconductivity at the University of Houston and the Robert A. Welch Foundation (Grant No. E-1146). L.H. would also like to acknowledge the support from the China Scholarship Council (No.201506095012). The numerical calculations were performed at the Center of Advanced Computing and Data Systems at the University of Houston.
\end{acknowledgments}\index{}

\begin{appendix}

\section{symmetries and normal state properties of the two models}

The nontrivial topology of the Weyl semimetal in the normal state also determines the topological properties of the LO phase. Here, we study the symmetries and topologies of the two models considered in this work.

\emph{Time-reversal-symmetry broken Weyl metal with a single pair of Weyl nodes.{\textemdash}} The model defined by Eq.(1) of the main text breaks the time-reversal symmetry $T=-is_{y}K$ ($K$ denotes the complex conjugation operation), since $T^{-1}h_{0}(\mathbf{k})T\ne h_{0}(-\mathbf{k})$. In the interpretation that the two degrees of freedom of the model represent the two spins of an electron, the model also breaks the inversion symmetry because $h_{0}(-\mathbf{k})\ne h_{0}(\mathbf{k})$. To be more specific, the time-reversal symmetry is broken because of the existence of Zeeman-like terms proportional to $t_{z}$ and $m$, and the breaking of the inversion symmetry is associated with the presence of a Rashba spin-orbit coupling term proportional to $t$. The model does not have the mirror reflection symmetries either. For example, the mirror reflection with respect to the plane perpendicular to $z$ (i.e., the $xy$ pane) is represented by $M_{z}=is_{z}$. It is broken by the model because
\begin{equation}
M_{z}^{\dagger}h_{0}(k_{x},k_{y},-k_{z})M_{z}\ne h_{0}(\mathbf{k}).
\end{equation}
The model, however, has a fourfold rotational symmetry around the $z$ axis
\begin{equation}
S^{\dagger}h_{0}[R_{\frac{\pi}{2}}(\mathbf{k})]S=h_{0}(\mathbf{k}),
\end{equation}
in which $R_{\frac{\pi}{2}}(\mathbf{k})=(k_{y},-k_{x},k_{z})$ and $S=e^{i\frac{\pi}{4}s_{z}}$. The model thus also has the twofold rotation symmetry around the $z$ axis
\begin{eqnarray}
S^{2\dagger}h_{0}[R_{\pi}(\mathbf{k})]S^{2}&=&\sigma_{z}h_{0}(-k_{x},-k_{y},k_{z})\sigma_{z}   \notag \\
&=&\sigma_{z}h_{0}(-\mathbf{k})\sigma_{z}=h_{0}(\mathbf{k}),
\end{eqnarray}
where we have used $h_{0}(k_{x},k_{y},-k_{z})=h_{0}(\mathbf{k})$ in the first equality of the second line. Because of this twofold rotation symmetry, the energy spectrum of the model is inversion symmetric with respect to $\mathbf{k}=(0,0,0)$ \cite{cho12}.

When $m$ is large, only one pair of Weyl nodes exists in the BZ, at $\mathbf{P}_{\pm}=(0,0,\pm Q)$. Introducing $f_{z}(k_{z})=t_{z}(\cos k_{z}-\cos Q)+2m$, we have $d_{3}(\mathbf{k})=f_{z}(k_{z})-m(\cos k_{x}+\cos k_{y})$. For a fixed value of $k_{z}$, the model defined by Eqs.(1) and (2) describes a quantum anomalous Hall insulator or a trivial insulator depending on the relative magnitude of $f_{z}(k_{z})$ and $m$. Explicitly, if the lower band $E_{-}(\mathbf{k})$ is fully occupied and the higher band $E_{+}(\mathbf{k})$ is completely empty, the Chern number (Hall conductance in unit of $e^{2}/h$) of the quasi-2D system with fixed $k_{z}$ turns out to be $C(k_{z})=\text{sgn}[f_{z}(k_{z})]\theta[2|m|-f_{z}(k_{z})]$ \cite{ludwig94}. $\text{sgn}(x)$ gives the sign of a real number $x$. The Heaviside step function $\theta(x)$ is one for $x>0$ and zero otherwise. The surface Fermi arcs of a Weyl semimetal can be understood as the edge states of the two-dimensional subsystems (labeled by $k_{z}$) with nontrivial $C(k_{z})$ \cite{hosur13}. For the parameters we considered in the main text ($t=-1$, $t_{z}=-2$, $m=1$, $Q=\frac{\pi}{4}$), we have $C=1$ for $k_{z}\in(-\frac{\pi}{4},\frac{\pi}{4})$ and $C=0$ for all other $k_{z}$. The monopole charge of the two Weyl nodes are thus $C_{-}=1$ for the $\mathbf{P}_{-}$ Weyl node and $C_{+}=-1$ for the $\mathbf{P}_{+}$ Weyl node. The total Berry flux through a Fermi pocket enclosing the $\mathbf{P}_{\alpha}$ Weyl node is $4\pi C_{\alpha}$ ($\alpha=\pm$).

The nontrivial topology of the Weyl metal in the normal state can also be understood from the spin configuration on the two Weyl pockets. From Eq.(2) of the main text, the spin configuration for states on the $\alpha$-th band defined by Eq.(3) of the main text is determined by ($<$$s_{x}$$>$,$<$$s_{y}$$>$,$<$$s_{z}$$>$)=$\alpha(d_{x},d_{y},d_{z})/|\mathbf{d}|$ ($\alpha=\pm$). For $\mu=0.2$, the conduction band $E_{+}(\mathbf{k})$ contributes to the Fermi surface. The corresponding spin configuration on a contour of the Fermi surface on the $k_{x}k_{z}$ plane is shown in Figure 8. The opposite monopole charges of the two Weyl nodes are reflected by the opposite windings in the spin configuration on the corresponding Weyl pockets.

\begin{figure}[!htb]\label{fig8} \centering
\includegraphics[width=8.5cm,height=5.5cm]{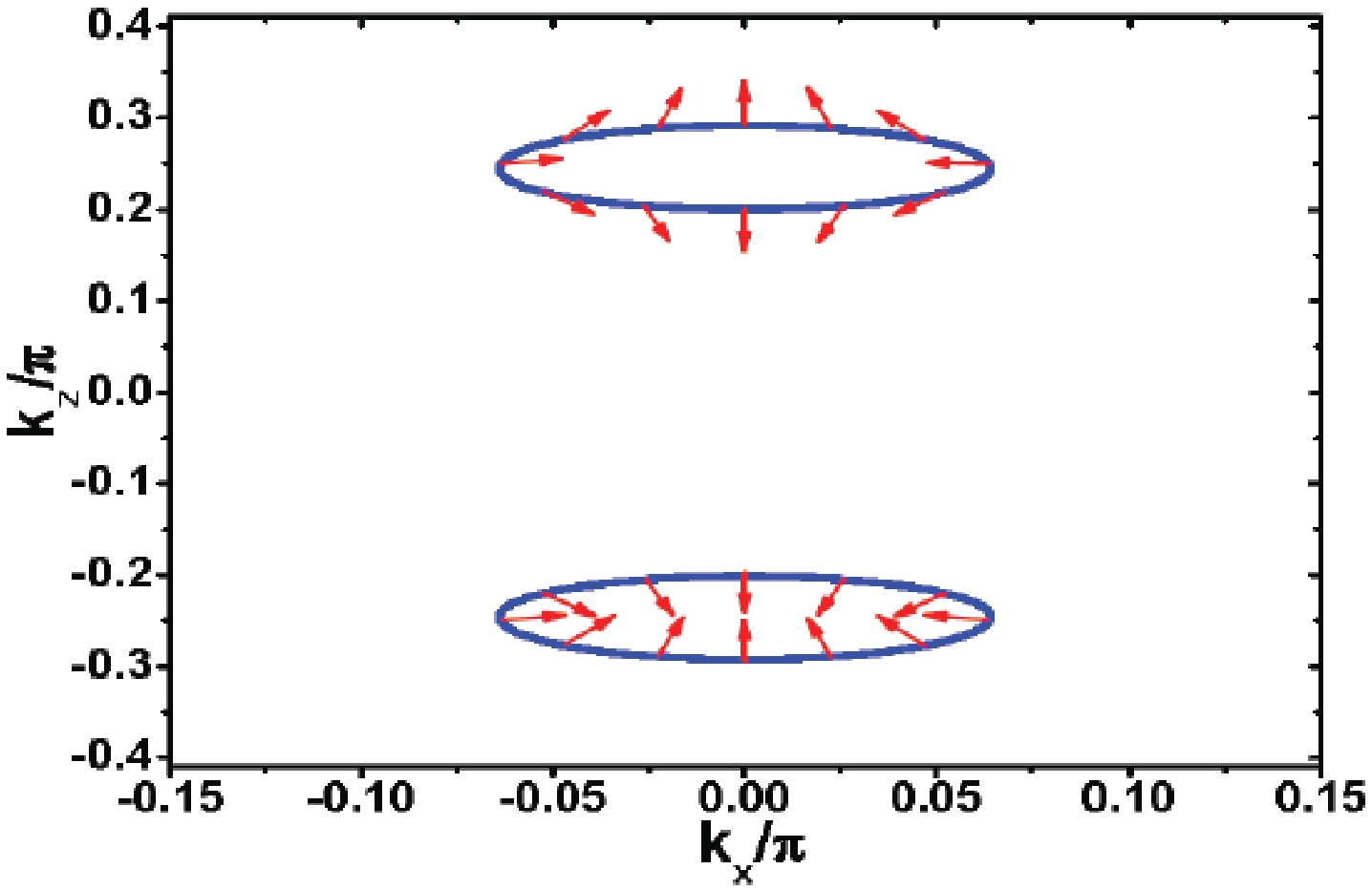}  \\
\caption{The spin configuration on the Fermi contour on the $k_{x}k_{z}$ plane ($k_{y}$=0), for the first model. $\mu=0.2$, $t=-1$, $t_{z}=-2$, $m=1$, $Q=\frac{\pi}{4}$. The red arrows label the direction of the spin for the corresponding states on the Fermi contour. The horizontal (vertical) component of the red arrow indicates the $x$-component ($z$-component) of the spin. $<$$s_{y}$$>$=$0$ for $k_{y}=0$.}
\end{figure}

\emph{Time-reversal-symmetric and inversion-asymmetric Weyl metal with two pairs of Weyl nodes.{\textemdash}}We regularize the model proposed by Hosur \emph{et al} to a cubic lattice \cite{hosur14}. Defining the basis as $\psi^{\dagger}_{\mathbf{k}}=[c^{\dagger}_{1\uparrow}(\mathbf{k}),c^{\dagger}_{2\uparrow}(\mathbf{k}) ,c^{\dagger}_{1\downarrow}(\mathbf{k}),c^{\dagger}_{2\downarrow}(\mathbf{k})]$ (1 and 2 label the two orbitals, $\uparrow$ and $\downarrow$ label the two spin states), the model is written as
\begin{equation}
H'_{1}=\sum\limits_{\mathbf{k}}\psi^{\dagger}_{\mathbf{k}}h'_{1}(\mathbf{k})\psi_{\mathbf{k}} =\sum\limits_{\mathbf{k}}\psi^{\dagger}_{\mathbf{k}}\begin{pmatrix} h_{\uparrow}'(\mathbf{k}) & 0 \\
0 & h_{\downarrow}'(\mathbf{k}) \end{pmatrix}\psi_{\mathbf{k}},
\end{equation}
where\cite{footnote1}
\begin{eqnarray}
h_{s}'(\mathbf{k})&=&\xi_{\mathbf{k}}\sigma_{0}+t_{z}'(\cos k_{z}-\cos Q)\sigma_{z}+\alpha_{s}t_{z}''\sin k_{z}\sigma_{z}   \notag \\
&&+t'(\alpha_{s}\sin k_{y}\sigma_{x}-\sin k_{x}\sigma_{y}).
\end{eqnarray}
$\alpha_{s}=1$ for $s=\uparrow$, and $\alpha_{s}=-1$ for $s=\downarrow$. $\sigma_{0}$ and $\sigma_{i}$ ($i=1,2,3$) are the unit matrix and pauli matrices in the orbital subspace. $\xi_{\mathbf{k}}=-2t_{1}(\cos k_{x}+\cos k_{y})-2t_{2}\cos k_{z}-\mu$. The four energy bands of the model are
\begin{equation}
E_{s\beta}'(\mathbf{k})=\xi_{\mathbf{k}}+\beta\sqrt{g^{2}_{sx}(\mathbf{k})+g^{2}_{sy}(\mathbf{k})+g'^{2}_{sz}(\mathbf{k})},
\end{equation}
where $\beta=\pm$, $g_{sx}(\mathbf{k})=t'\alpha_{s}\sin k_{y}$, $g_{sy}(\mathbf{k})=-t'\sin k_{x}$, and $g_{sz}'(\mathbf{k})=t_{z}'(\cos k_{z}-\cos Q)+\alpha_{s}t_{z}''\sin k_{z}$. The Weyl nodes are determined by $g_{sx}(\mathbf{k})=g_{sy}(\mathbf{k})=g_{sz}'(\mathbf{k})=0$. Besides the Weyl points along $(0,0,k_{z})$, there are clearly also Weyl points along $(\pi,0,k_{z})$, $(0,\pi,k_{z})$, and $(\pi,\pi,k_{z})$. To retain only the Weyl nodes along $(0,0,k_{z})$, we add the following term to the model
\begin{equation}
M(2-\cos k_{y}-\cos k_{y})s_{0}\sigma_{z}.
\end{equation}
The $s$-spin part of the model thus becomes
\begin{equation}
h_{s}(\mathbf{k})=h_{s}'(\mathbf{k})+M(2-\cos k_{x}-\cos k_{y})\sigma_{z},
\end{equation}
which leads to the model defined by Eq.(5) of the main text. Redefining $g_{sz}(\mathbf{k})=g_{sz}'(\mathbf{k})+M(2-\cos k_{x}-\cos k_{y})$, the four energy bands become
\begin{equation}
E_{s\beta}(\mathbf{k})=\xi_{\mathbf{k}}+\beta\sqrt{g^{2}_{sx}(\mathbf{k})+g^{2}_{sy}(\mathbf{k})+g^{2}_{sz}(\mathbf{k})}.
\end{equation}
The Weyl nodes are now determined by
\begin{equation}
g_{sx}(\mathbf{k})=g_{sy}(\mathbf{k})=g_{sz}(\mathbf{k})=0.
\end{equation}
For sufficiently large $M$, the above condition leads to
\begin{equation}
k_{x}=k_{y}=g_{sz}(0,0,k_{z})=0.
\end{equation}
When $t_{z}''=0$, we have a pair of Dirac nodes at $(0,0,\pm Q)$. As we increase $t_{z}''$, each Dirac node is separated into two Weyl nodes. For one particular set of parameters
\begin{equation}
Q=\frac{\pi}{2},   \hspace{0.5cm}   t_{z}''=t_{z}',
\end{equation}
the two pairs of Weyl nodes are at $\pm\mathbf{Q}_{1}=\pm(0,0,\frac{\pi}{4})$ and $\pm\mathbf{Q}_{2}=\pm(0,0,\frac{3\pi}{4})$. Among the four Weyl nodes, the $-\mathbf{Q}_{1}\equiv\mathbf{P}_{2}$ and $\mathbf{Q}_{2}\equiv\mathbf{P}_{4}$ nodes are associated with spin-$\uparrow$ electrons, and the $-\mathbf{Q}_{2}\equiv\mathbf{P}_{1}$ and $\mathbf{Q}_{1}\equiv\mathbf{P}_{3}$ nodes are associated with spin-$\downarrow$ electrons.

To understand the topology of the normal phase, we rewrite $g_{sz}(\mathbf{k})=f_{s}(k_{z})-M(\cos k_{x}+\cos k_{y})$. For a fixed $k_{z}$, we consider the low-energy band of the two spin subsystems which constitutes the Weyl node. The $k_{z}$-resolved Chern number for spin-$s$ electron is thus \cite{ludwig94}
\begin{equation}
C_{s}(k_{z})=\text{sgn}[f_{s}(k_{z})]\theta[2|M|-|f_{s}(k_{z})|].
\end{equation}
For the parameters we choose in the main text ($t_{1}=t_{2}=0$, $t'=2$, $t_{z}'=t_{z}''=\sqrt{2}$, $M=2$, and $Q=\pi/2$), we have $C_{\uparrow}(k_{z}\in[-\pi,-\pi/4))=C_{\uparrow}(k_{z}\in(3\pi/4,\pi])=C_{\downarrow}(k_{z}\in[-\pi,-3\pi/4))=C_{\downarrow}(k_{z}\in(\pi/4,\pi])=1$ and $C_{s}(k_{z})=0$ otherwise. The monopole charges of the four Weyl nodes are thus $C_{1}=C_{2}=-C_{3}=-C_{4}=-1$, which are separately the total Berry flux through a Fermi pocket enclosing the corresponding Weyl node. Similar to the time-reversal-symmetry broken Weyl metal, the Fermi arc surface states in the normal phase can be understood as the chiral edge states of the quasi-2D subsystems stacked along $k_{z}$.

Similar to the discussion for the previous model, the nontrivial topology of the Weyl metal described by the present model can be understood from the orbital configuration on the Weyl pockets. From the above discussions, the orbital configuration for states on the $E_{s\beta}(\mathbf{k})$ band defined by Eq.(A9) is determined by ($<$$\sigma_{x}$$>$,$<$$\sigma_{y}$$>$,$<$$\sigma_{z}$$>$)=$\beta(g_{sx},g_{sy},g_{sz})/\sqrt{g^{2}_{sx}+g^{2}_{sy}+g^{2}_{sz}}$ ($\beta=\pm$; $\alpha_s=1$ for $s=\uparrow$,  $\alpha_s=-1$ for $s=\downarrow$). For $\mu=0.5$, the conduction bands $E_{s+}(\mathbf{k})$ contribute to the Fermi surface. The corresponding orbital configuration on a contour of the Fermi surface on the $k_{x}k_{z}$ plane is shown in Figure 9. The monopole charges of the four Weyl nodes obtained above are consistent with the patterns of winding in the orbital configuration on the corresponding Weyl pockets.

\begin{figure}[!htb]\label{fig9} \centering
\includegraphics[width=8.5cm,height=7.5cm]{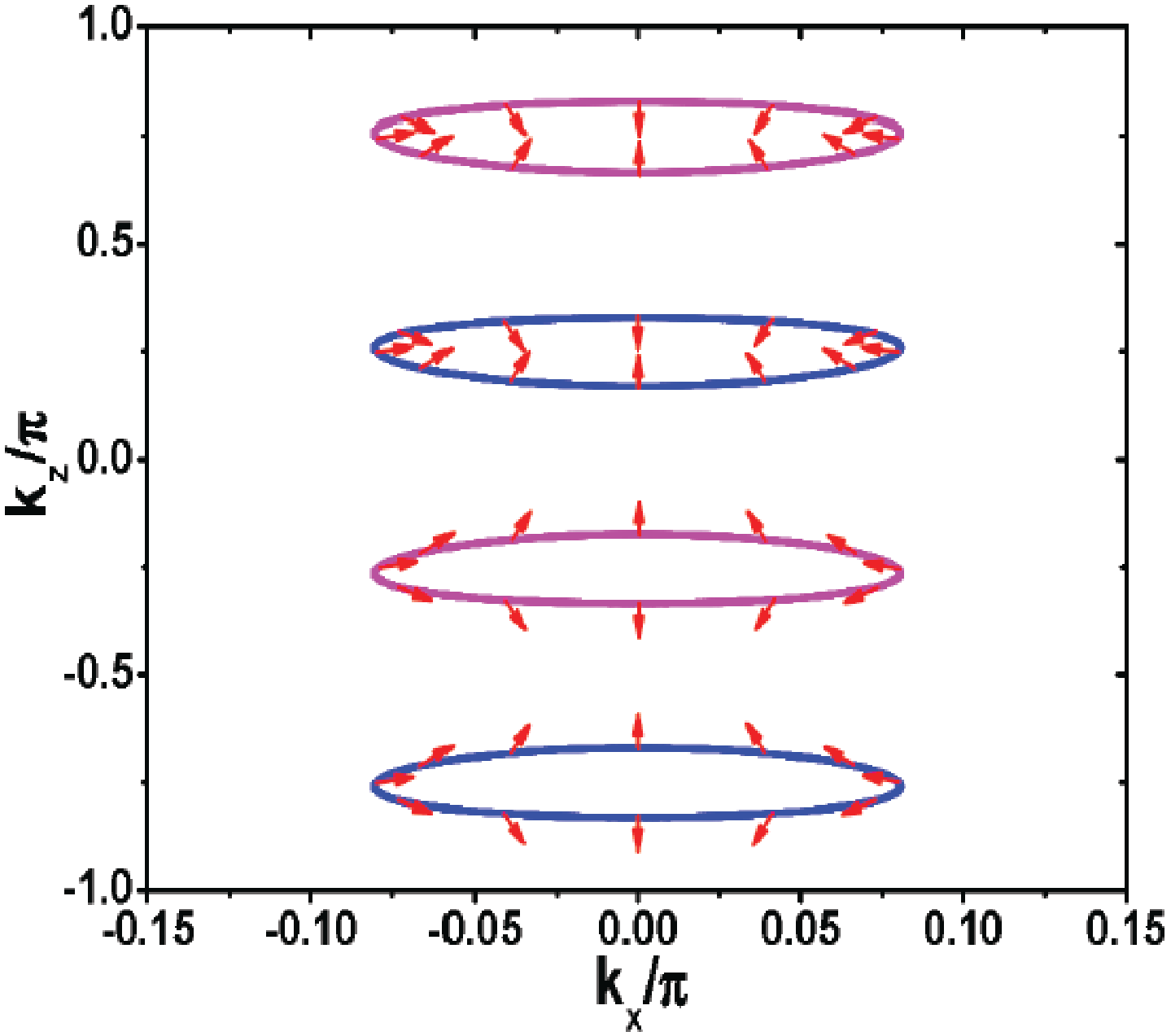}  \\
\caption{The orbital configuration on the Fermi contour on the $k_{x}k_{z}$ plane ($k_{y}$=0), for the second model. $\mu=0.5$, $t_{1}=t_{2}=0$, $t'=2$, $t_{z}'=t_{z}''=\sqrt{2}$, $M=2$, and $Q=\pi/2$. The magenta pockets are spin-up ($s=\uparrow$) pockets. The blue pockets are spin-down ($s=\downarrow$) pockets. The red arrows label the direction of the orbital for the corresponding states on the Fermi contour. The horizontal (vertical) component of the red arrow indicates the $y$-component ($z$-component) of the orbital. $<$$\sigma_{x}$$>$=$0$ for $k_{y}=0$.}
\end{figure}

In the normal phase, the important symmetries of the model include four-fold rotational symmetry $S=e^{i\frac{\pi}{4}s_{z}\sigma_{z}}$, the spin rotation symmetry $s_{z}$, and the time-reversal symmetry $T=-is_{y}\sigma_{0}K$ \cite{hosur14}. Explicitly, we have
\begin{equation}
S^{\dagger}h_{1}[R_{\frac{\pi}{2}}(\mathbf{k})]S=h_{1}(\mathbf{k}),
\end{equation}
\begin{equation}
s^{\dagger}_{z}h_{1}(\mathbf{k})s_{z}=h_{1}(\mathbf{k}),
\end{equation}
and
\begin{equation}
T^{-1}h_{1}(\mathbf{k})T=h_{1}(-\mathbf{k}).
\end{equation}

\section{formulae for numerical calculation of energy spectra for a thin film: RLO}

In the RLO picture of the LO phase, the pairing term is introduced only for states within an energy shell centering at the Fermi surface. The states outside this shell are not influenced by the formation of pairing. For bulk properties related to the pairing term, we can ignore completely the states outside the pairing shell. When considering the energy spectra of a thin film, however, it is necessary to go beyond the pairing shell to reproduce the Fermi arc structure in the results. This is achieved in the present work by treating the two type of states differently in the following manner.

\subsection{Time-reversal-symmetry broken Weyl metal with a single pair of Weyl pockets}

Considering the $s$-wave spin-singlet LO phase defined in the main text \cite{cho12,bednik15}
\begin{equation}
H^{RLO}=\frac{1}{2}\sum\limits_{\mathbf{q},\alpha}\Delta_{\alpha} \phi^{(\alpha)\dagger}_{\mathbf{q}}is_{y}[\phi^{(\alpha)\dagger}_{-\mathbf{q}}]^{\text{T}}+\text{H.c.},
\end{equation}
where $\Delta_{\alpha}$ ($\alpha=\pm$) are the pairing amplitudes (taken as real numbers) for states close to the two Weyl pockets, $\mathbf{q}$ is the wave vector relative to the Weyl point $\mathbf{P}_{+}$ or $\mathbf{P}_{-}$, and $\text{H.c.}$ means taking the Hermitian conjugate of the previous expression. For $\mu>0$ which corresponds to electron doping, the $\mathbf{q}$-summation is restricted to within an energy shell $|E_{+}(\mathbf{q}+\mathbf{P}_{\alpha})|<\omega_{c}$ and $|E_{+}(-\mathbf{q}+\mathbf{P}_{\alpha})|<\omega_{c}$ surrounding the $\alpha$-th Weyl node, where $\omega_{c}$ is the cutoff energy. $\omega_{c}$ is set to be of the same order of magnitude as the pairing amplitudes $\Delta_{\alpha}$.

For the RLO picture of the LO phase, the pairing is on one hand restricted to the neighborhood of the Fermi surface and thus can be considered as local in the momentum space, and on the other hand it is extended throughout the whole system in the real space. For example, transforming the $z$ direction of the model to the real space, the pairing amplitude coupling the $n$-th and the $n'$-th layer is written for a set of fixed $(q_{x},q_{y})$ as
\begin{equation}
[e^{iQ(n+n')}\pm e^{-iQ(n+n')}](\frac{1}{N_{z}}\sum\limits_{q_{z}(q_{x},q_{y})}e^{iq_{z}(n-n')})i\Delta s_{y},
\end{equation}
where $N_{z}$ is the number of layers along $z$ direction, the summation over $q_{z}$ is restricted only to the neighborhood of the Fermi surface and so depends on the wave vectors $q_{x}$ and $q_{y}$ in a nontrivial manner. For a rational $Q$, each $n$ is coupled to an extensive set of layers in an oscillatory manner determined by the factor $[e^{iQ(n+n')}\pm e^{-iQ(n+n')}]$. This extensive coupling between the layers makes invalid the standard transfer matrix method (iterative Green's function method) of obtaining the surface Green's functions. As such, a straightforward numerical diagonalization is mandatory if we want to study the properties of surface states of a film.

We consider a thin film with two surfaces perpendicular to the $x$-axis. Define the number of layers as $N_{x}$, the $x$-direction should be treated in the real space. $k_{y}$ and $k_{z}$ are still good quantum numbers and take values in the surface BZ. To explore possible surface states connecting the two Weyl nodes, we study the energy spectra of the film along a line in the surface BZ running parallel to the $k_{z}$ direction. That is, we fix $k_{y}$ and study the dependence of the energy spectrum on $k_{z}$.

As we vary $k_{z}$ from $-\pi$ to $\pi$, we can put the wave vectors we encounter into three groups. This is performed in terms of the normal state dispersion $E_{+}(\mathbf{k})$, by supplementing $(k_{y},k_{z})$ by the third component $k_{x}$ of the 3D BZ for the bulk material. The first group contains wave vectors for which there is no $k_{x}\in[-\pi,\pi)$ that  makes $|E_{+}(\mathbf{k})|<\omega_{c}$. For wave vectors in this group, we do not introduce the pairing term. However, we still introduce the Nambu basis in the usual manner and define $\tilde{\phi}^{\dagger}_{\mathbf{k}}=[\phi^{\dagger}_{\mathbf{k}},\phi^{\text{T}}_{-\mathbf{k}}]$. In such a manner, the energy spectra for all wave vectors would still have the same number of states. For these wave vectors, the model in the original 3D BZ is simply
\begin{equation}
H_{Nambu}=H_{0}^{Nambu}=\frac{1}{2}\sum\limits_{\mathbf{k}}\tilde{\phi}^{\dagger}_{\mathbf{k}}\begin{pmatrix} H_{0}(\mathbf{k}) & 0  \\
0 & -H^{\text{T}}_{0}(-\mathbf{k}) \end{pmatrix}\tilde{\phi}_{\mathbf{k}}.
\end{equation}
For a thin film, the model is obtained by simply making the partial Fourier transformation of the $x$-direction from $k_{x}$ to $n_{x}$, which yields
\begin{eqnarray}
&&H_{0}^{Nambu}=\frac{1}{2}\sum\limits_{n_{x}k_{y}k_{z}}\tilde{\phi}^{\dagger}_{n_{x}k_{y}k_{z}}[\tilde{h}_{0}(k_{y},k_{z})\tilde{\phi}_{n_{x}k_{y}k_{z}}   \notag \\ &&+\tilde{h}_{+}\tilde{\phi}_{n_{x}+1,k_{y}k_{z}}+\tilde{h}_{-}\tilde{\phi}_{n_{x}-1,k_{y}k_{z}}.
\end{eqnarray}
We have defined
\begin{eqnarray}
\tilde{h}_{0}(k_{y},k_{z})&=&t\sin k_{y}\tau_{z}s_{y}+[t_{z}(\cos k_{z}-\cos Q)   \notag \\
&&+m(2-\cos k_{y})]\tau_{z}s_{z}-\mu\tau_{z}s_{0},
\end{eqnarray}
\begin{equation}
\tilde{h}_{+}=-\frac{1}{2}(it\tau_{0}s_{x}+m\tau_{z}s_{z}),
\end{equation}
and $\tilde{h}_{-}=\tilde{h}^{\dagger}_{+}$.

The second group of wave vectors consist of $(k_{y},k_{z})$ lying within the pairing shell surrounding $\mathbf{P}_{+}$. For these $(k_{y},k_{z})$, on one hand there exists $k_{x}$ that makes $|E_{+}(\mathbf{k})|<\omega_{c}$ and $|E_{+}(2\mathbf{P}_{+}-\mathbf{k})|<\omega_{c}$, and on the other hand $|k_{z}-Q|<|k_{z}-(-Q)|$. Supplementing $(k_{y},k_{z})$ with the values of $k_{x}$ that keeps $|E_{+}(\mathbf{k})|<\omega_{C}$, the intranode LO pairing is expressed in terms of the following Nambu basis $\tilde{\phi}^{\dagger}_{\mathbf{k}}=[\phi^{\dagger}_{\mathbf{k}},\phi^{\text{T}}_{2\mathbf{P}_{+}-\mathbf{k}}]$. The two diagonal $2\times2$ block matrices constitute
\begin{equation}
H_{0}^{Nambu}=\frac{1}{2}\sum\limits_{\mathbf{k}}\tilde{\phi}^{\dagger}_{\mathbf{k}}\begin{pmatrix} H_{0}(\mathbf{k}) & 0  \\
0 & -H^{\text{T}}_{0}(2\mathbf{P}_{+}-\mathbf{k}) \end{pmatrix}\tilde{\phi}_{\mathbf{k}}.
\end{equation}
For this part of the model, the summation over $k_{x}$ spans the whole range of $[-\pi,\pi)$. Making Fourier transformation from $k_{x}$ to $x$, and introducing the new Nambu basis $\tilde{\phi}^{\dagger}_{n_{x}k_{y}k_{z}}=[\phi^{\dagger}_{n_{x}k_{y}k_{z}},\phi^{\text{T}}_{n_{x},-k_{y},2Q-k_{z}}]$, it becomes
\begin{eqnarray}
H_{0}^{Nambu}&=&\frac{1}{2}\sum\limits_{n_{x},k_{y}k_{z}}\tilde{\phi}^{\dagger}_{n_{x}k_{y}k_{z}}[\tilde{h}'_{0}(k_{y},k_{z})\tilde{\phi}_{n_{x}k_{y}k_{z}}   \notag \\ &&+\tilde{h}'_{+}\tilde{\phi}_{n_{x}+1,k_{y}k_{z}}+\tilde{h}'_{-}\tilde{\phi}_{n_{x}-1,k_{y}k_{z}}.
\end{eqnarray}
We have $\tilde{h}'_{\pm}=\tilde{h}_{\pm}$ and
\begin{equation}
\tilde{h}'_{0}(k_{y},k_{z})=\begin{pmatrix} h'_{0}(k_{y},k_{z}) & 0  \\
0 & -h'^{\text{T}}_{0}(-k_{y},2Q-k_{z}) \end{pmatrix},
\end{equation}
where $h'_{0}(k_{y},k_{z})=t\sin k_{y}s_{y}+[t_{z}(\cos k_{z}-\cos Q)+m(2-\cos k_{y})]s_{z}-\mu s_{0}$.
The pairing term is
\begin{equation}
H^{RLO}=\frac{1}{2}\sum\limits_{\mathbf{k}}\tilde{\phi}^{\dagger}_{\mathbf{k}} \Delta_{+}\tau_{y}s_{y}\tilde{\phi}_{\mathbf{k}},
\end{equation}
where the summation over $k_{x}$ is now subjected to the constraint of $|E_{+}(\mathbf{k})|<\omega_{c}$ and $|E_{+}(2\mathbf{P}_{+}-\mathbf{k})|<\omega_{c}$. Making the Fourier transformation from $k_{x}$ to $n_{x}$, the above pairing term is expressed in terms of the above Nambu basis defined in the mixed $(n_{x},k_{y}k_{z})$ space as
\begin{equation}
H_{pair}^{RLO}=\frac{1}{2}\sum\limits_{k_{y}k_{z}}\sum\limits_{n_{1}n_{2}}\tilde{\phi}^{\dagger}_{n_{1}k_{y}k_{z}} \Delta_{+}\gamma_{n_{1}n_{2}}(k_{y},k_{z})\tau_{y}s_{y}\tilde{\phi}_{n_{2}k_{y}k_{z}}.
\end{equation}
The pairing amplitude between an arbitrary pair of layers, $n_{1}$ and $n_{2}$, is determined by
\begin{equation}
\gamma_{n_{1}n_{2}}(k_{y},k_{z})=\frac{1}{N_{x}}\sum\limits_{k_{x}}e^{ik_{x}(n_{1}-n_{2})},
\end{equation}
where the summation over $k_{x}$ is now restricted by $|E_{+}(k_{x},k_{y}k_{z})|<\omega_{c}$  and $|E_{+}(-k_{x},-k_{y},2Q-k_{z})|<\omega_{c}$, and will span only a finite region of $[-\pi,\pi)$. Therefore, the pairing is highly nonlocal in $n_{x}$. This is a manifestation of the Heisenberg uncertainty relation that the more localized is the physics in the momentum space, the more extended it will be in the real space. Note that, the long-range coupling between different layers occurs only through the pairing term. The terms corresponding to the normal state electronic structure only couples nearest-neighboring layers.

In a completely similar manner as the procedure taken for the second group of wave vectors, we can study the third group of wave vectors which consist of $(k_{y},k_{z})$ lying within the pairing shell surrounding $\mathbf{P}_{-}$. For these $(k_{y},k_{z})$, on one hand there exists $k_{x}$ that makes $|E_{+}(\mathbf{k})|<\omega_{c}$  and $|E_{+}(2\mathbf{P}_{-}-\mathbf{k})|<\omega_{c}$, and on the other hand $|k_{z}-Q|>|k_{z}-(-Q)|$. The formulae for this group of wave vectors are simply adapted from those for the second group by substituting $\mathbf{P}_{-}$ for $\mathbf{P}_{+}$ and $\Delta_{-}$ for $\Delta_{+}$.

\subsection{Time-reversal-symmetric and inversion-asymmetric Weyl metal with two pairs of Weyl pockets}

For the LO phase of the Weyl metal described by Eqs.(5)-(7), we consider the following pairing term
\begin{equation}
H_{\nu\alpha}^{LO}=\frac{1}{2}\sum\limits_{\mathbf{q}}\psi^{(\alpha)\dagger}_{\mathbf{q}}\Delta_{\nu\alpha}s_{\nu}i\sigma_{y} [\psi^{(\alpha)\dagger}_{-\mathbf{q}}]^{\text{T}}+\text{H.c.},
\end{equation}
where $\nu=0$ or 3, the wave vector dependence of $\Delta_{0\alpha}$ and $\Delta_{3\alpha}$ have been neglected. In the picture of RLO, the summation over $\mathbf{q}$ is restricted to the pairing energy shell of width $\omega_{c}$ centering at the Fermi surface.

The formulae that will be used to calculate the energy spectra of a thin film of this phase can be derived in the same manner as that for the RLO state of the Weyl metal with broken time-reversal symmetry. Consider a multilayer sample with $N_{x}$ layers and two surfaces perpendicular to the $x$ axis. Again, $k_{y}$ and $k_{z}$ are good quantum numbers, whereas the $x$ coordinate will be treated in real space. The energy spectra will still be calculated along a line in the surface BZ parallel to $k_{z}$ (i.e., with fixed $k_{y}$). As we vary $k_{z}$ along the line from $-\pi$ to $\pi$, we can put the wave vectors we encounter into five groups. The first group is outside all the four pairing energy shells and thus not subject to the pairing term. The remaining four groups correspond separately to states belonging to one of the four pairing energy shells.

The first group of wave vectors are those $(k_{y},k_{z})$ for which no $k_{x}$ makes $|E_{s+}(k_{x},k_{y},k_{z})|<\omega_{C}$, $s=\uparrow$ and $\downarrow$. No pairing term exists for these wave vectors. We define the Nambu basis in the usual manner $\tilde{\psi}^{\dagger}_{\mathbf{k}}=[\psi^{\dagger}_{\mathbf{k}},\psi^{\text{T}}_{-\mathbf{k}}]$. Making Fourier transformations from $k_{x}$ to $n_{x}$, we get
\begin{eqnarray}
&&H_{0}^{Nambu}=\frac{1}{2}\sum\limits_{n_{x}k_{y}k_{z}}\tilde{\psi}^{\dagger}_{n_{x}k_{y}k_{z}}[\tilde{h}'_{0}(k_{y},k_{z})\tilde{\psi}_{n_{x}k_{y}k_{z}}   \notag \\ &&+\tilde{h}'_{+}\tilde{\psi}_{n_{x}+1,k_{y}k_{z}}+\tilde{h}'_{-}\tilde{\psi}_{n_{x}-1,k_{y}k_{z}}].
\end{eqnarray}
Here, we have
\begin{equation}
\tilde{h}'_{0}(k_{y},k_{z})=\begin{pmatrix} h'_{1}(k_{y},k_{z}) & 0 \\
0 & -h'^{\text{T}}_{1}(-k_{y},-k_{z}) \end{pmatrix},
\end{equation}
\begin{equation}
\tilde{h}'_{+}=\begin{pmatrix} h_{xp} & 0 \\
0 & -h_{xp} \end{pmatrix},
\end{equation}
and $\tilde{h}'_{-}=\tilde{h}'^{\dagger}_{+}$. The $4\times4$ matrices in the above definitions are
\begin{eqnarray}
&&h'_{1}(k_{y},k_{z})=\xi'_{\mathbf{k}}\sigma_{0}s_{0}+t'\sin k_{y}\sigma_{x}s_{z}+\sigma_{z}[M(2-\cos k_{y})  \notag \\
&&+t'_{z}(\cos k_{z}-\cos Q)+t''_{z}\sin k_{z}s_{z}],
\end{eqnarray}
where $\xi'_{\mathbf{k}}=-2t_{1}\cos k_{y}-2t_{2}\cos k_{z}-\mu$, and
\begin{equation}
h_{xp}=-t_{1}\sigma_{0}+\frac{it'}{2}\sigma_{y}-\frac{1}{2}M\sigma_{z}.
\end{equation}

The second to the fifth groups of wave vectors can be treated in the same manner. Taking the second group within the pairing energy shell surrounding the Weyl node $\mathbf{P}_{1}$ as an example. In order for a wave vector $(k_{y},k_{z})$ to belong to this group, firstly there should be one or several $k_{x}$ that makes $|E_{\downarrow+}(k_{x},k_{y},k_{z})|<\omega_{c}$ and $|E_{\downarrow+}(-k_{x},-k_{y},2P_{1z}-k_{z})|<\omega_{c}$, secondly it should be closer to $\mathbf{P}_{1}$ than to the remaining three Weyl nodes which demands $|k_{z}-P_{1z}|<|k_{z}-P_{iz}|$ ($i=2,3,4$). For wave vectors within the pairing energy shell surrounding $\mathbf{P}_{1}$, the Nambu basis for the bulk state can be taken as $\tilde{\psi}^{\dagger}_{\mathbf{k}}=[\psi^{\dagger}_{\mathbf{k}},\psi^{\text{T}}_{2\mathbf{P}_{1}-\mathbf{k}}]$. Similar to the first group of wave vectors, the BdG Hamiltonian other than the pairing term can be transformed to the real space along $x$ easily, which gives
\begin{eqnarray}
&&H_{0}^{Nambu}=\frac{1}{2}\sum\limits_{n_{x}k_{y}k_{z}}\tilde{\psi}^{\dagger}_{n_{x}k_{y}k_{z}}[\tilde{h}''_{0}(k_{y},k_{z})\tilde{\psi}_{n_{x}k_{y}k_{z}}   \notag \\ &&+\tilde{h}'_{+}\tilde{\psi}_{n_{x}+1,k_{y}k_{z}}+\tilde{h}'_{-}\tilde{\psi}_{n_{x}-1,k_{y}k_{z}}],
\end{eqnarray}
where
\begin{equation}
\tilde{h}'_{0}(k_{y},k_{z})=\begin{pmatrix} h'_{1}(k_{y},k_{z}) & 0 \\
0 & -h'^{\text{T}}_{1}(-k_{y},2P_{1z}-k_{z}) \end{pmatrix}.
\end{equation}
The RLO pairing term is written as
\begin{equation}
H_{\nu}^{RLO}=\frac{1}{2}\sum\limits_{\mathbf{k}}\tilde{\psi}^{\dagger}_{\mathbf{k}} \Delta_{1}\begin{pmatrix} 0 & is_{\nu}\sigma_{y} \\
-is_{\nu}\sigma_{y} & 0 \end{pmatrix} \tilde{\psi}_{\mathbf{k}},
\end{equation}
where the summation over $k_{x}$ is now subjected to the constraints of $|E_{\downarrow+}(\mathbf{k})|<\omega_{c}$ and $|E_{\downarrow+}(2\mathbf{P}_{1}-\mathbf{k})|<\omega_{c}$. Making the Fourier transformation from $k_{x}$ to $n_{x}$, the above pairing term is expressed in terms of the above Nambu basis defined in the mixed $(n_{x},k_{y}k_{z})$ space as
\begin{equation}
H_{pair}^{RLO}=\frac{1}{2}\sum\limits_{k_{y}k_{z}}\sum\limits_{n_{1}n_{2}}\tilde{\psi}^{\dagger}_{n_{1}k_{y}k_{z}} \Delta_{1}\begin{pmatrix} 0 & is_{\nu}\sigma_{y} \\
-is_{\nu}\sigma_{y} & 0 \end{pmatrix} \tilde{\psi}_{n_{2}k_{y}k_{z}}.
\end{equation}
$\gamma_{n_{1}n_{2}}(k_{y},k_{z})$, the pairing amplitude between an arbitrary pair of layers $n_{1}$ and $n_{2}$, has the same expression as Eq.(B12).
The formulae for $(k_{y},k_{z})$ belonging to the third to the fifth groups of wave vectors are determined in the same manner as above and thus will not be listed here.

\section{formulae for the numerical calculation of energy spectra for a thin film: ELO}

In the ELO picture, the pairing term defined within each Fermi pocket is extended to span the whole BZ \cite{zhou15,cho12}. That is, the wave vector summations in Eqs.(4) and (13) are extended throughout the whole BZ. In this case, we can take advantage of the summation identity
\begin{equation}
\sum\limits_{\mathbf{k}}e^{i\mathbf{k}\cdot(\mathbf{R}_{m}-\mathbf{R}_{n})}=N\delta_{\mathbf{R}_{m},\mathbf{R}_{n}},
\end{equation}
and transform the pairing term to a very simple form in the real space.

\subsection{Time-reversal-symmetry broken Weyl metal with a single pair of Weyl pockets}

For the ELO state of time-reversal-symmetry broken Weyl metal with a single pair of Weyl nodes, the pairing term defined in Eq.(4) in the real space reads
\begin{eqnarray}
H^{ELO}&=&\frac{1}{2}\sum\limits_{\mathbf{R}}\phi^{\dagger}(\mathbf{R})is_{y}[\phi^{\dagger}(\mathbf{R})]^{\text{T}}  \notag \\ &&\cdot[\Delta_{+}e^{i\mathbf{R}\cdot(2\mathbf{P}_{+})}+\Delta_{-}e^{i\mathbf{R}\cdot(2\mathbf{P}_{-})}]+\text{H.c.}
\end{eqnarray}
Since the two Fermi pockets close to the two Weyl nodes are of the same size for the present model, it is reasonable to assume that $|\Delta_{+}|=|\Delta_{-}|$. We further assume that $\Delta_{\pm}$ are all real numbers, so that there are basically two cases $\Delta_{-}=\Delta_{+}$ or $\Delta_{-}=-\Delta_{+}$ \cite{cho12}. Since $\mathbf{P}_{-}=-\mathbf{P}_{+}$, it is clear that the pairing in the combined form show a sinusoidal oscillation in space. Take the plus sign in the above bracket, the spatial dependence of the pairing term is simply $e^{i\mathbf{R}\cdot(2\mathbf{P}_{+})}+e^{i\mathbf{R}\cdot(2\mathbf{P}_{-})}=2\cos(2\mathbf{R}\cdot\mathbf{P}_{+}) =2\cos(2n_{z}Q)$. $n_{z}$ is the label for the layer number along the $z$ direction. The lattice constants along the three directions will be taken as length units. In the ELO interpretation, the pairing is thus localized in space and shows an oscillatory dependence on spatial coordinates. In another word, the unit cell of the system is effectively enlarged along the $z$ direction which is the direction the two Weyl nodes aligns with each other.

For the sake of simplicity, we take a simple rational value for $Q$ as $Q=\pi/4$. Cases with more general $Q$ can be studied similarly. For $Q=\pi/4$, the ELO pairing enlarges the lattice parameter along $z$ direction as four times that for the normal phase. Considering the case $\Delta_{+}=\Delta_{-}=\Delta_{0}$, the four sites in the enlarged unit cell are associated with the pairing amplitudes of $\Delta_{1}=2\Delta_{0}$, $\Delta_{2}=0$, $\Delta_{3}=-2\Delta_{0}$, and $\Delta_{4}=0$. Therefore, we introduce the basis vectors for the four sites in an enlarged unit cell separately as $p_{\alpha s}(\mathbf{i})$, in which $\alpha=1,...,4$ labels the four sites in unit cell, $s=\uparrow$ or $\downarrow$ labels the two spin states on a site, $\mathbf{i}$ labels the unit cells.

In this extended unit cell, the ELO pairing term has a very simple form in the mixed $(n_{x},k_{y},\tilde{k}_{z})$ space
\begin{eqnarray}
&&H^{\text{ELO}}=\Delta_{0}\sum\limits_{n_{x}}\sum\limits_{k_{y}\tilde{k}_{z}}\sum\limits_{\alpha\beta}  \notag \\ &&[p^{\dagger}_{1\alpha}(n_{x},k_{y},\tilde{k}_{z})(is_{y})_{\alpha\beta}p^{\dagger}_{1\beta}(n_{x},-k_{y},-\tilde{k}_{z})-   \notag \\
&&p^{\dagger}_{3\alpha}(n_{x},k_{y},\tilde{k}_{z})(is_{y})_{\alpha\beta}p^{\dagger}_{3\beta}(n_{x},-k_{y},-\tilde{k}_{z})]+\text{H.c.},
\end{eqnarray}
where the tilde in $\tilde{k}_{z}$ is added to remind us that the lattice constant along $z$ axis has enlarged four times compared to the normal phase. The normal part of the model can also be easily transformed to the mixed space after introducing the extended unit cell with four sites.

It is interesting to notice that, in the ELO picture, the finite momentum pairing translates to spatial modulation in the pairing amplitude. In the enlarged unit cell, though the pairing amplitude has a site dependence, the pairing has a BCS-like form in the sense that only states with opposite wave vector are coupled together by the superconducting term. The differences from a true BCS pairing are manifested through a subsystem of the whole system where no pairing term is formed.

\subsection{Time-reversal-symmetric and inversion-asymmetric Weyl metal with two pairs of Weyl pockets}

In the ELO picture of the LO pairing in TRS and IB Weyl metal, the $\mathbf{q}$-summation for each Weyl node is extended to the whole BZ. Every wave vector thus participate in four separate pairings simultaneously, and therefore states in all four Weyl nodes are coupled together in a highly nontrivial manner.

Performing Fourier transformation to the two wave-vector-independent pairing channels defined in Eq.(13), we have
\begin{equation}
H_{\nu}^{ELO}=\frac{1}{2}\sum\limits_{\mathbf{i}}\tilde{\psi}^{\dagger}(\mathbf{R}_{\mathbf{i}})
\begin{pmatrix} 0 & is_{\nu}\sigma_{y}f(\mathbf{R}_{\mathbf{i}})   \\
-is_{\nu}\sigma_{y}f^{\ast}(\mathbf{R}_{\mathbf{i}}) & 0 \end{pmatrix}
\tilde{\psi}(\mathbf{R}_{\mathbf{i}}),
\end{equation}
where $\nu=0$ or $3$ indicates the two pairing channels and
\begin{equation}
f(\mathbf{R}_{\mathbf{i}})=\sum\limits_{\alpha}\Delta_{\nu\alpha}e^{2i\mathbf{R}_{\mathbf{i}}\cdot\mathbf{P}_{\alpha}}
\end{equation}
is the pairing amplitude on site $\mathbf{R}_{\mathbf{i}}$. In the special case of $\mathbf{P}_{\alpha}=(0,0,P_{\alpha z})$ and $P_{\alpha z}=(2\alpha-5)\pi/4$ ($\alpha=1,\ldots,4$) that we focus on, $f(\mathbf{R}_{\mathbf{i}})$ varies with the $z$ coordinate in a period of four lattice sites. The pairing amplitudes on the four consecutive layers in an order of increasing $z$ coordinate are $f_{\nu1}=\sum_{\alpha}\Delta_{\nu\alpha}$, $f_{\nu2}=i(\Delta_{\nu1}-\Delta_{\nu2}+\Delta_{\nu3}-\Delta_{\nu4})$, $f_{\nu3}=-f_{\nu1}$, and $f_{\nu4}=-f_{\nu2}$.

Similar to the case for the ELO state of the TRS Weyl metal, we enlarge the unit cell to contain four sites aligned along the $z$ axis. The four component creation operators on the four sites in the unit cell $\mathbf{R}_{i}$ are defined as $p^{\dagger}_{l}(\mathbf{R}_{i})$ ($l=1,...,4$). The ELO pairing term is thus written as
\begin{equation}
H_{\nu}^{ELO}=\frac{1}{2}\sum\limits_{\mathbf{i}}\sum\limits_{l}p^{\dagger}_{l}(\mathbf{R}_{\mathbf{i}})
f_{\nu l}is_{\nu}\sigma_{y}[p^{\dagger}_{l}(\mathbf{R}_{\mathbf{i}})]^{\text{T}}+\text{H.c.}
\end{equation}
For a film with two surfaces perpendicular to the $x$-axis, the pairing term is easily transformed to the mixed $(n_{x},k_{y},\tilde{k}_{z})$ space as
\begin{eqnarray}
H_{\nu}^{ELO}&=&\frac{1}{2}\sum\limits_{n_{x}k_{y}\tilde{k}_{z}}\sum\limits_{l}p^{\dagger}_{l}(n_{x},k_{y},\tilde{k}_{z})
f_{\nu l}is_{\nu}\sigma_{y}                                       \notag \\
&& [p^{\dagger}_{l}(n_{x},-k_{y},-\tilde{k}_{z})]^{\text{T}}+\text{H.c.}
\end{eqnarray}
The normal part of the model can also be easily transformed to the mixed $(n_{x},k_{y},\tilde{k}_{z})$ space after introducing the enlarged unit cell with four sites.

\section{analytical studies of surface Andreev bound states}

One peculiar feature of the surface Andreev bound states (SABSs) shown in Figs. 2(a)-2(d), 4(a), and 4(b) is that they disperse along $k_{z}$ in the same direction on both of the two surfaces. This is in contrast to the usual counter-propagating boundary states. In this section, we make an analytical study on the SABSs. We will focus on the time-reversal-symmetry broken Weyl metal with a single pair of Weyl pockets in the normal phase. The procedures can be readily extended to the model with two pairs of Weyl nodes in the normal phase. The analytical analysis below confirms the numerical results shown in the main text for the SABSs.

\subsection{Insights from the low-energy effective model}

Defining $\tilde{E}_{\nu}(\mathbf{k})=\nu\sqrt{d_{1}^{2}(\mathbf{k})+d_{2}^{2}(\mathbf{k})+d_{3}^{2}(\mathbf{k})}$ ($\nu=\pm$), the eigenenergies shown in Eq.(3) are rewritten as
\begin{equation}
E_{\nu}(\mathbf{k})=\tilde{E}_{\nu}(\mathbf{k})-\mu.
\end{equation}
For a general wave vector $\mathbf{k}$, we can take the eigenvector for the $E_{\nu}(\mathbf{k})$ eigenstate as
\begin{equation}
\begin{pmatrix} u_{\nu\uparrow}(\mathbf{k}) \\
u_{\nu\downarrow}(\mathbf{k})\end{pmatrix}=\frac{1}{D_{\nu}(\mathbf{k})}\begin{pmatrix}
\tilde{E}_{\nu}(\mathbf{k})+d_{3}(\mathbf{k}) \\
d_{+}(\mathbf{k}) \end{pmatrix},
\end{equation}
where we have introduced $d_{\pm}(\mathbf{k})=d_{1}(\mathbf{k})\pm id_{2}(\mathbf{k})$ and $D_{\nu}(\mathbf{k})=\sqrt{2\tilde{E}_{\nu}(\mathbf{k})[\tilde{E}_{\nu}(\mathbf{k})+d_{3}(\mathbf{k})]}$.

In consistency with the parameters used in the main text, we assume $\mu>0$ and the Fermi surface consists of two small and well-separated Weyl pockets. Let us focus on the weak-coupling picture and consider the LO paring formed in the Weyl pocket centering around $\mathbf{P}_{\alpha}$ ($\alpha=\pm$). In the eigenbasis defined by Eq.(D2), the LO pairing defined by Eq.(4) is projected to the following effective pairing between the single-particle states $\mathbf{\mathbf{q}}+\mathbf{P}_{\alpha}$ and $\mathbf{-\mathbf{q}}+\mathbf{P}_{\alpha}$ close to the Fermi surface
\begin{equation}
\frac{1}{2}\sum\limits_{\mathbf{q}}\tilde{\Delta}_{\alpha}(\mathbf{q})a^{\dagger}_{\alpha}(\mathbf{q})a^{\dagger}_{\alpha}(-\mathbf{q})+\text{H.c.},
\end{equation}
where the single-particle operator $a^{\dagger}_{\alpha}(\mathbf{q})$ are defined in Sec.III of the main text. Since $\mathbf{q}$ is small, we are justified to approximate $d_{i}(\pm\mathbf{q}+\mathbf{P}_{\alpha})$ ($i=x,y,z$) as polynomials of $q_{i}$ ($i=x,y,z$). The pairing amplitude is thus obtained as
\begin{equation}
\tilde{\Delta}_{\alpha}(\mathbf{q})=-f(\mathbf{q})\frac{\text{sgn}(t)(q_{x}-iq_{y})}{\sqrt{q_{x}^{2}+q_{y}^{2}}}\Delta_{\alpha},
\end{equation}
where
\begin{equation}
f(\mathbf{q})=[1-\frac{t_{z}^{2}\sin^{2}Q q_{z}^{2}[t_{z}\cos Q q_{z}^{2}-m(q_{x}^{2}+q_{y}^{2})]^{2}}{4[t^{2}(q_{x}^{2}+q_{y}^{2})+t_{z}^{2}\sin^{2}Q q_{z}^{2}]^{2}}]^{-\frac{1}{2}}
\end{equation}
is an even function of $\mathbf{q}$ and satisfies $f(\mathbf{q})\in[1,(1-\frac{\mu^{2}\cos^{2}Q}{4t_{z}^{2}\sin^{4}Q})^{-\frac{1}{2}}]$, where we have used the linear approximation to the dispersion to obtain the Fermi momentum along $q_{z}$. $f(\mathbf{q})$ takes the maximal value on the line of $q_{x}=q_{y}=0$. Note that the effective pairing $\tilde{\Delta}_{\alpha}(\mathbf{q})$ depends on the choice of the eigenbasis for the single-particle states $a^{\dagger}_{\alpha}(\mathbf{q})$. In Ref.\cite{rwang16}, another basis set was chosen and the effective pairing adopts a slightly different form, besides the omission of the $f(\mathbf{q})$ factor. An important feature of the effective pairing is that $\tilde{\Delta}_{\alpha}(\mathbf{q})$ depends on $\alpha$ only through $\Delta_{\alpha}$. As a result, the two cases $\Delta_{-}=\pm \Delta_{+}$ differ by a relative sign in the effective pairing.

The low-energy effective model for the LO pairing in the Weyl pocket centering around $\mathbf{P}_{\alpha}$ is thus
\begin{equation}
h_{\alpha}(\mathbf{q})=\begin{pmatrix} E_{+}(\mathbf{q}+\mathbf{P}_{\alpha}) & \tilde{\Delta}_{\alpha}(\mathbf{q}) \\
\tilde{\Delta}_{\alpha}^{\ast}(\mathbf{q}) & -E_{+}(-\mathbf{q}+\mathbf{P}_{\alpha}) \end{pmatrix},
\end{equation}
which is expressed in the Nambu basis $[a^{\dagger}_{\alpha}(\mathbf{q}), a_{\alpha}(-\mathbf{q})]$. The model has a particle-hole symmetry
\begin{equation}
\Xi^{-1}h_{\alpha}(\mathbf{q})\Xi=-h_{\alpha}(-\mathbf{q}),
\end{equation}
where $\Xi=\tau_{1}K$ is the particle-hole operator. Here, we have regarded $\mathbf{P}_{\alpha}$ as the center of momentum for the $\alpha$-th Weyl pocket. By focusing separately on the off-diagonal pairing term and the diagonal terms, we can arrive at two important conclusions on the properties of the LO phase.

The pairing term, as has been noticed before \cite{rwang16,bednik15}, is chiral in the $q_{x}q_{y}$ plane. On the other hand, it is an even function of $q_{z}$ and depends only weakly on $q_{z}$. On the line of $q_{x}=q_{y}=0$, the effective pairing is completely independent of $q_{z}$. Therefore, from the pairing term alone, we would expect chiral surface states along $q_{y}$ and no dispersion of the surface state along $q_{z}$, on the two surfaces perpendicular to $x$.

The above expectation is natural for conventional BCS-like pairing in systems with a centrosymmetric Fermi surface. The band structure close to the Weyl node, however, is not exactly symmetric with respect to the Weyl node. Specifically, $E_{+}(\mathbf{q}+\mathbf{P}_{\alpha})\ne E_{+}(-\mathbf{q}+\mathbf{P}_{\alpha})$ for $q_{z}\ne0$ and thus the Weyl cones tilt along $q_{z}$. As a result, the two diagonal terms of $h_{\alpha}(\mathbf{q})$ are not opposite to each other and the quasiparticle spectrum is not symmetric with respect to $E=0$ for wave vectors with $q_{z}\ne0$ \cite{hao1617}. Note that this is not in conflict with the particle-hole symmetry, which connects states at $\mathbf{q}$ and states at $-\mathbf{q}$. To quantify the above effect, we rewrite the two diagonal terms of $h_{\alpha}(\mathbf{q})$ as
\begin{eqnarray}
&&\frac{E_{+}(\mathbf{q}+\mathbf{P}_{\alpha})-E_{+}(-\mathbf{q}+\mathbf{P}_{\alpha})}{2}\tau_{0}  \notag \\ &&+\frac{E_{+}(\mathbf{q}+\mathbf{P}_{\alpha})+E_{+}(-\mathbf{q}+\mathbf{P}_{\alpha})}{2}\tau_{3}.
\end{eqnarray}
The second term is an even function of $\mathbf{q}$. It has the standard form of the diagonal terms in a conventional BCS superconductor with a spectrum locally symmetric with respect to $E=0$. The coefficient of the first term quantifies the deviation of the center of the quasiparticle spectrum from $E=0$. It is approximately
\begin{equation}
\alpha\frac{t_{z}\cos Q q_{z}^{2}-m(q_{x}^{2}+q_{y}^{2})}{2\sqrt{t^{2}(q_{x}^{2}+q_{y}^{2})+(t_{z}\sin Q q_{z})^{2}}}t_{z}\sin Q q_{z},
\end{equation}
where $\alpha=\pm$ for the two Weyl pockets. Eq.(D9) has two important properties. Firstly, it has a factor $\alpha$ and has opposite sign for the two pockets. Secondly, it is an \emph{odd function} of $q_{z}$. From these properties of Eq.(D9), we can understand qualitatively the SABSs shown in Figs.2(a) to 2(d) (which can be extended readily to the cases of Figs.4(a) and 4(b)). The presence of this odd in $q_{z}$ term makes possible the existence of odd in $q_{z}$ dispersion of the SABSs traversing the superconducting gap. On the other hand, since this term is proportional to $\tau_{0}$ and thus non-chiral, the SABSs on the two surfaces disperse along $q_{z}$ in the same direction. In addition, the $\alpha$ factor in this term renders the SABSs for the two Weyl pockets to disperse along the opposite direction.

\subsection{Explicit construction of the surface Andreev bound states}

Having gained some insights from the low-energy effective model, we try to construct explicitly the wave functions and dispersions of the surface Andreev bound states (SABSs). Working in the original basis, the model for the RLO pairing within the $\alpha$-th Weyl pocket is approximated as
\begin{eqnarray}
&&h_{\alpha}(\mathbf{q})=t(q_{x}\tau_{0}\sigma_{1}+q_{y}\tau_{3}\sigma_{2})-\alpha t_{z}\sin Q q_{z}\tau_{0}\sigma_{3}-\mu\tau_{3}\sigma_{0}   \notag \\  &&-\Delta\tau_{2}\sigma_{2}+\frac{1}{2}[m(q_{x}^{2}+q_{y}^{2})-t_{z}\cos Q q_{z}^{2}]\tau_{3}\sigma_{3},
\end{eqnarray}
where we have dropped the subscript $\alpha$ on $\Delta$ and regard $\Delta$ as a real number. Consider the SABSs on the two $yz$ surfaces of a film grown along the $x$ direction. They are solutions of the differential equation
\begin{equation}
h_{\alpha}(-i\partial_{x},q_{y},q_{z})\varphi(x,q_{y},q_{z})=E\varphi(x,q_{y},q_{z}),
\end{equation}
where $E$ is the eigenenergy and $\varphi(x,q_{y},q_{z})$ is the corresponding eigenvector.

For simplicity, we first solve the differential equation for $q_{y}=q_{z}=0$ and then get the eigenstates for nonzero $q_{y}$ or $q_{z}$ by perturbation. For $q_{z}=0$ the quasiparticle spectrum is symmetric with respect to $E=0$. The SABSs, if they exist, should have zero-energy solutions. We thus look for solutions of the form
\begin{equation}
h_{\alpha}(-i\partial_{x},0,0)e^{\lambda x}\eta_{\lambda}=0.
\end{equation}
The eigenfunction for $\lambda$ is
\begin{equation}
\text{det}[h_{\alpha}(-i\lambda,0,0)]=(\frac{m^{2}}{4}\lambda^{4}-t^{2}\lambda^{2}-\Delta^{2}-\mu^{2})^2-(2t\Delta\lambda)^{2}=0,
\end{equation}
which factors into two fourth-order algebraic equations and are exactly solvable. Denoting the four solutions of $\lambda$ from $\frac{m^{2}}{4}\lambda^{4}-t^{2}\lambda^{2}+2t\Delta\lambda-\Delta^{2}-\mu^{2}=0$ as $\lambda_{\beta}$ ($\beta=1,...,4$). The four solutions of $\lambda$ from $\frac{m^{2}}{4}\lambda^{4}-t^{2}\lambda^{2}-2t\Delta\lambda-\Delta^{2}-\mu^{2}=0$ are their opposites, $\lambda_{\beta}'=-\lambda_{\beta}$ ($\beta=1,...,4$). Suppose we have obtained $\lambda_{\beta}$ ($\beta=1,...,4$), the two sets of eigenvectors $\eta_{\beta}$ and $\eta_{\beta}'$ which correspond separately to $\lambda_{\beta}$ and $\lambda_{\beta}'$ can be solved as
\begin{equation}
\eta_{\beta}=\frac{1}{N_{\beta}}\begin{pmatrix} i(\frac{m}{2}\lambda_{\beta}^{2}-\mu)  \\  \Delta-t\lambda_{\beta}  \\ \frac{m}{2}\lambda_{\beta}^{2}-\mu   \\    i(\Delta-t\lambda_{\beta}) \end{pmatrix},
\end{equation}
\begin{equation}
\eta_{\beta}'=\frac{1}{N_{\beta}'}\begin{pmatrix} -i(\frac{m}{2}\lambda_{\beta}'^{2}-\mu)  \\  \Delta+t\lambda_{\beta}'  \\ \frac{m}{2}\lambda_{\beta}'^{2}-\mu   \\    -i(\Delta+t\lambda_{\beta}') \end{pmatrix},
\end{equation}
where $\beta=1,...,4$, $N_{\beta}=\sqrt{2[|\frac{m}{2}\lambda_{\beta}^{2}-\mu|^{2}+|\Delta-t\lambda_{\beta}|^{2}]}$ and $N_{\beta}'=\sqrt{2[|\frac{m}{2}\lambda_{\beta}'^{2}-\mu|^{2}+|\Delta+t\lambda_{\beta}'|^{2}]}=N_{\beta}$.

To construct the zero energy SABSs localized on the left and right boundary of the film, we regroup the eight solutions for $\lambda$ into one set $\tilde{\lambda}_{\beta}$ ($\beta=1,...,4$) with negative real parts and another set $\tilde{\lambda}_{\beta}'=-\tilde{\lambda}_{\beta}$ ($\beta=1,...,4$) with positive real parts. The eigenvectors are also regrouped and denoted as $\tilde{\eta}_{\beta}$ and $\tilde{\eta}_{\beta}'$ ($\beta=1,...,4$), respectively.

The surface states on the left surface of the film satisfy the boundary conditions
\begin{equation}
\varphi_{L}(x=0)=\varphi_{L}(x=+\infty)=0.
\end{equation}
In writing down the second equality, we have ignored the long-range pairing correlation intrinsic to the RLO picture. $\varphi_{L}(x)$ thus should be constructed as a linear combination of $\tilde{\eta}_{\beta}$ ($\beta=1,...,4$),
\begin{equation}
\varphi_{L}(x)=\sum\limits_{\beta=1}^{4}a_{\beta}\tilde{\eta}_{\beta}e^{\tilde{\lambda}_{\beta}x},
\end{equation}
where the coefficients $a_{\beta}$ ($\beta=1,...,4$) are determined by the boundary condition $\varphi_{L}(x=0)=0$.
The surface states on the right surface of the film satisfy the boundary conditions
\begin{equation}
\varphi_{R}(x=0)=\varphi_{R}(x=-\infty)=0.
\end{equation}
Again, we have ignored the long-range pairing correlation intrinsic to the RLO picture in writing down the second equality. $\varphi_{R}(x)$ should be constructed as a linear combination of $\tilde{\eta}_{\beta}'$ ($\beta=1,...,4$),
\begin{equation}
\varphi_{R}(x)=\sum\limits_{\beta=1}^{4}a_{\beta}'\tilde{\eta}_{\beta}'e^{\tilde{\lambda}_{\beta}'x},
\end{equation}
where the coefficients $a_{\beta}'$ ($\beta=1,...,4$) are determined by the boundary condition $\varphi_{R}(x=0)=0$. Once the general solution for $a_{\beta}$ and $a_{\beta}'$ ($\beta=1,...,4$) are obtained from the boundary conditions, we can further normalize them by requiring $\int_{x=0}^{x=+\infty}\varphi_{L}^{\dagger}(x)\varphi_{L}(x)dx=\int_{x=-\infty}^{x=0}\varphi_{R}^{\dagger}(x)\varphi_{R}(x)dx=1$. For the parameters used in the main text for Figs. 1(a) and 1(c), there is one and only one linearly independent set of solutions for $a_{\beta}$ and $a_{\beta}'$ ($\beta=1,...,4$). The zero-energy SABSs are thus nondegenerate Majorana fermions.

\begin{figure}[!htb]\label{fig10} \centering
\includegraphics[width=8.6cm,height=11.6cm]{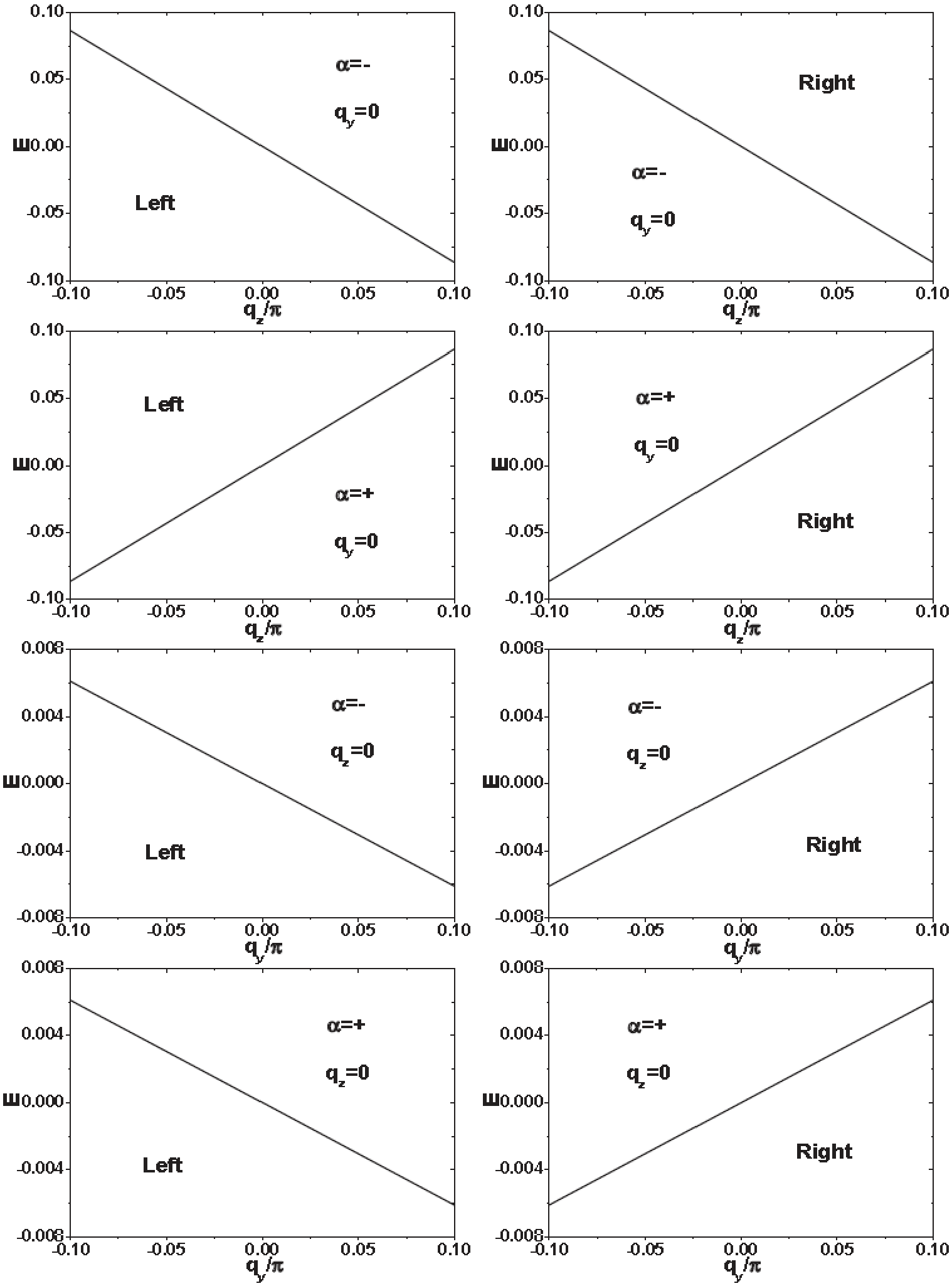}  \\
\caption{The dispersion of the surface Andreev bound states (SABSs), on the left (with `Left' on the figures) and right (with `Right' on the figures) $yz$ surfaces of a film grown along $x$. $\alpha=-$ and $\alpha=+$ indicates the Weyl pocket to which the SABSs correspond. $\Delta_{+}=\Delta_{-}=0.02$.}
\end{figure}

After solving the zero-energy SABSs for $q_{y}=q_{z}=0$, we can calculate the dispersion of the SABSs at finite $q_{y}$ or $q_{z}$ by taking the following term as perturbation
\begin{equation}
h_{\alpha}(q_{y},q_{z})=h_{\alpha}(\mathbf{q})-h_{\alpha}(q_{x},q_{y}=0,q_{z}=0).
\end{equation}
The dispersion of the SABSs on the left surface is thus
\begin{eqnarray}
E^{L}_{\alpha}(q_{y},q_{z})&=&\int_{x=0}^{x=+\infty}dx\varphi_{L}^{\dagger}(x)h_{\alpha}(q_{y},q_{z})\varphi_{L}(x) \notag \\
&=&\sum\limits_{\alpha',\beta'}\frac{-a_{\alpha'}^{\ast}a_{\beta'}}{\tilde{\lambda}_{\alpha'}^{\ast}+\tilde{\lambda}_{\beta'}} \tilde{\eta}_{\alpha'}^{\dagger}h_{\alpha}(q_{y},q_{z})\tilde{\eta}_{\beta'}.
\end{eqnarray}
Similarly, the dispersion of the SABSs on the right surface is
\begin{equation}
E^{R}_{\alpha}(q_{y},q_{z})=\sum\limits_{\alpha',\beta'}\frac{a_{\alpha'}'^{\ast}a_{\beta'}'}{\tilde{\lambda}_{\alpha'}'^{\ast}+\tilde{\lambda}_{\beta'}'} \tilde{\eta}_{\alpha'}'^{\dagger}h_{\alpha}(q_{y},q_{z})\tilde{\eta}_{\beta'}'.
\end{equation}
The numerical results for the dispersions of the SABSs are shown in Fig.10, for $\Delta_{+}=\Delta_{-}=0.02$. The results confirm the qualitative aspects of the SABSs in Fig.2(a) and Fig. 2(c). In addition, they also confirm the analysis in the previous section on the chiral nature of the SABSs along $q_{y}$, and that the effective pairings for the two Weyl nodes are the same for $\Delta_{+}=\Delta_{-}$.

\section{mean-field theory}

\subsection{Difference between RLO and ELO in terms of mean-field theory}

In this section, we illustrate the difference between the RLO and ELO pictures from the point of view of mean-field theory to the pairing transition. For simplicity, we focus on the model defined by Eq.(1) of the main text, which has a single pair of Weyl nodes in the normal phase. Consider the following phenomenological pairing interaction \cite{cho12}
\begin{equation}
H_{int}=V_{0}\sum\limits_{\mathbf{i}}n_{\mathbf{i}}n_{\mathbf{i}}+V_{1}\sum\limits_{<\mathbf{i},\mathbf{j}>}n_{\mathbf{i}}n_{\mathbf{j}},
\end{equation}
which consists of an on-site term proportional to $V_{0}$ and a nearest-neighboring (NN) interaction term proportional to $V_{1}$. $n_{\mathbf{i}}=\sum_{\sigma=\uparrow,\downarrow}c^{\dagger}_{\mathbf{i}\sigma}c_{\mathbf{i}\sigma}$ is the number operator for electrons on site $\mathbf{i}$. $H_{int}$ can be written in the momentum space as
\begin{equation}
H_{int}=\sum\limits_{\mathbf{q}}V(\mathbf{q})n_{\mathbf{q}}n_{-\mathbf{q}}+V_{0}n_{\mathbf{q}=\mathbf{0}}.
\end{equation}
$n_{\mathbf{q}}=\sum_{\mathbf{k}\sigma}c^{\dagger}_{\mathbf{k}\sigma}c_{\mathbf{k+q},\sigma}$. $n_{\mathbf{q}=\mathbf{0}}$ is the operator for the total number of electrons in the system.
\begin{equation}
V(\mathbf{q})=\frac{1}{N}[V_{0}+V_{1}(\cos q_{x}+\cos q_{y}+\cos q_{z})],
\end{equation}
where $N$ is the total number of unit cells in the simple cubic lattice, which is also the number of wave vectors in the first BZ.

The mean-field decoupling of $H_{int}$ can be carried out in two manners, in the real space by working with Eq.(E1) or in the momentum space in terms of Eq.(E2). In the real-space approach, the pairing correlation is introduced to every site by defining mean-fields (superconducting order parameters) like $<c_{\mathbf{i}\sigma}c_{\mathbf{i}\sigma'}>$ and $<c_{\mathbf{i}\sigma}c_{\mathbf{i+\boldsymbol{\delta}}\sigma'}>$, where $\mathbf{i}$+$\boldsymbol{\delta}$ is an NN site of $\mathbf{i}$. When Fourier transformed to the momentum space, the pairing correlation runs through the whole BZ. If we are considering the LO phase, by assuming a sinusoidally varying superconducting order parameter, the \emph{local-in-real-space} mean-field decoupling naturally gives the ELO state.

When working with Eq.(E2), we can of course introduce mean-fields in a manner exactly identical to the Fourier transformed local-in-real-space pairing. However, in the spirit of the conventional BCS weak-coupling pairing, the pairing term is also commonly restricted close to the Fermi surface, within a pairing shell much narrower than the chemical potential. This \emph{local-in-momentum-space} mean-field decoupling is most commonly used in making analytical study of the mean-field pairing instability \cite{cho12,bednik15,lu16}. It is exactly what we meant by RLO when applying the local-in-momentum-space mean-field decoupling to Eq.(E2) to the LO channel.

Which of the two pictures are correct depends on the nature of the pairing interaction. If the underlying pairing mechanism is the electron-phonon coupling, then the pairing interaction is attractive only within a shell around the Fermi surface of the width the Debye frequency. In this case, it is natural to work with Eq.(E2) in the momentum space form and restrict the summation over momentum to the neighborhood of the Fermi surface. On the other hand, if the underlying pairing mechanism is strong electron-electron interaction, the real-space pairing picture can also be more relevant. An example is the $t$-$J$ model arising from strong on-site Coulomb repulsion, which has been applied to high temperature superconducting cuprates \cite{baskaran87,kotliar88} and also to the iron group superconductors \cite{si08,seo08}. Note that, while the underlying pairing mechanism can be strong correlation, the effective pairing interaction may not indeed be very large. If the pairing interaction under discussion is of this origin, it is more convenient to start from the real-space form of the interaction, e.g. Eq.(E1), and introduce local-in-real-space mean-field decoupling.

Our purpose in the present work is to point out the fundamental differences between the two pictures when they are applied to the LO state of Weyl metals. Which kind of pairing interaction is the correct pairing interaction for a specific Weyl metal is however beyond the scope of this study. We simply point out that the Weyl metals can be found in vastly different systems. Some systems are clearly not dominated by strong electron correlation, such as the heterostructure consisting of periodically stacked layers of topological insulator thin film and normal insulator thin film \cite{burkov11,meng12}. In these systems, the pairing is most likely mediated by the electron-phonon interaction and the local-in-momentum-space (RBZP) picture seems more reasonable. On the other hand, there are also systems where the effect of electron correlation is strong and crucial in determining the true ground state, such as the pyrochlore iridates \cite{wan11,yang11} and the Weyl-Kondo semimetal \cite{lai16,dzsaber16}. In these systems, the local-in-real-space (EBZP) picture might instead be more appropriate.

\subsection{Comparison of BCS phase and LO phase in the weak-coupling limit}

Presently, there exist some controversy in the literature on the leading weak-coupling pairing instability, the BCS state versus the LO state, in the model with one pair of Weyl nodes \cite{cho12,bednik15}. We clarify this issue by mean-field calculation of transition temperature and zero-temperature ground state energy for the model defined by Eqs.(1)-(3) of the main text and the pairing interactions defined by Eqs.(E1)-(E3).

For the sake of simplicity, we first focus on the ideal case for which the Fermi surface consists of two small spherical Weyl pockets and then analyze possible effects of more realistic parameters. For a wave vector $\mathbf{k}=\mathbf{q}+\mathbf{P}_{\alpha}$ close to $\mathbf{P}_{\alpha}$ ($\alpha=\pm$), we can expand approximately $d_{1}\simeq tq_{x}$, $d_{2}\simeq tq_{y}$, and $d_{3}\simeq -\alpha t_{z}\sin Q q_{x}$. Assuming further that $t_{z}\sin Q=t$, the Fermi velocity is isotropic around each Weyl node, the two Weyl pockets are spherical and have the same size. The annihilation operator for the state belonging to the $\alpha$-th Weyl pocket can be written as $c_{\sigma}(\mathbf{k})=c_{\sigma}(\mathbf{q}+\mathbf{P}_{\alpha})=c_{\alpha,\mathbf{q},\sigma}$. We assume that the Fermi level lies in the conduction band, that is $\mu>0$. In the weak-coupling approximation, the pairing interaction is nonzero only if all states involved in the interaction are in the neighborhood of the Fermi surface, $|\xi_{\mathbf{q}}-\mu|<\omega_{c}$. $\xi_{\mathbf{q}}=|t|q$. The energy cutoff $\omega_{c}$ is assumed to be much smaller than $\mu$. In this convention, the weak-coupling pairing interactions responsible for the RLO state and for the BCS state are separately \begin{eqnarray}
&&H_{int}^{\text{RLO}}=\sum\limits_{\alpha,\sigma_{1},\sigma_{2}}\sum\limits_{\mathbf{q}_{1},\mathbf{q}_{2}}V(\mathbf{q}_{1}-\mathbf{q}_{2})   \notag \\
&& c^{\dagger}_{\alpha,\mathbf{q}_{1},\sigma_{1}}c^{\dagger}_{\alpha,-\mathbf{q}_{1},\sigma_{2}} c_{\alpha,-\mathbf{q}_{2},\sigma_{2}}c_{\alpha,\mathbf{q}_{2},\sigma_{1}},
\end{eqnarray}
and
\begin{eqnarray}
&&H_{int}^{\text{BCS}}=\sum\limits_{\alpha_{1},\alpha_{2}}\sum\limits_{\sigma_{1},\sigma_{2}}\sum\limits_{\mathbf{q}_{1},\mathbf{q}_{2}} V(\mathbf{q}_{1}+\mathbf{P}_{\alpha_{1}}-\mathbf{q}_{2}-\mathbf{P}_{\alpha_{2}})    \notag \\ && c^{\dagger}_{\alpha_{1},\mathbf{q}_{1},\sigma_{1}}c^{\dagger}_{-\alpha_{1},-\mathbf{q}_{1},\sigma_{2}}
c_{-\alpha_{2},-\mathbf{q}_{2},\sigma_{2}}c_{\alpha_{2},\mathbf{q}_{2},\sigma_{1}}.
\end{eqnarray}
Here and after in this section, the summations over the relative momenta (i.e., $\mathbf{q}_{1}$ and $\mathbf{q}_{2}$) are always restricted to a shell of width $2\omega_{c}$ around the Fermi surface.

Following previous works, we focus on the spin-singlet pairing channel for both the RLO state and the BCS state \cite{cho12,bednik15,zhou15}. Remembering that the relative momenta $|\mathbf{q}|\simeq0$, we can make the approximations $\cos q_{\alpha}\simeq1$ ($\alpha=x,y,z$) and $\cos(q_{1z}-q_{2z}\pm 2Q)\simeq \cos(2Q)$. The pairing interactions can thus be reformulated as \cite{cho12}
\begin{equation}
H_{int}^{\text{RLO}}=\frac{V_{LO}}{2N}\sum\limits_{\alpha,\mathbf{q}_{1},\mathbf{q}_{2}}\hat{X}^{\dagger}_{\alpha}(\mathbf{q}_{1})\hat{X}_{\alpha}(\mathbf{q}_{2}), \notag
\end{equation}
\begin{equation}
\hat{X}_{\alpha}(\mathbf{q})=[\phi_{-\mathbf{q}+\mathbf{P}_{\alpha}}]^{\text{T}}(-i\sigma_{y})\phi_{\mathbf{q}+\mathbf{P}_{\alpha}},    
\end{equation}
\begin{equation}
V_{LO}=V_{0}+3V_{1}, \notag
\end{equation}
for the RLO state, and
\begin{equation}
H_{int}^{\text{BCS}}=\frac{2V_{BCS}}{N}\sum\limits_{\mathbf{q}_{1},\mathbf{q}_{2}}\hat{X}^{\dagger}(\mathbf{q}_{1})\hat{X}(\mathbf{q}_{2}), \notag
\end{equation}
\begin{equation}
\hat{X}(\mathbf{q})=[\phi_{-\mathbf{q}+\mathbf{P}_{-}}]^{\text{T}}(-i\sigma_{y})\phi_{\mathbf{q}+\mathbf{P}_{+}},   
\end{equation}
\begin{equation}
V_{BCS}=V_{0}+\frac{5+\cos(2Q)}{2}V_{1}, \notag
\end{equation}
for the BCS state. $N$ is the total number of unit cells in the system, which is equal to the total volume since we have set the lattice constants to the length units. Defining the mean-field order parameters
\begin{equation}
\Delta_{\alpha}=\frac{V_{LO}}{N}\sum\limits_{\mathbf{q}}<\hat{X}_{\alpha}(\mathbf{q})>,
\end{equation}
and
\begin{equation}
\Delta=\frac{V_{BCS}}{N}\sum\limits_{\mathbf{q}}<\hat{X}(\mathbf{q})>,
\end{equation}
the pairing interactions are approximately decoupled into
\begin{equation}
H_{int}^{\text{LO}}=\frac{1}{2}\sum\limits_{\alpha,\mathbf{q}}[\Delta^{\ast}_{\alpha}\hat{X}_{\alpha}(\mathbf{q}) +\Delta_{\alpha}\hat{X}^{\dagger}_{\alpha}(\mathbf{q})]-\frac{N}{2V_{LO}}\sum\limits_{\alpha}|\Delta_{\alpha}|^{2},
\end{equation}
and
\begin{equation}
H_{int}^{\text{BCS}}=2\sum\limits_{\mathbf{q}}[\Delta^{\ast}\hat{X}(\mathbf{q}) +\Delta\hat{X}^{\dagger}(\mathbf{q})] -\frac{2N}{V_{BCS}}|\Delta|^{2}.
\end{equation}
Taking the eigenvector defined in Eq.(D2) for the single-particle states in the normal phase, we can project the pairing terms in the mean-field Hamiltonian to the conduction band. For the RLO state, we have
\begin{equation}
H_{int}^{\text{LO}}\simeq\frac{1}{2}\sum\limits_{\alpha,\mathbf{q}}[\tilde{\Delta}_{\alpha}(\mathbf{q})a^{\dagger}_{\alpha}(\mathbf{q})a^{\dagger}_{\alpha}(-\mathbf{q})+\text{H.c.}] -\frac{N}{2V_{LO}}\sum\limits_{\alpha}|\Delta_{\alpha}|^{2},
\end{equation}
where the effective pairing amplitude
\begin{equation}
\tilde{\Delta}_{\alpha}(\mathbf{q})=-\frac{\text{sgn}(t)(q_{x}-iq_{y})}{\sqrt{q_{x}^{2}+q_{y}^{2}}}\Delta_{\alpha},
\end{equation}
For the BCS state, we have
\begin{equation}
H_{int}^{\text{BCS}}=2\sum\limits_{\mathbf{q}}[\tilde{\Delta}(\mathbf{q})a^{\dagger}_{+}(\mathbf{q})a^{\dagger}_{-}(-\mathbf{q})+\text{H.c.}] -\frac{2N}{V_{BCS}}|\Delta|^{2},
\end{equation}
where
\begin{equation}
\tilde{\Delta}(\mathbf{q})=-\frac{\text{sgn}(t)(q_{x}-iq_{y})}{\sqrt{q_{x}^{2}+q_{y}^{2}+q_{z}^{2}}}\Delta.
\end{equation}

The formula determining the transition temperature $T_{c}$ for the RLO state and the BCS state turn out to have the same form
\begin{equation}
1=-\frac{V}{2\pi^{2}|t|^3}\int_{0}^{\omega_{c}}d\xi\frac{\xi^{2}+\mu^{2}}{\xi}\text{tanh}\frac{\xi}{2k_{B}T_{c}}.
\end{equation}
The effective pairing interaction is $V=V_{LO}$ for the RLO state and is $V=\frac{4}{3}V_{BCS}$ for the BCS state. The additional factor of $\frac{4}{3}$ for the BCS state is a combination of two factors. Firstly, as was pointed out by Cho \emph{et al} the nodal nature of the BCS state [Eq.(E15)] in comparison to the fully-gapped nature of the RLO state [Eq.(E13)] gives a factor of $2/3$ because \cite{cho12}
\begin{equation}
\frac{2}{3}=\int_{0}^{\pi}\sin^{3}\theta d\theta/\int_{0}^{\pi}\sin\theta d\theta.
\end{equation}
Secondly, as was emphasized by Bednik \emph{et al} and is also clear from Eqs.(E4)-(E7), the phase space responsible for the BCS state is twice the phase space responsible for the RLO state \cite{bednik15}.

In the parameter regions where Eq.(E16) has solution for $T_{c}$, $T_{c}$ increases as $-V>0$ increases. Therefore, for purely on-site pairing interaction ($V_{0}<0$ and $V_{1}=0$), the BCS state is favored. For purely nearest-neighbor pairing interaction ($V_{0}=0$ and $V_{1}<0$), on the other hand, the conclusion depends on the value of $Q$: Assuming $Q\in(0,\pi)$, the RLO state is favored when $Q\in(\pi/3,2\pi/3)$, and the BCS state is favored for other values of $Q$. While the on-site pairing interaction is usually considered as natural for phonon-mediated pairing interaction \cite{cho12,bednik15}, the inter-site pairing interaction (e.g., the terms proportional to $V_{1}$) can also be more natural if the underlying pairing mechanism is strong electron correlation. Examples of the latter case include the $t-J$ type of models applied to cuprates and iron pnictides \cite{baskaran87,kotliar88,si08,seo08}. Suppose we can still make the weak-coupling approximation, the RLO state can also be the leading pairing instability if $Q\in(\pi/3,2\pi/3)$.

We now compare the mean-field energies of the RLO state and the BCS state at zero temperature. The full mean-field Hamiltonian is
\begin{equation}
H_{\text{BCS/RLO}}=\sum\limits_{\alpha}\sum\limits_{\mathbf{q}}\tilde{\xi}_{\mathbf{q}}a^{\dagger}_{\alpha}(\mathbf{q})a_{\alpha}(\mathbf{q})+H_{int}^{\text{BCS/RLO}},
\end{equation}
where $\tilde{\xi}_{\mathbf{q}}=\xi_{\mathbf{q}}-\mu=|t|q-\mu$. The mean-field energy is defined as the expectation value of the full mean-field Hamiltonian with respect to the ground state wave function. For the weak-coupling phase, it is enough to focus on the states within the pairing shell.
We have
\begin{equation}
\frac{E_{\text{RLO}}}{N}=\frac{1}{2\pi^{2}|t|^{3}}\int_{-\omega_{c}}^{\omega_{c}}d\xi(\xi+\mu)^{2}[\xi-\sqrt{\xi^{2}+|\Delta_{\alpha}|^{2}}] -\frac{|\Delta_{\alpha}|^{2}}{V_{LO}},
\end{equation}
for the RLO state, where $|\Delta_{\alpha}|$ is independent of $\alpha$ and is determined by
\begin{equation}
1=-\frac{V_{LO}}{4\pi^{2}|t|^{3}}\int_{-\omega_{c}}^{\omega_{c}}d\xi\frac{\xi^{2}+\mu^{2}}{\sqrt{\xi^{2}+|\Delta_{\alpha}|^{2}}}.
\end{equation}
For the BCS state we have
\begin{equation}
\frac{E_{\text{BCS}}}{N}=\frac{1}{2\pi^{2}|t|^{3}}\int_{-\omega_{c}}^{\omega_{c}}d\xi(\xi+\mu)^{2}f(\xi)-\frac{2|\Delta|^{2}}{V_{BCS}},
\end{equation}
where $f(\xi)=\xi-\int_{-1}^{1}dy\sqrt{\xi^{2}+4(1-y^{2})|\Delta|^{2}}$ and the BCS pairing amplitude is determined by
\begin{equation}
1=-\frac{V_{BCS}}{4\pi^{2}|t|^{3}}\int_{-\omega_{c}}^{\omega_{c}}d\xi\int_{-1}^{1}dy\frac{(\xi^{2}+\mu^{2})(1-y^{2})}{\sqrt{\xi^{2}+4(1-y^{2})|\Delta|^{2}}}.
\end{equation}

\begin{figure}[!htb]\label{fig11} \centering
\includegraphics[width=8.5cm,height=7.5cm]{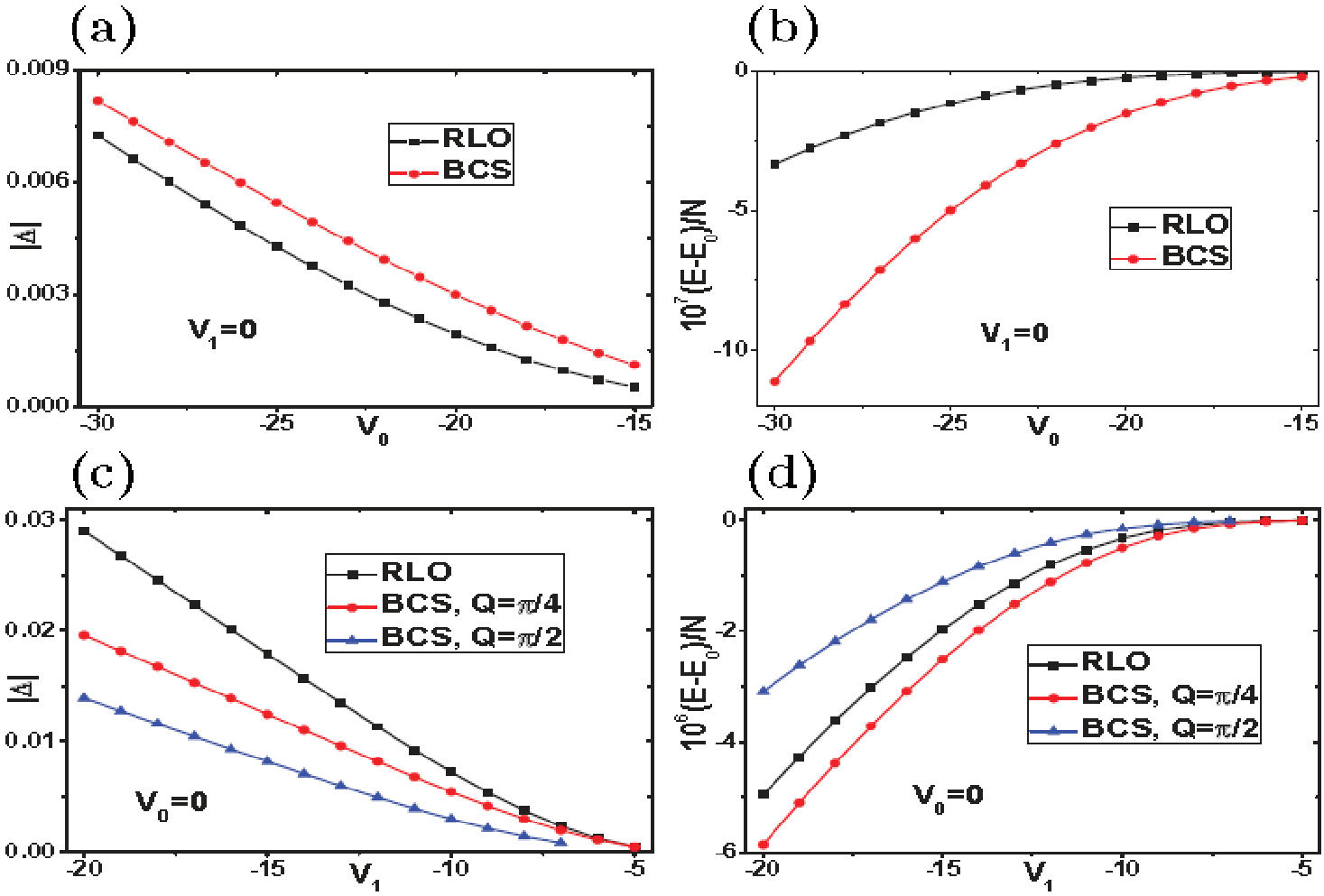} \\
\caption{(Color online) Comparison of mean-field energies and order parameter amplitudes for the RLO state and BCS state, in the weak-coupling limit. $\mu=0.5$, $\omega_{c}=0.05$, and $t=1$ are adopted in all calculations. (a) and (b) are results as a function of $V_{0}$ by setting $V_{1}=0$. (c) and (d) are results as a function of $V_{1}$ by setting $V_{0}=0$. The results in (a) and (b) are independent of $Q$. In (c) and (d), the results for the RLO state are also independent of $Q$. Two typical values of $Q$ are considered for the BCS state in (c) and (d). Two points are missing in (c) and (d) for the BCS state with $Q=\pi/2$, because no solutions for $\Delta$ are found there.}
\end{figure}

In the absence of pairing (i.e. $\Delta_{\alpha}=\Delta=0$), we have
\begin{equation}
\frac{E_{\text{BCS}}}{N}=\frac{E_{\text{RLO}}}{N}\equiv\frac{E_{0}}{N}.
\end{equation}
The numerical results for typical parameters are shown in Fig.11. With only $V_{0}$ nonzero, the BCS state is clearly the ground state [Fig.11(b)]. On the other hand, when only $V_{1}$ is nonzero, the conclusion depends on the value of $Q$. As is shown in Fig.11(d), while for $Q=\frac{1}{4}\pi$ the ground state is still the BCS state, for $Q=\frac{1}{2}\pi$ the ground state change to the RLO state. According to the previous analysis of the leading pairing instability, $\frac{1}{2}\pi$ ($\frac{1}{4}\pi$) is within (outside of) the range of $(\pi/3,2\pi/3)$ where the RLO state is the leading pairing instability with higher $T_{c}$. Therefore, we have seen that calculation of the zero-temperature mean-field energy has yielded the same conclusion as the calculation of $T_{c}$. That is, the true ground state depends on the pairing mechanism. The RLO state could be the ground state if $V_{1}$ or a pairing interaction of similar nature is the underlying pairing mechanism.

In the above analysis of this subsection, we have assumed that the Weyl pockets are inversion symmetric with respect to the Weyl nodes. By this assumption and using the same $\omega_{c}$, the number of single-particle states contributing to the BCS state is equal to the number of single-particle states contributing to the RLO state, which facilitates the comparison between the two states. Upon incorporating the tilting of the Weyl pockets along $k_{z}$, the number of single-particle states contributing to the RLO state becomes smaller than the number of single-particle states contributing to the BCS state, as was first pointed out by Bednik \emph{et al} \cite{bednik15}. This factor is thus adverse to the RLO state. But for a pairing interaction such as Eq.(E1) with $V_{0}=0$ and $V_{1}<0$, and for $Q$ lying in an appropriate range, the RLO state still has a chance to be the ground state.

\end{appendix}



\begin{references}
\bibitem{footnote1} We have exchanged $k_{x}$ and $k_{y}$ in the term proportional to $t'$, which ensures that the model has a four-fold rotational symmetry around the $z$ axis.
\bibitem{note2} Note that, a model which breaks both inversion symmetry and time-reversal symmetry not definitely lead to $E_{n}(\mathbf{k})\ne E_{n}(-\mathbf{k})$. For example, the model defined by Eq.(1) breaks the inversion symmetry and time-reversal symmetry (see Appendix A for more details), but the band structure defined by Eq.(3) is centrosymmetric with respect to the center of the BZ.
\end{references}
\end{document}